\documentclass[sigconf, nonacm]{acmart}

\usepackage[compress]{cleveref}
\usepackage{bbm}
\usepackage{xspace}
\usepackage{subcaption}
\usepackage[short]{optidef}
\usepackage{array}
\usepackage[noend]{algpseudocode}
\usepackage{algorithm}
\usepackage{microtype} %
\usepackage{bm}
\usepackage{setspace}

\crefname{section}{Section}{Sections}
\Crefname{section}{Section}{Sections}
\crefname{figure}{Figure}{Figures}
\Crefname{figure}{Figure}{Figures}
\crefname{subfigure}{Figure}{Figures}
\Crefname{subfigure}{Figure}{Figures}
\crefrangelabelformat{subfigure}{#3#1#4--#5(\crefstripprefix{#1}{#2}#6}

\newcommand{\maureen}[1]{\textcolor{orange}{[[Maureen: #1]]}}

\newcommand{\system}{VOCALExplore\xspace}
\newcommand{\method}[1]{\textsc{#1}}

\newcommand{\rd}{\textsc{R3D}\xspace}
\newcommand{\mvit}{\textsc{MViT}\xspace}
\newcommand{\random}{\textsc{Random}\xspace}
\newcommand{\clip}{\textsc{CLIP}\xspace}
\newcommand{\clippool}{\textsc{CLIP (Pooled)}\xspace}
\newcommand{\coreset}{\textsc{Coreset}\xspace}
\newcommand{\clustermargin}{\textsc{Cluster-Margin}\xspace}
\newcommand{\ve}{\textsc{VE-sample}\xspace}
\newcommand{\vecm}{\textsc{VE-sample (CM)}\xspace}
\newcommand{\deer}{\textsc{Deer}\xspace}

\newcommand{\ktw}{\textsc{K20}\xspace}
\newcommand{\ktwsk}{\textsc{K20 (skew)}\xspace}
\newcommand{\bdd}{\textsc{BDD}\xspace}
\newcommand{\bears}{\textsc{Bears}\xspace}
\newcommand{\charades}{\textsc{Charades}\xspace}

\newcommand{\nsqueeze}[1]{\textls[-10]{#1}}

\newcommand{\revision}[2]{#2}

\setcopyright{none}
\renewcommand\footnotetextcopyrightpermission[1]{}
\AtBeginDocument{%
  \providecommand\BibTeX{{%
    \normalfont B\kern-0.5em{\scshape i\kern-0.25em b}\kern-0.8em\TeX}}}

\begin{document}

\title{VOCALExplore: Pay-as-You-Go \\ Video Data Exploration and Model Building}
\subtitle{Technical Report\vspace{-0.5em}}
\author{Maureen Daum$^1$, Enhao Zhang$^1$, Dong He$^1$, Stephen Mussmann$^1$, \\ Brandon Haynes$^2$, Ranjay Krishna$^1$, and Magdalena Balazinska$^1$}

\affiliation{%
  \institution{$^1$University of Washington,\quad \{mdaum, enhaoz,
    donghe, mussmann, ranjay, magda\}@cs.washington.edu}
  \city{}
  \country{}}

\affiliation{%
  \institution{$^2$Microsoft Gray Systems Lab,\quad brandon.haynes@microsoft.com}
  \city{}
  \country{}}

\renewcommand{\shortauthors}{}

\maketitle

\newcommand{\architectureFigure}{
    \begin{figure*}[t!]
        \centering
        \includegraphics[width=0.7\textwidth]{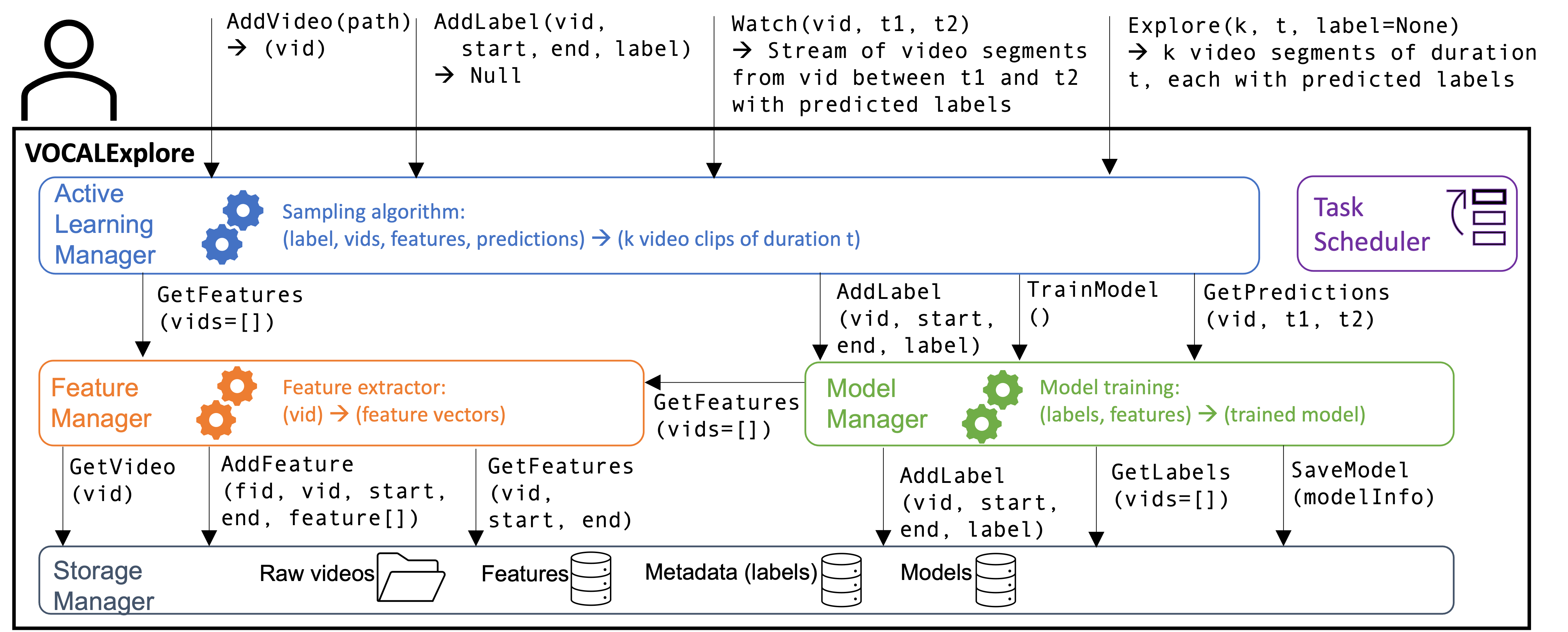}
        \vspace{-1em}
        \caption{\system architecture. The Task Scheduler coordinates the activities of the various managers.}
        \label{fig:architecture}
        \vspace{-1em}
    \end{figure*}
}

\newcommand{\workflowFigure}{
    \begin{figure*}[t!]
        \centering
        \includegraphics[width=0.95\textwidth]{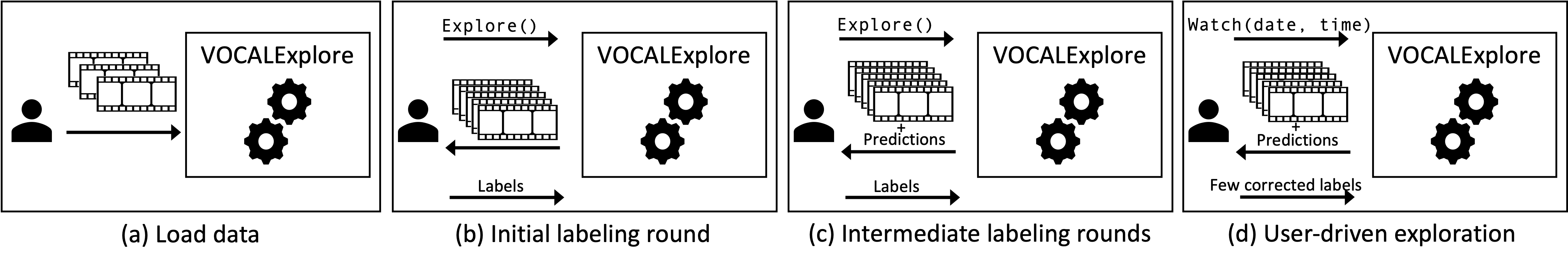}
        \vspace{-1.5em}
        \caption{\system workflow.}
        \label{fig:workflow}
    \end{figure*}
}

\newcommand{\smoothingKineticsUniformCurves}{
    \begin{figure}[t!]
        \centering
        \includegraphics[width=0.95\columnwidth]{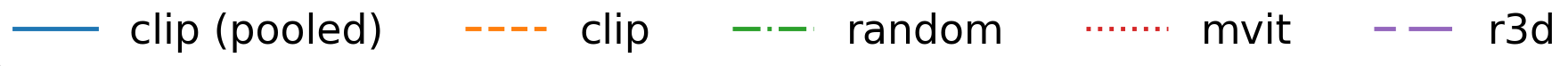}

        \includegraphics[width=0.95\columnwidth]{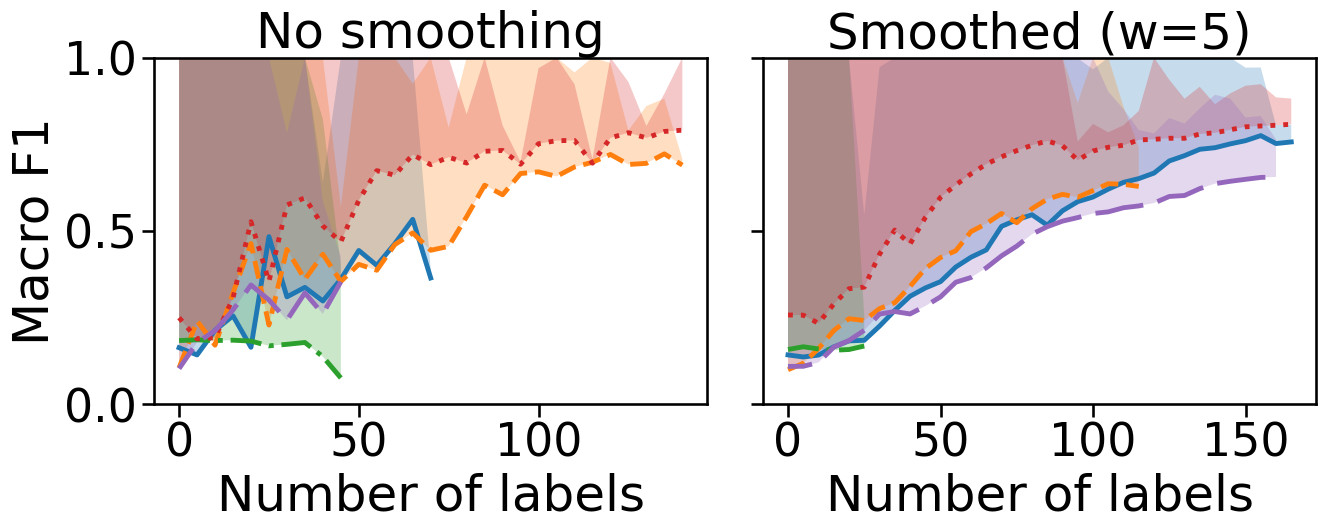}
        \caption{Feature selection progress for the K7m4-10 unskewed dataset showing one example without and one example with smoothing. Lower bounds are shown as the solid lines, and upper bounds are marked by the shaded area. Feature selection was performed with $C=5$ and $T=50$, and performance was estimated with the macro F1 score. \maureen{I think it's confusing that I'm showing graphs for two different runs; I should update the right graph to be a smoothed version of the left one.}}
        \label{fig:smoothingKineticsUniformCurves}
    \end{figure}
}

\newcommand{\featureSelectionStepFigure}{
    \begin{figure}[t!]
        \centering
        \includegraphics[width=0.95\columnwidth]{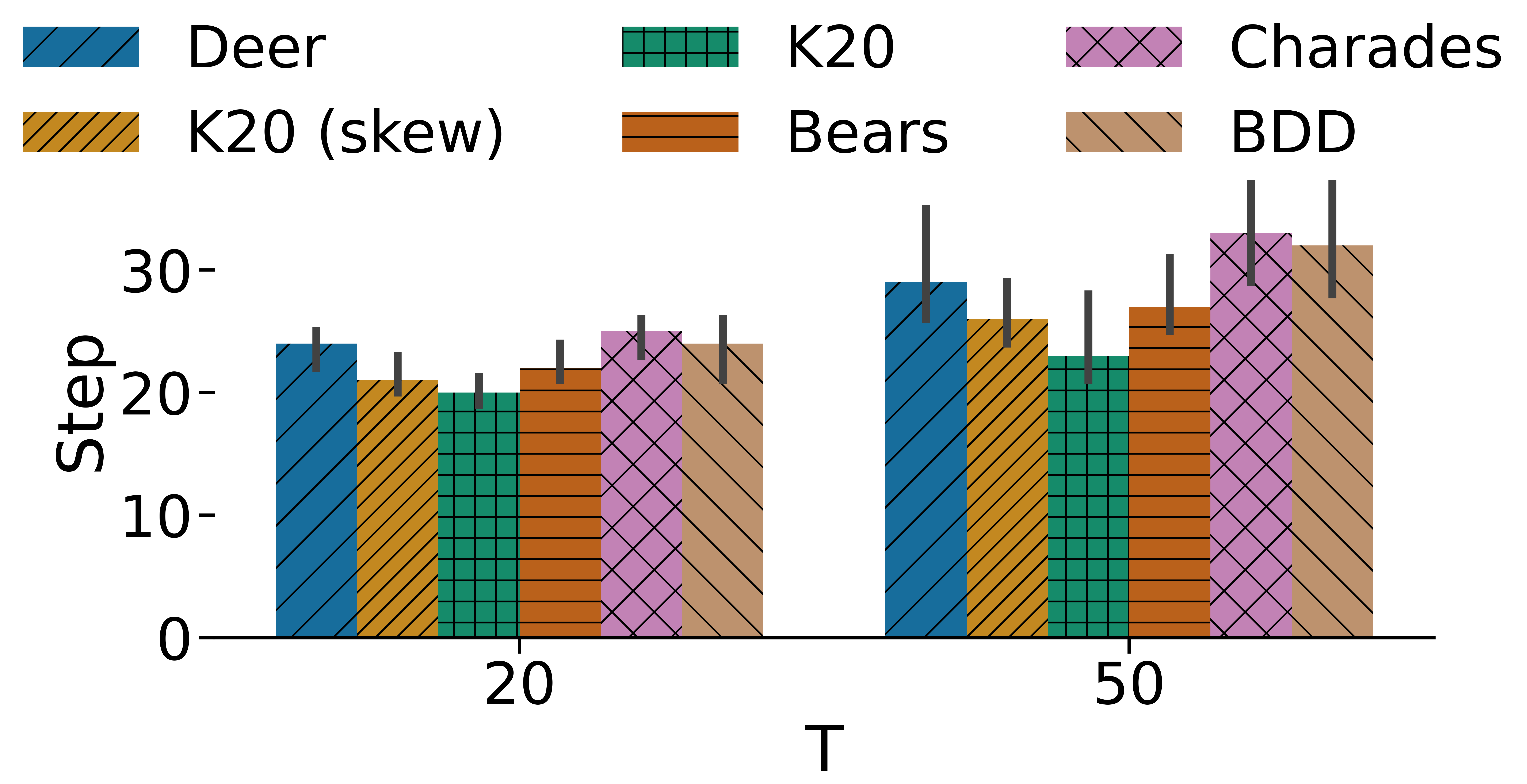}
        \caption{Median feature selection step when $C=5$ and $w=5$. Error bars show the interquartile range. \system converges to a single feature within a small number of steps.}
        \label{fig:featureSelectionStepFigure}
    \end{figure}
}

\newcommand{\nclassesSeenDeerK}{
    \begin{figure}[t!]
        \centering
        \begin{subfigure}{0.45\columnwidth}
            \includegraphics[width=\textwidth]{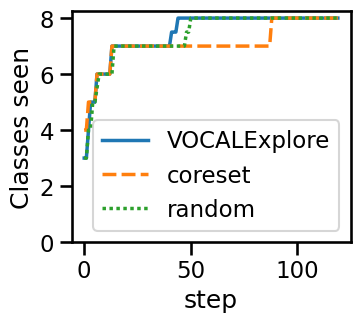}
            \caption{Deer}
            \label{subfig:deernclasses}
        \end{subfigure}
        \begin{subfigure}{0.45\columnwidth}
            \includegraphics[width=\textwidth]{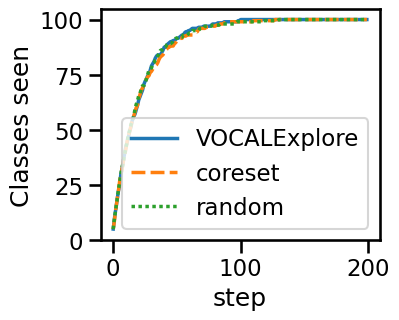}
            \caption{K7M4-100}
            \label{subfig:k7m4nclasses}
        \end{subfigure}
        \caption{Number of classes seen by each \method{Explore} step.}
    \end{figure}
}

\newcommand{\nclassesSeenKTen}{
    \begin{figure}[t!]
        \centering
        \includegraphics[width=\columnwidth]{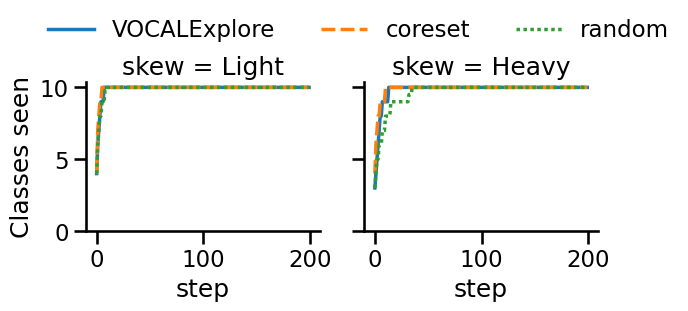}
        \caption{Number of classes seen by each \method{Explore} step for K7M4-10.}
        \label{fig:k7m410classes}
    \end{figure}
}

\newcommand{\nclassesSeenKTwenty}{
    \begin{figure}[t!]
        \centering
        \includegraphics[width=\columnwidth]{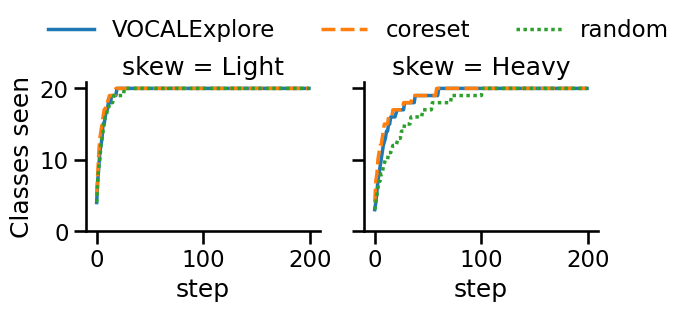}
        \caption{Number of classes seen by each \method{Explore} step for K7M4-20.}
        \label{fig:k7m420classes}
    \end{figure}
}

\newcommand{\labelcountsDeer}{
    \begin{figure}[t!]
        \centering
        \includegraphics[width=\columnwidth]{figures/labelcounts-deer.png}
        \caption{Number of examples of each activity type returned from \method{Explore} for the deer dataset.}
        \label{fig:labelcountsdeer}
    \end{figure}
}

\newcommand{\labelapDeer}{
    \begin{figure}[t!]
        \centering
        \includegraphics[width=\columnwidth]{figures/labelap-deer.png}
        \caption{Average precision for activities in the deer dataset.}
        \label{fig:labelapdeer}
    \end{figure}
}

\newcommand{\aggacc}{
    \begin{figure}[t!]
        \centering
        \begin{subfigure}{\columnwidth}
            \includegraphics[width=\textwidth]{figures/aggacc-k7m4.png}
            \caption{K7M4-100}
            \label{subfig:aggacck7m4}
        \end{subfigure}

        \begin{subfigure}{\columnwidth}
            \includegraphics[width=\textwidth]{figures/aggacc-k7m4-10.png}
            \caption{K7M4-10}
            \label{subfig:aggacck7m410}
        \end{subfigure}

        \begin{subfigure}{\columnwidth}
            \includegraphics[width=\textwidth]{figures/aggacc-k7m4-20.png}
            \caption{K7M4-20}
            \label{subfig:aggacck7m420}
        \end{subfigure}
        \caption{Aggregate accuracy}
        \label{fig:aggacc}
    \end{figure}
}

\newcommand{\featureRankingFigure}{
    \begin{figure*}[t!]
        \centering
        \includegraphics[width=0.45\textwidth]{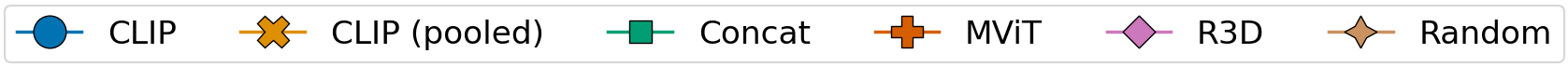}

        \begin{subfigure}{0.25\textwidth}
            \includegraphics[width=\textwidth]{figures/aggf1-randomifuniform-deer.png}
            \vspace{-2em}
            \caption{\textsc{Deer}}
            \label{subfig:aggf1deer}
        \end{subfigure}
        \begin{subfigure}{0.25\textwidth}
            \includegraphics[width=\textwidth]{figures/aggf1-randomifuniform-k7m4-20-noskew.png}
            \vspace{-2em}
            \caption{\textsc{K20}}
            \label{subfig:aggf1k20}
        \end{subfigure}
        \begin{subfigure}{0.25\textwidth}
            \includegraphics[width=\textwidth]{figures/aggf1-randomifuniform-k7m4-20-zipf2.png}
            \vspace{-2em}
            \caption{\textsc{K20 (skew)}}
            \label{subfig:aggf1k20skew}
        \end{subfigure}

        \vspace{-2em}
        \caption{Macro F1 score when using the \ve sampling method showing that the best feature varies across datasets. ``Concat'' refers to concatenating all of the features into a single feature vector.}
        \label{fig:featureRankingFigure}
    \end{figure*}
}

\newcommand{\figureSampleSelectionFigure}{
    \begin{figure}[t!]
        \centering
        \includegraphics[width=0.6\columnwidth]{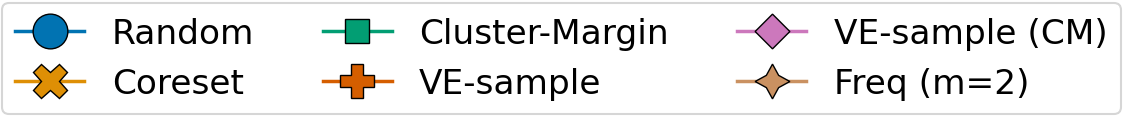}

        \begin{subfigure}{0.48\columnwidth}
            \includegraphics[width=\textwidth]{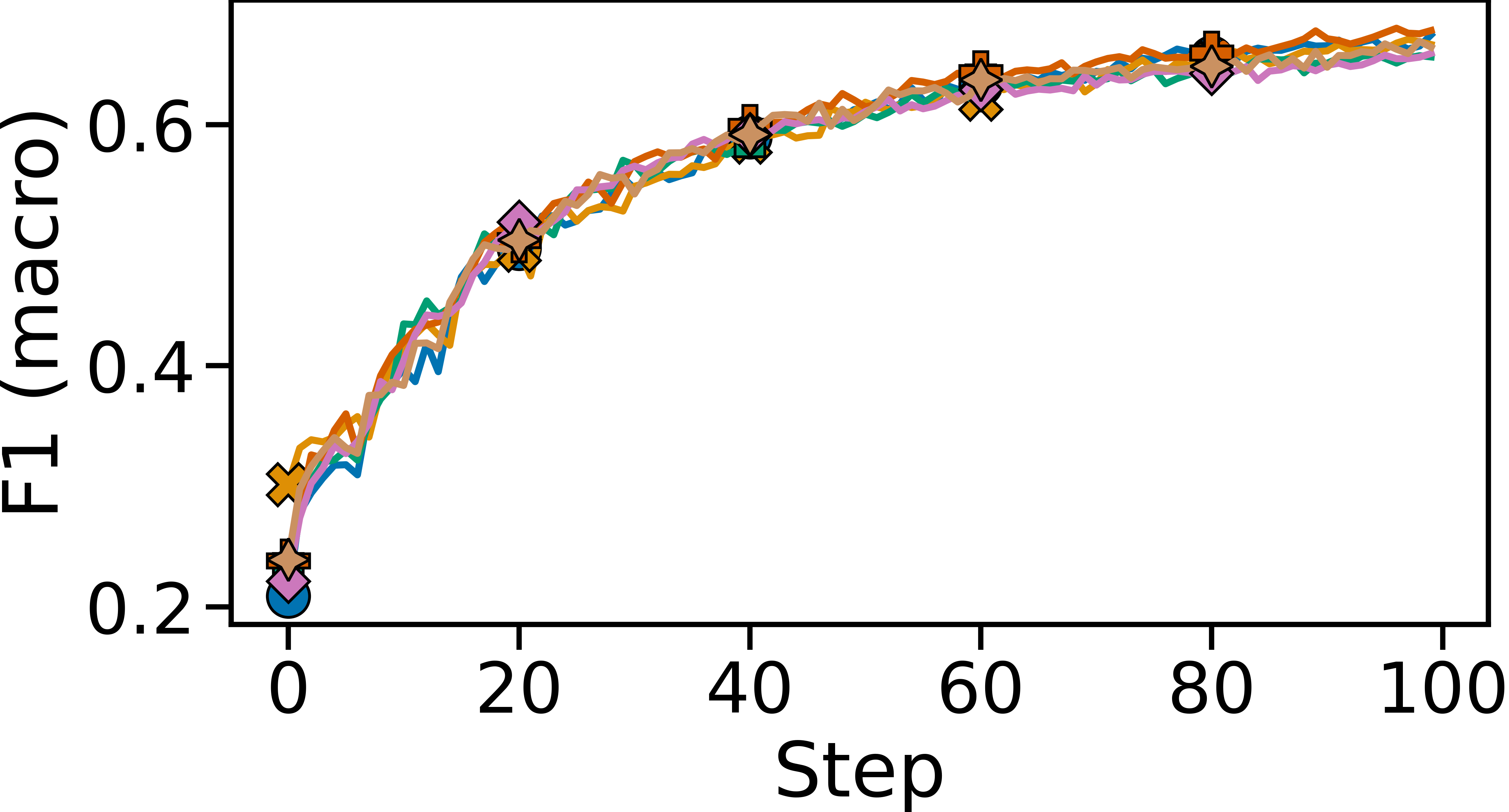}
            \vspace{-2em}
            \caption{\deer F1}
        \end{subfigure}
        \begin{subfigure}{0.48\columnwidth}
            \includegraphics[width=\textwidth]{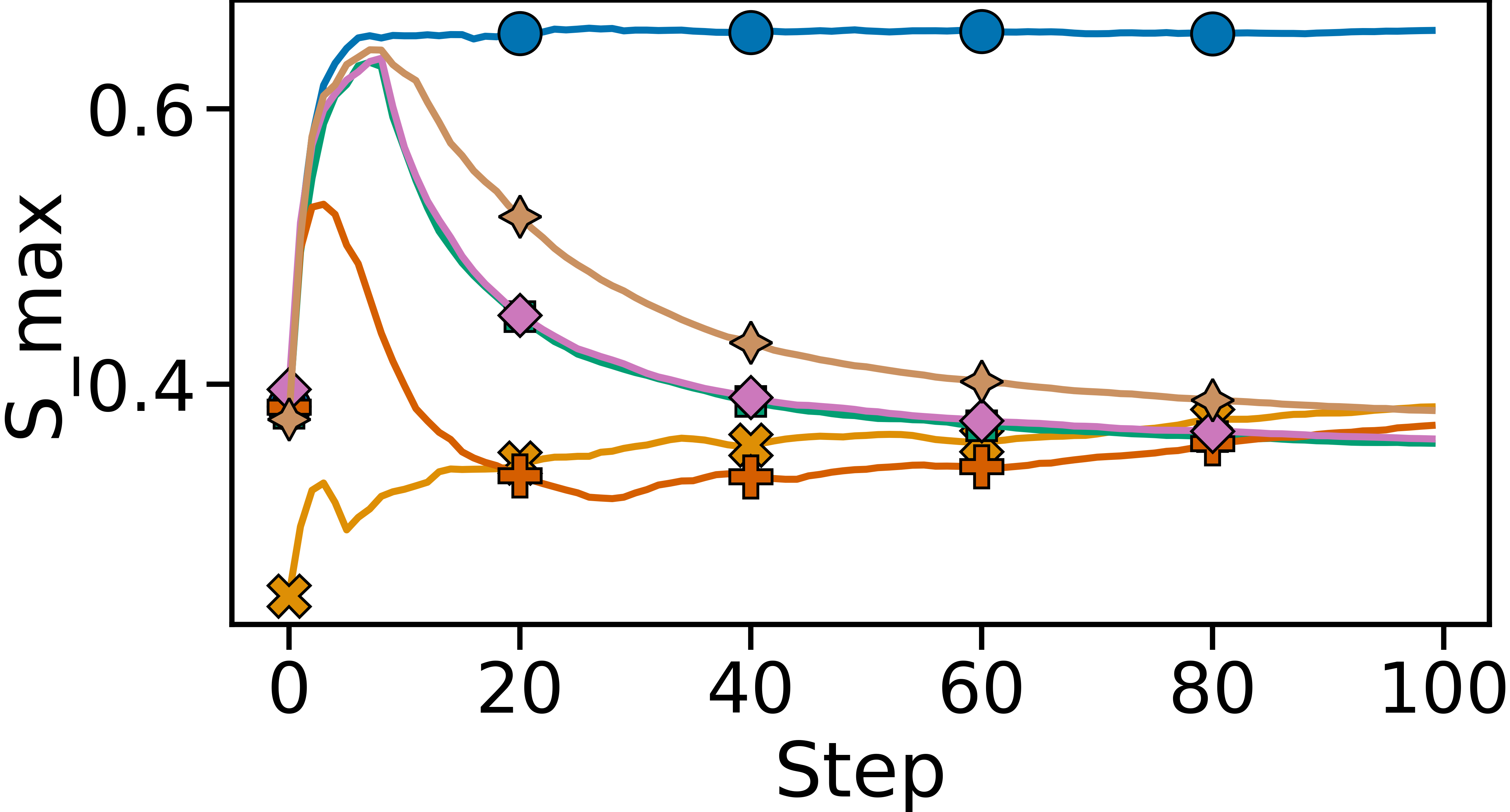}
            \vspace{-2em}
            \caption{\deer $S_{max}$}
        \end{subfigure}

        \begin{subfigure}{0.48\columnwidth}
            \includegraphics[width=\textwidth]{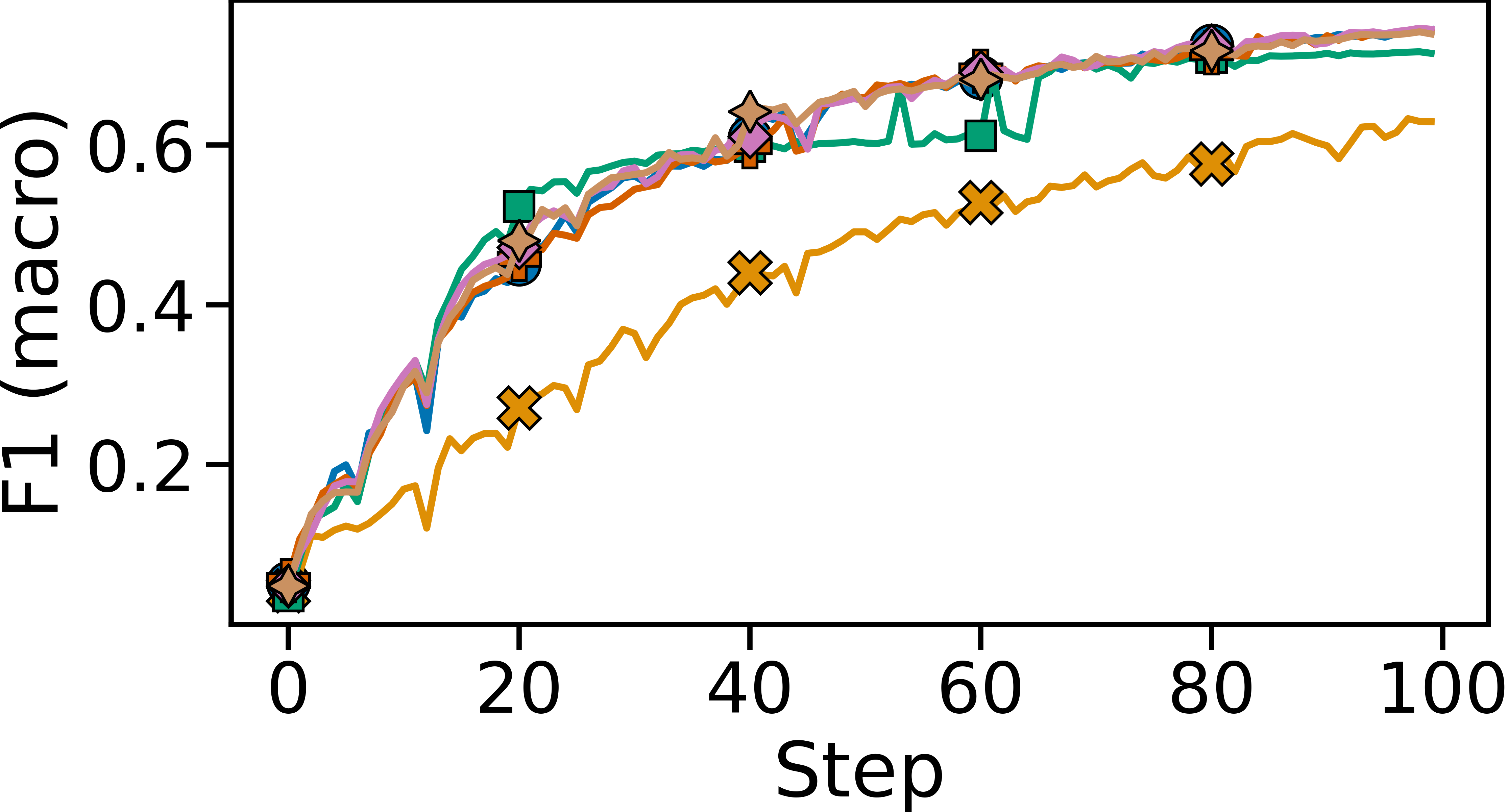}
            \vspace{-2em}
            \caption{\ktw F1}
        \end{subfigure}
        \begin{subfigure}{0.48\columnwidth}
            \includegraphics[width=\textwidth]{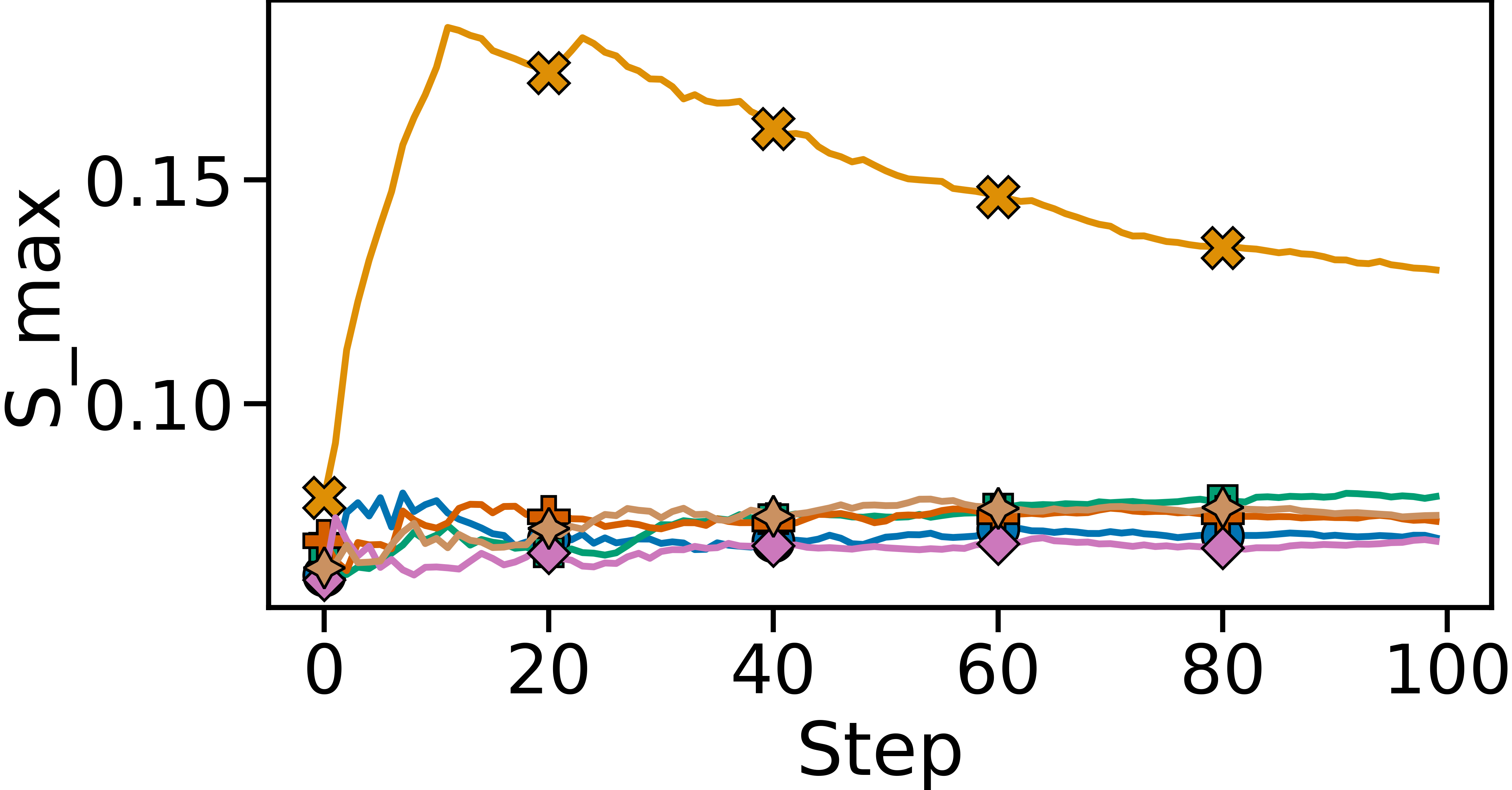}
            \vspace{-2em}
            \caption{\ktw $S_{max}$}
        \end{subfigure}

        \begin{subfigure}{0.48\columnwidth}
            \includegraphics[width=\textwidth]{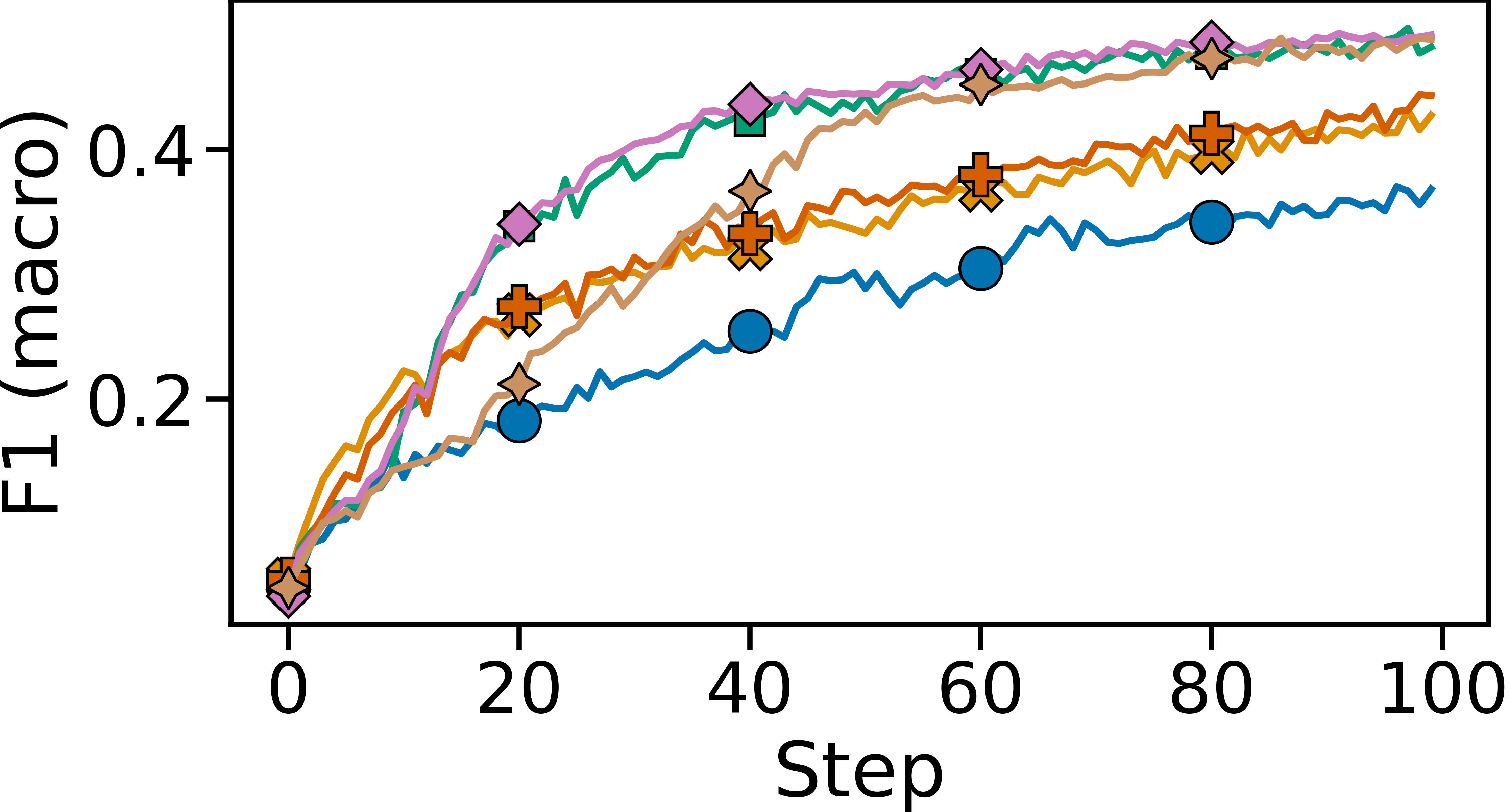}
            \vspace{-2em}
            \caption{\ktwsk F1}
        \end{subfigure}
        \begin{subfigure}{0.48\columnwidth}
            \includegraphics[width=\textwidth]{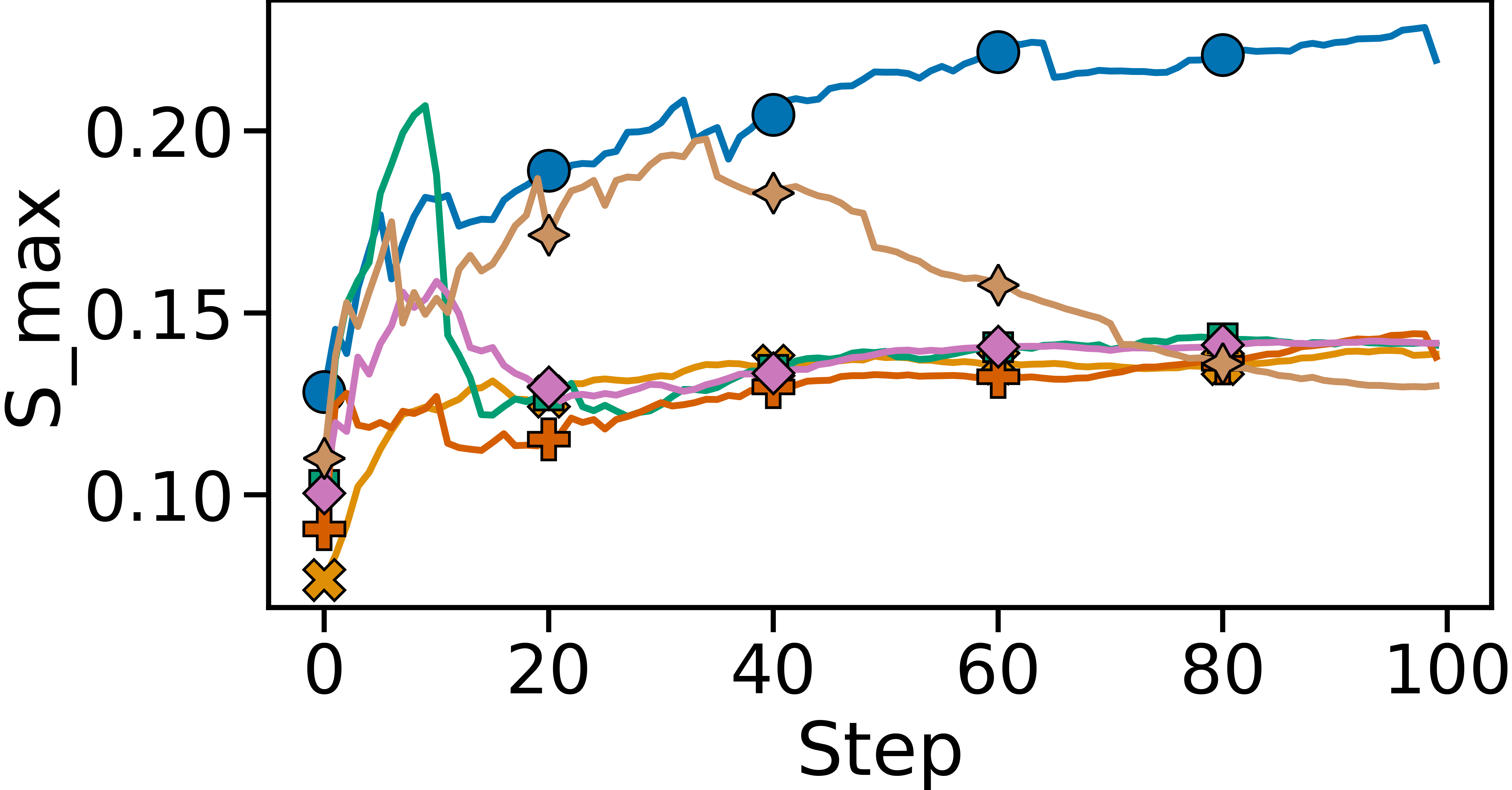}
            \vspace{-2em}
            \caption{\ktwsk $S_{max}$}
        \end{subfigure}

        \begin{subfigure}{0.48\columnwidth}
            \includegraphics[width=\textwidth]{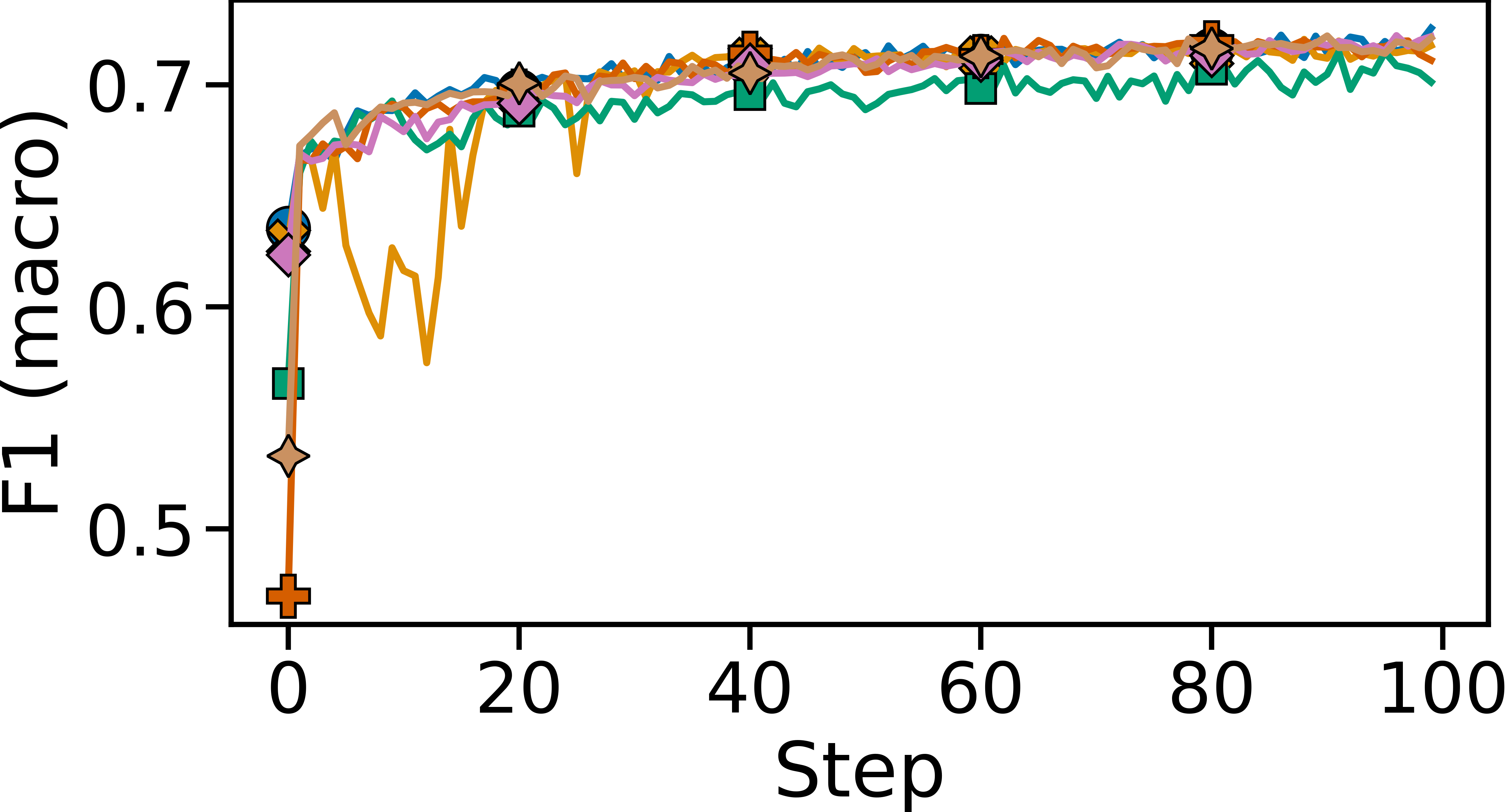}
            \vspace{-2em}
            \caption{\bears F1}
        \end{subfigure}
        \begin{subfigure}{0.48\columnwidth}
            \includegraphics[width=\textwidth]{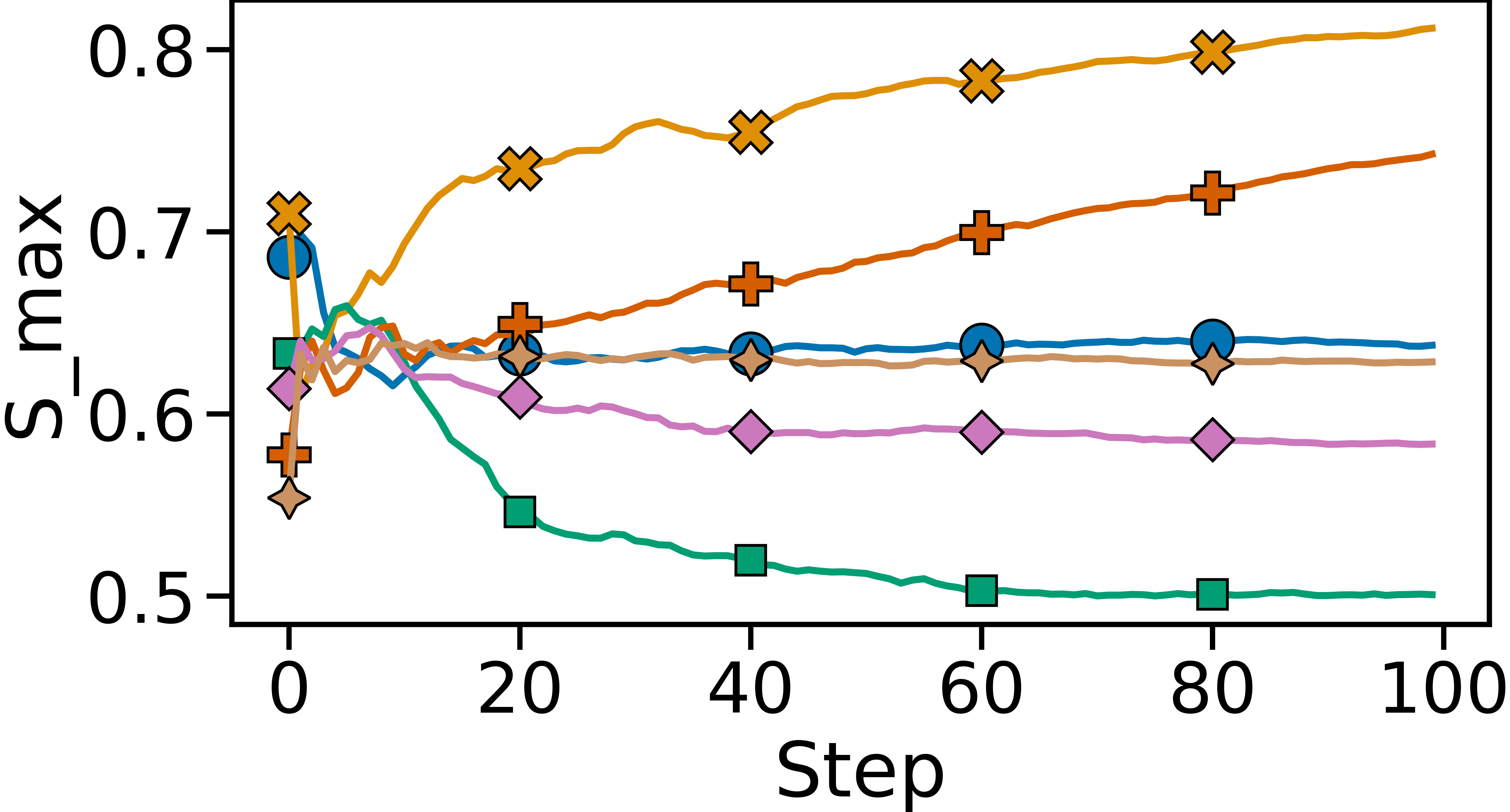}
            \vspace{-2em}
            \caption{\bears $S_{max}$}
        \end{subfigure}

        \begin{subfigure}{0.48\columnwidth}
            \includegraphics[width=\textwidth]{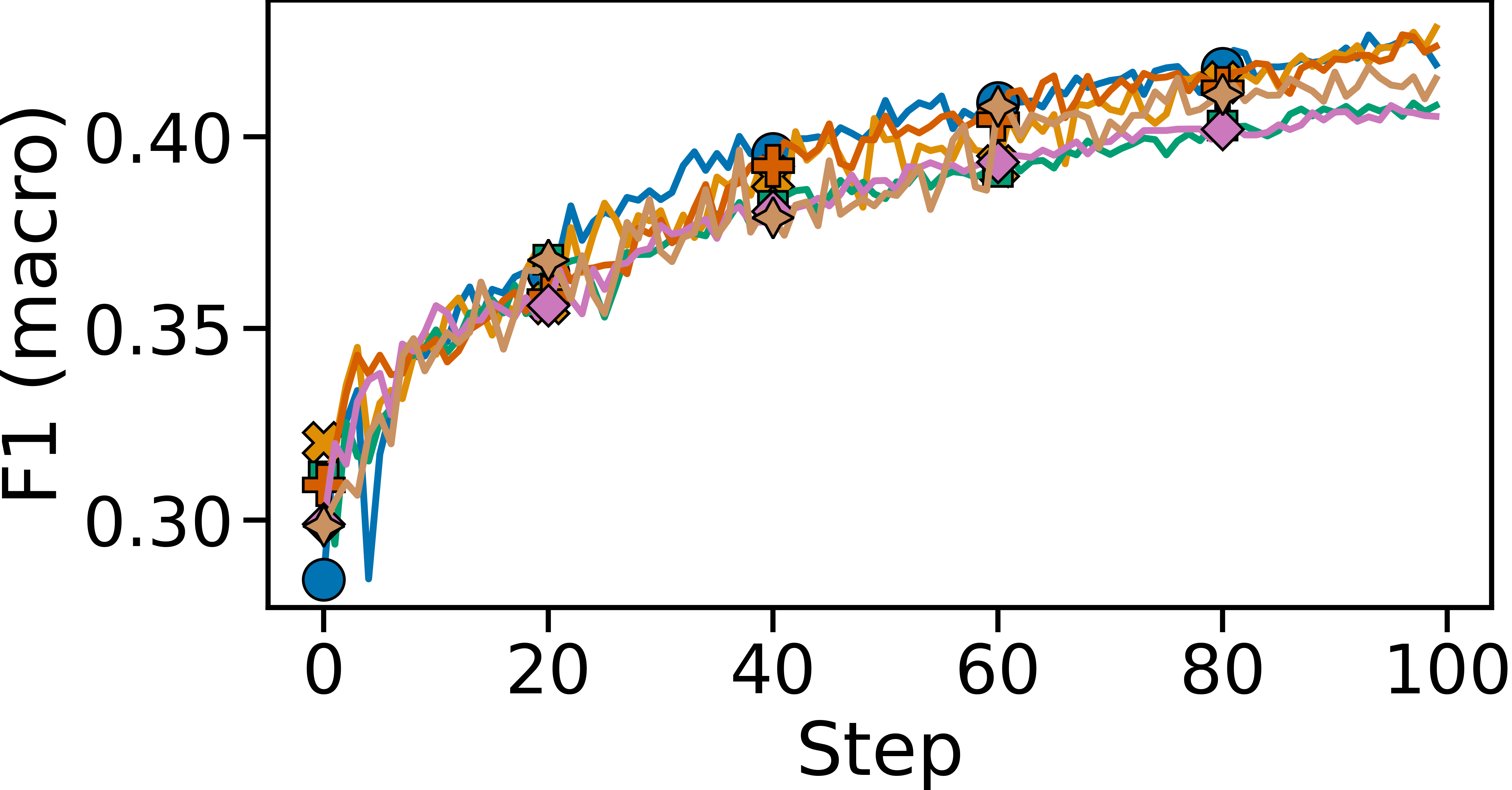}
            \vspace{-2em}
            \caption{\bdd F1}
        \end{subfigure}
        \begin{subfigure}{0.48\columnwidth}
            \includegraphics[width=\textwidth]{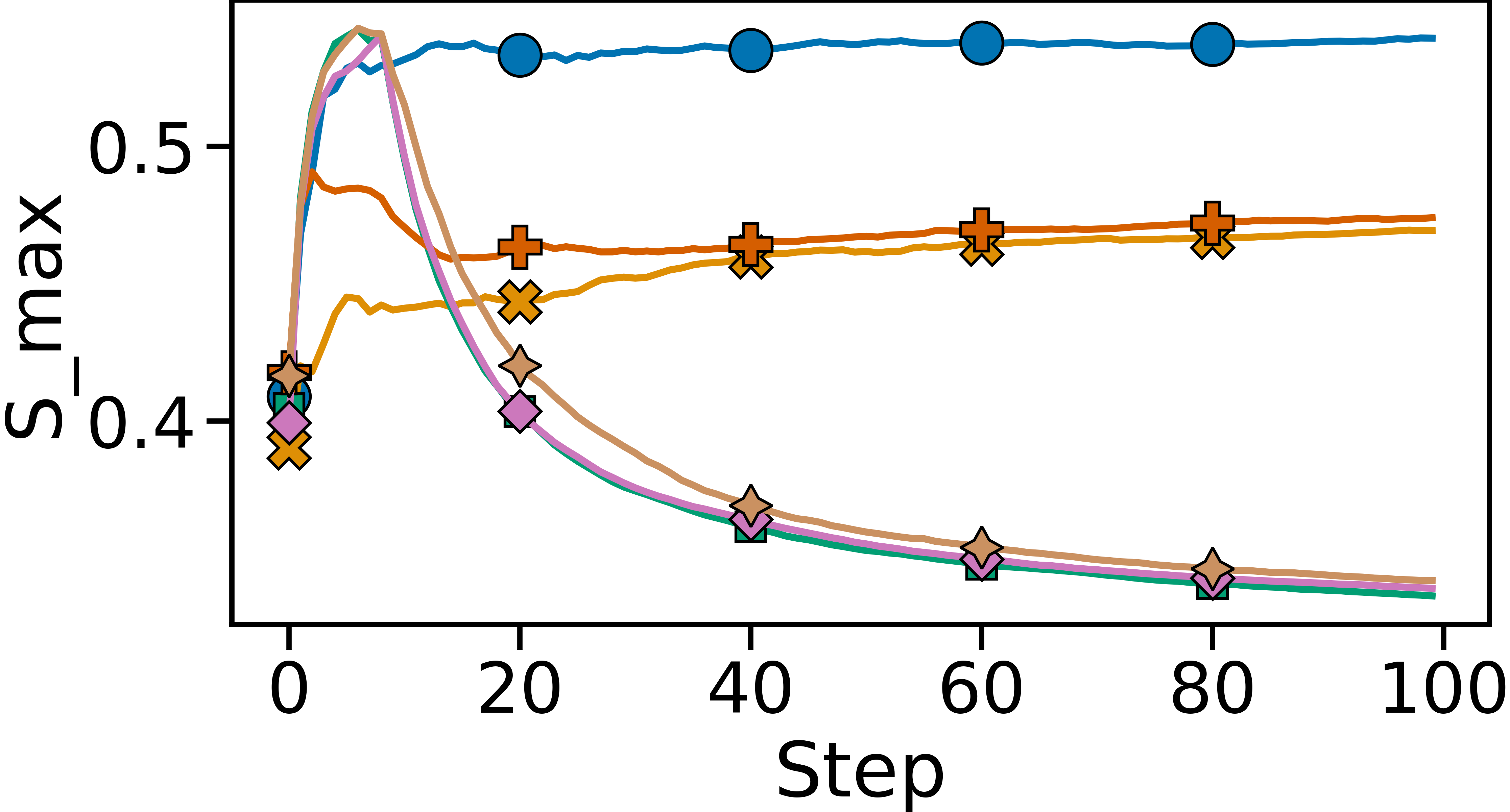}
            \vspace{-2em}
            \caption{\bdd $S_{max}$}
        \end{subfigure}

        \begin{subfigure}{0.48\columnwidth}
            \includegraphics[width=\textwidth]{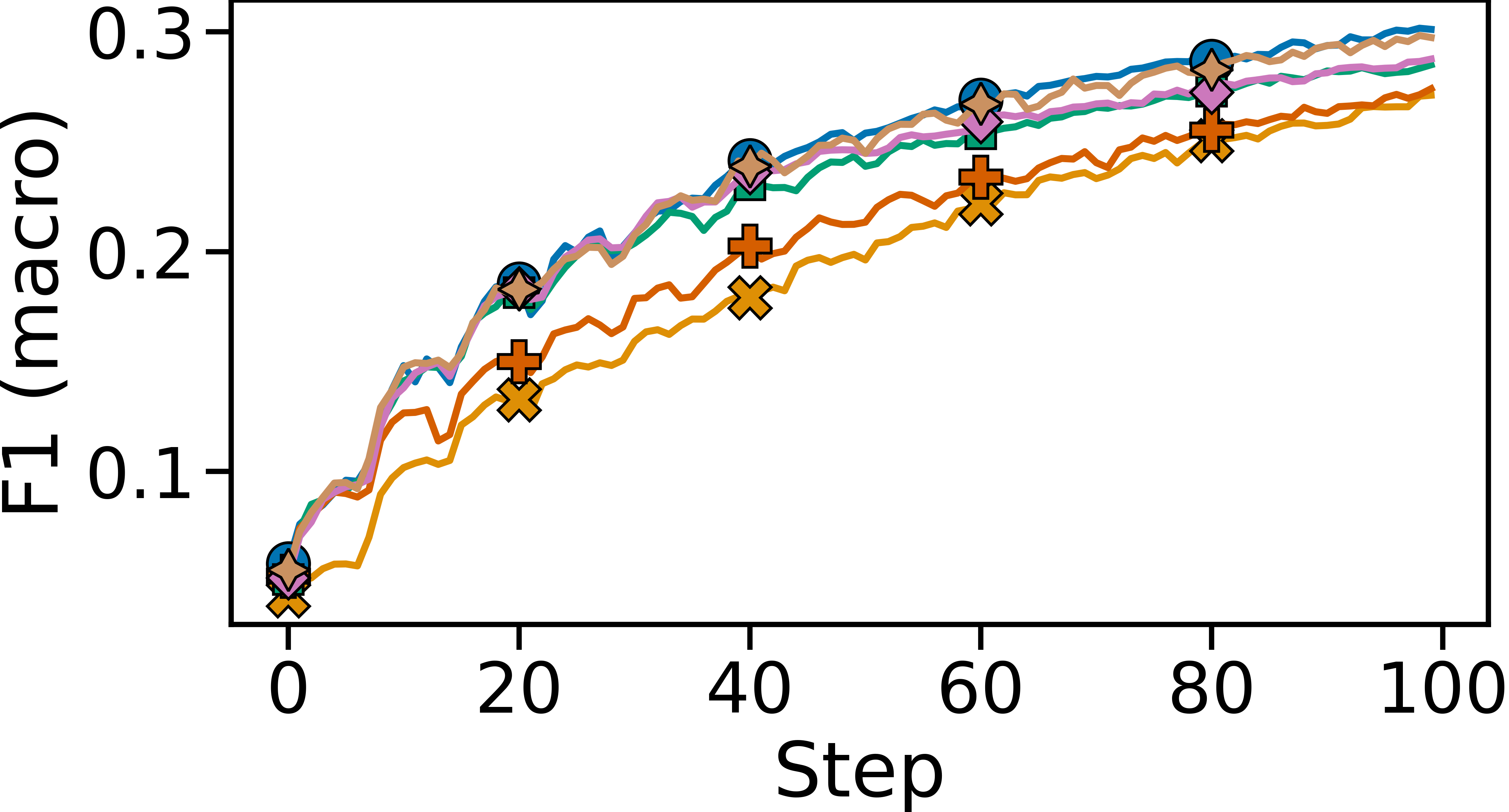}
            \vspace{-2em}
            \caption{\charades F1}
        \end{subfigure}
        \begin{subfigure}{0.48\columnwidth}
            \includegraphics[width=\textwidth]{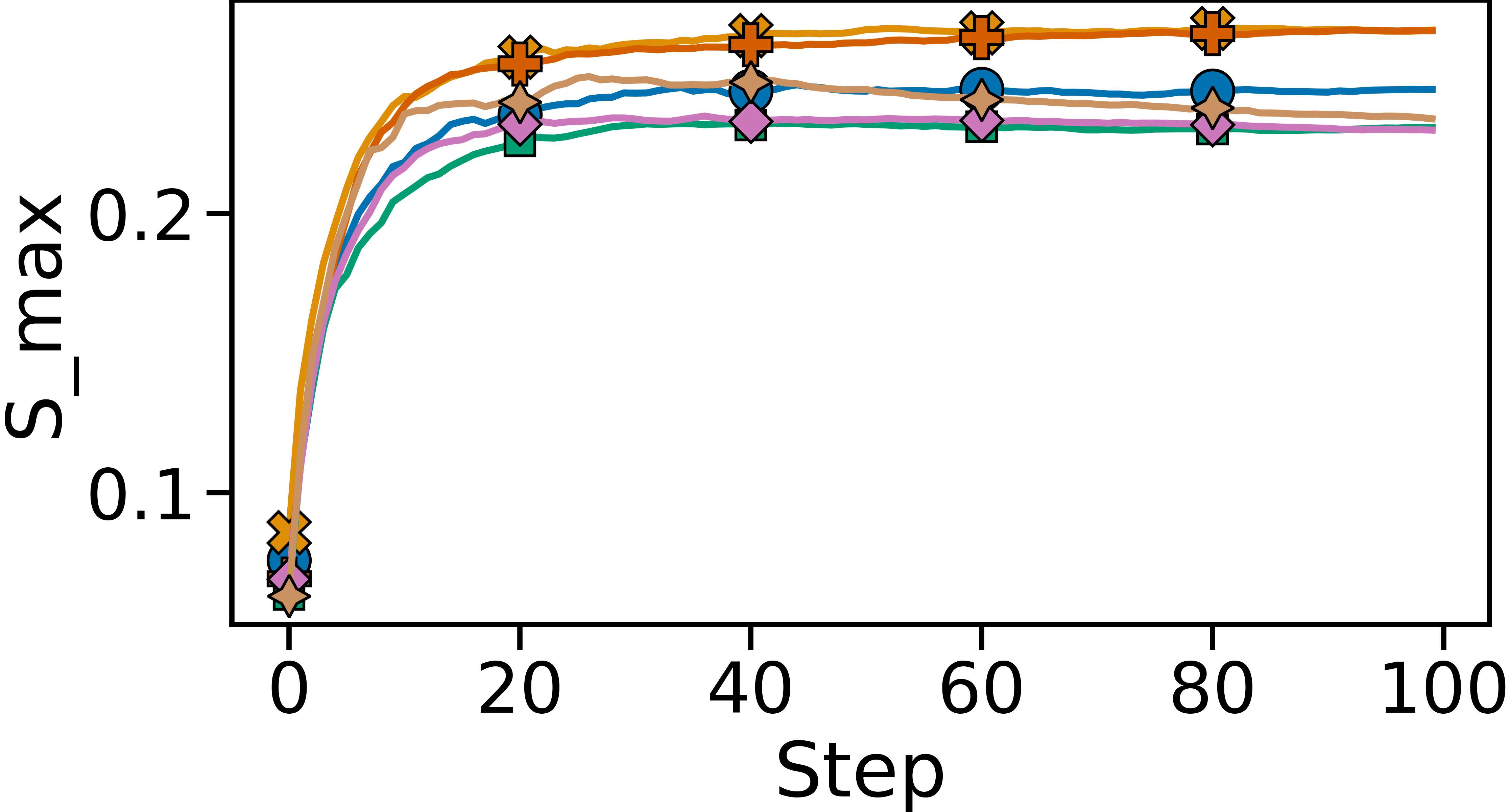}
            \vspace{-2em}
            \caption{\charades $S_{max}$}
        \end{subfigure}

        \vspace{-1em}
        \caption{\revision{MR-6}{\system's data sampling method yields models with the highest F1 scores and samples from a diverse set of classes ($S_{max}$, lower is better) across datasets.}
        }
        \label{fig:sampleSelectionFigure}
        \vspace{-1em}
    \end{figure}
}

\newcommand{\featureSelectionFFigure}{
    \begin{figure}[t!]
            \centering
            \includegraphics[width=\columnwidth]{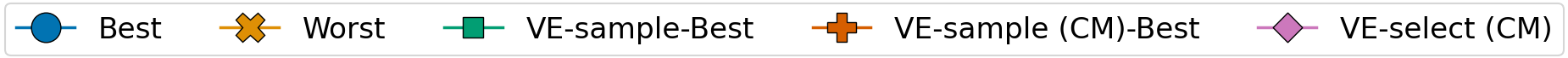}

            \begin{subfigure}{0.48\columnwidth}
                \includegraphics[width=\textwidth]{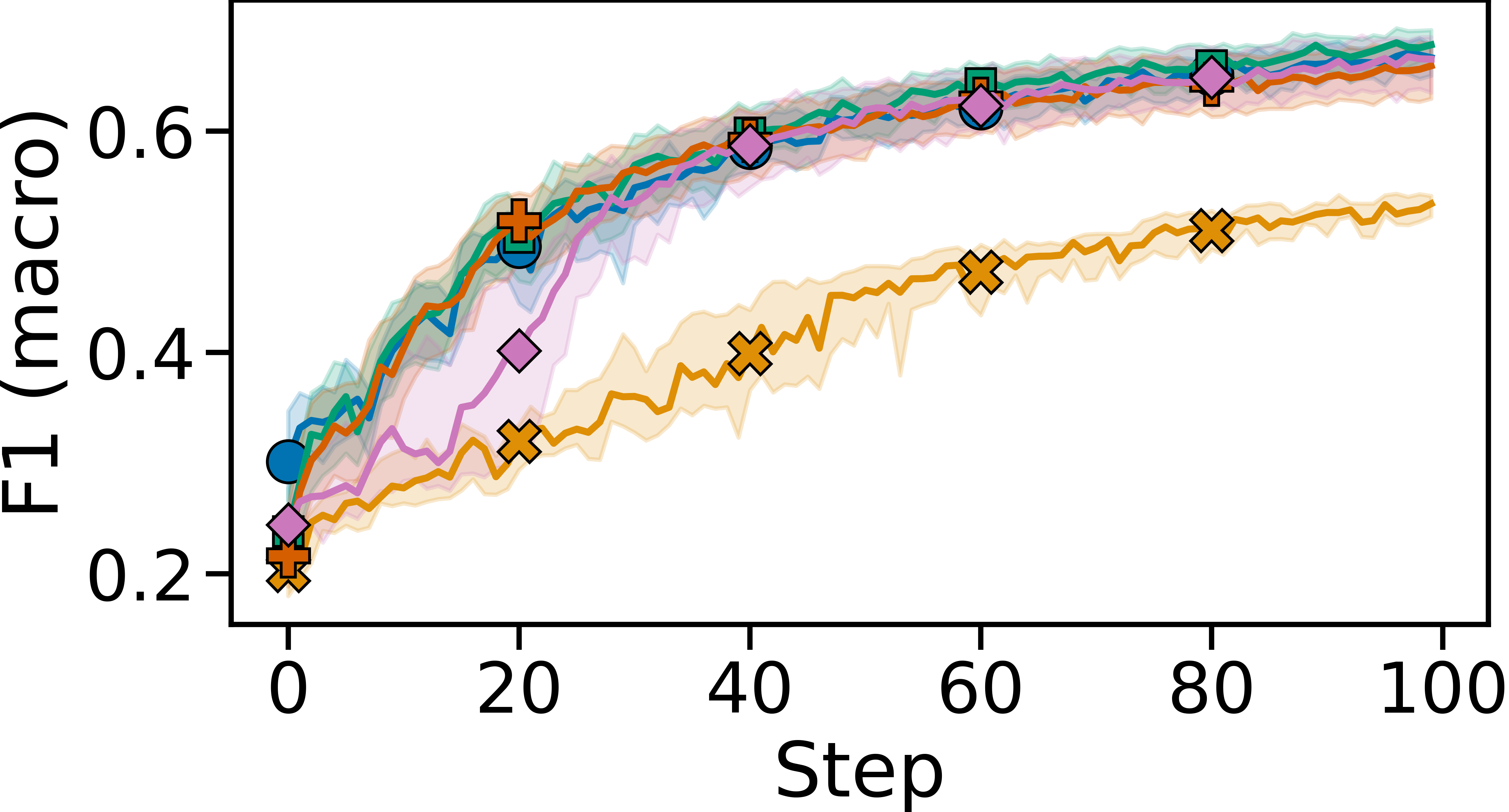}
                \vspace{-2em}
                \caption{\textsc{Deer}}
                \label{subfig:featureselectionf1deer}
            \end{subfigure}
            \begin{subfigure}{0.48\columnwidth}
                \includegraphics[width=\textwidth]{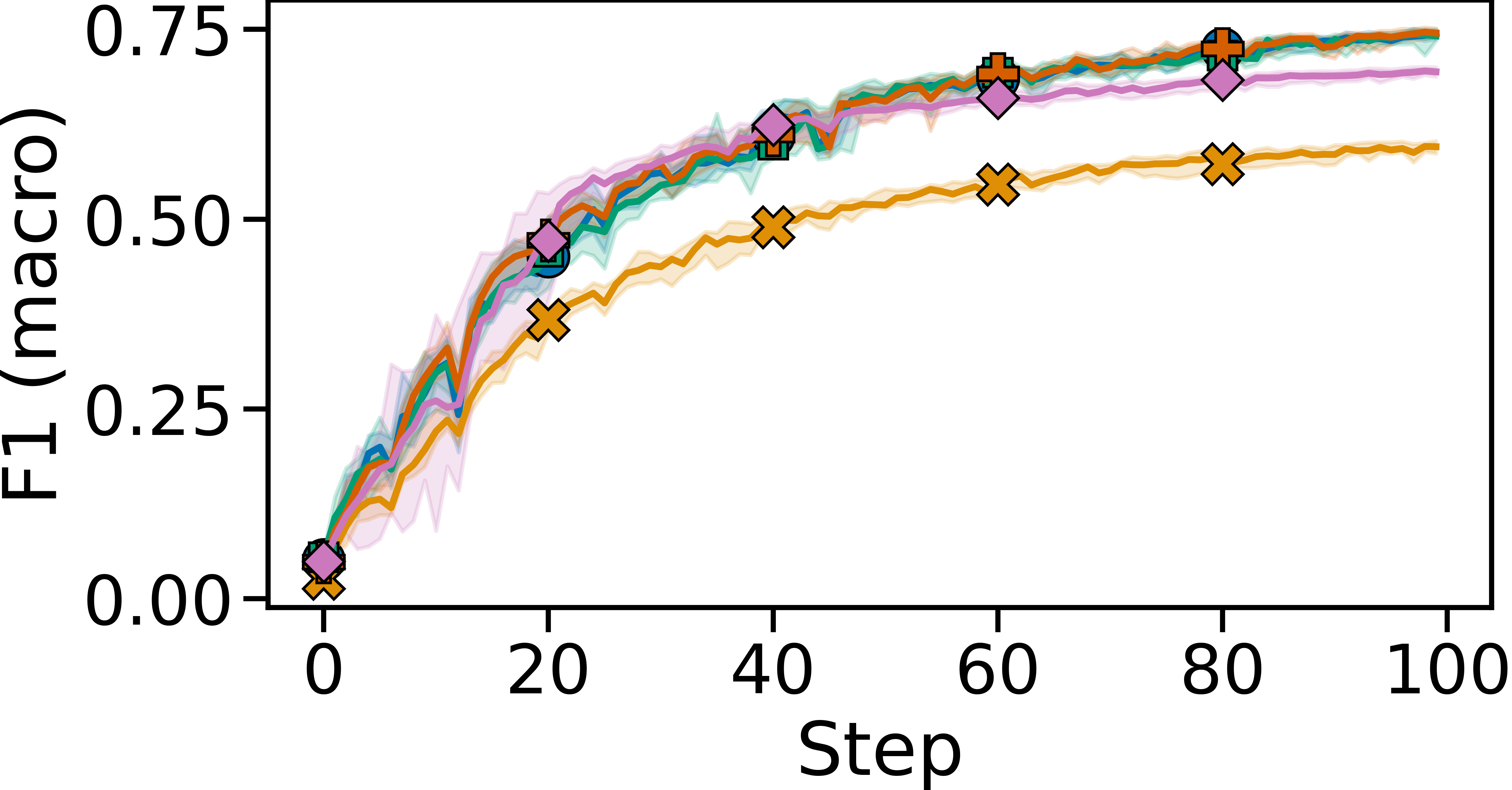}
                \vspace{-2em}
                \caption{\textsc{K20}}
                \label{subfig:featureselectionf1k20}
            \end{subfigure}

            \begin{subfigure}{0.48\columnwidth}
                \includegraphics[width=\textwidth]{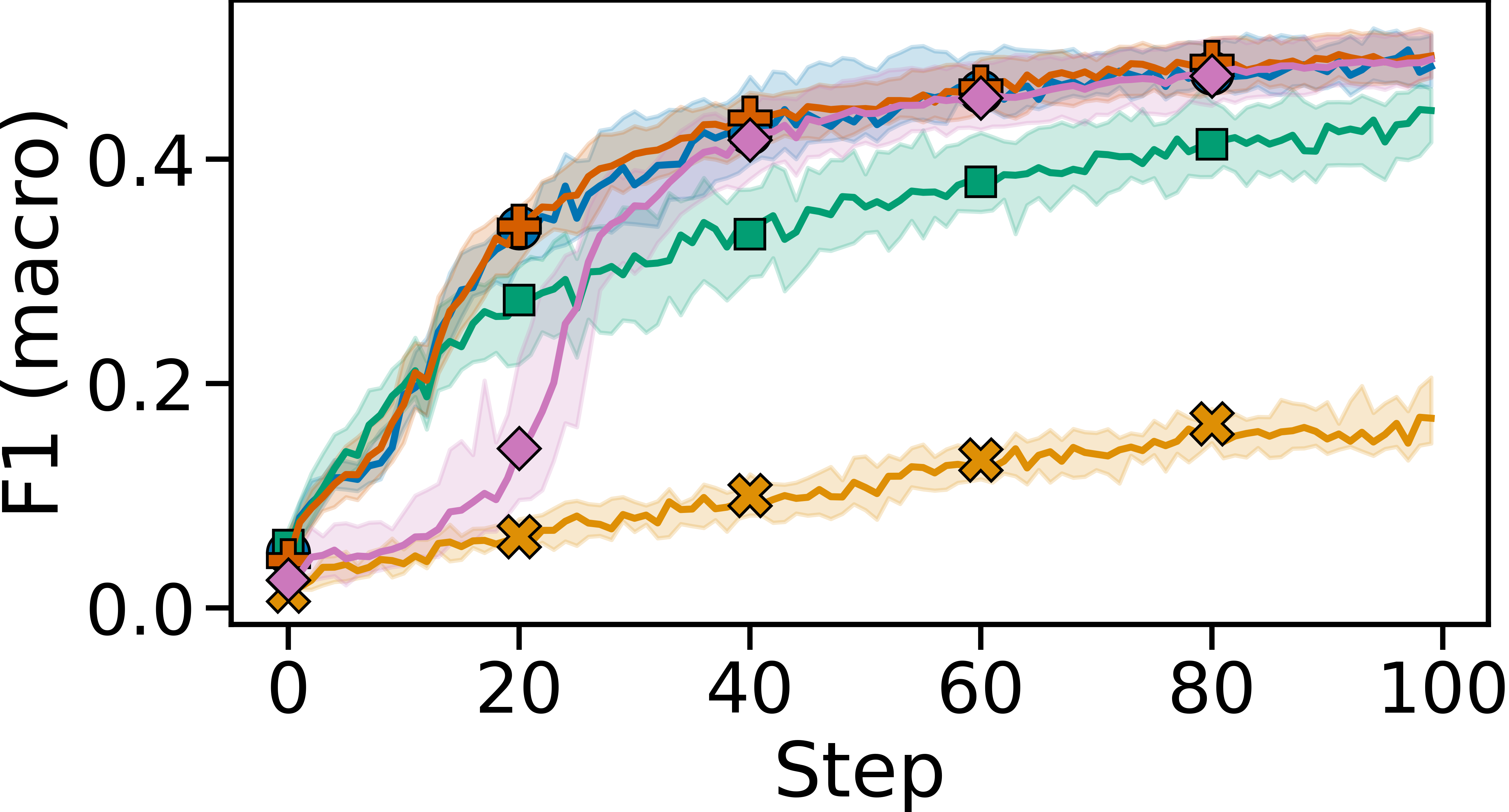}
                \vspace{-2em}
                \caption{\textsc{K20 (skew)}}
                \label{subfig:featureselectionf1k20skew}
            \end{subfigure}
            \begin{subfigure}{0.48\columnwidth}
                \includegraphics[width=\textwidth]{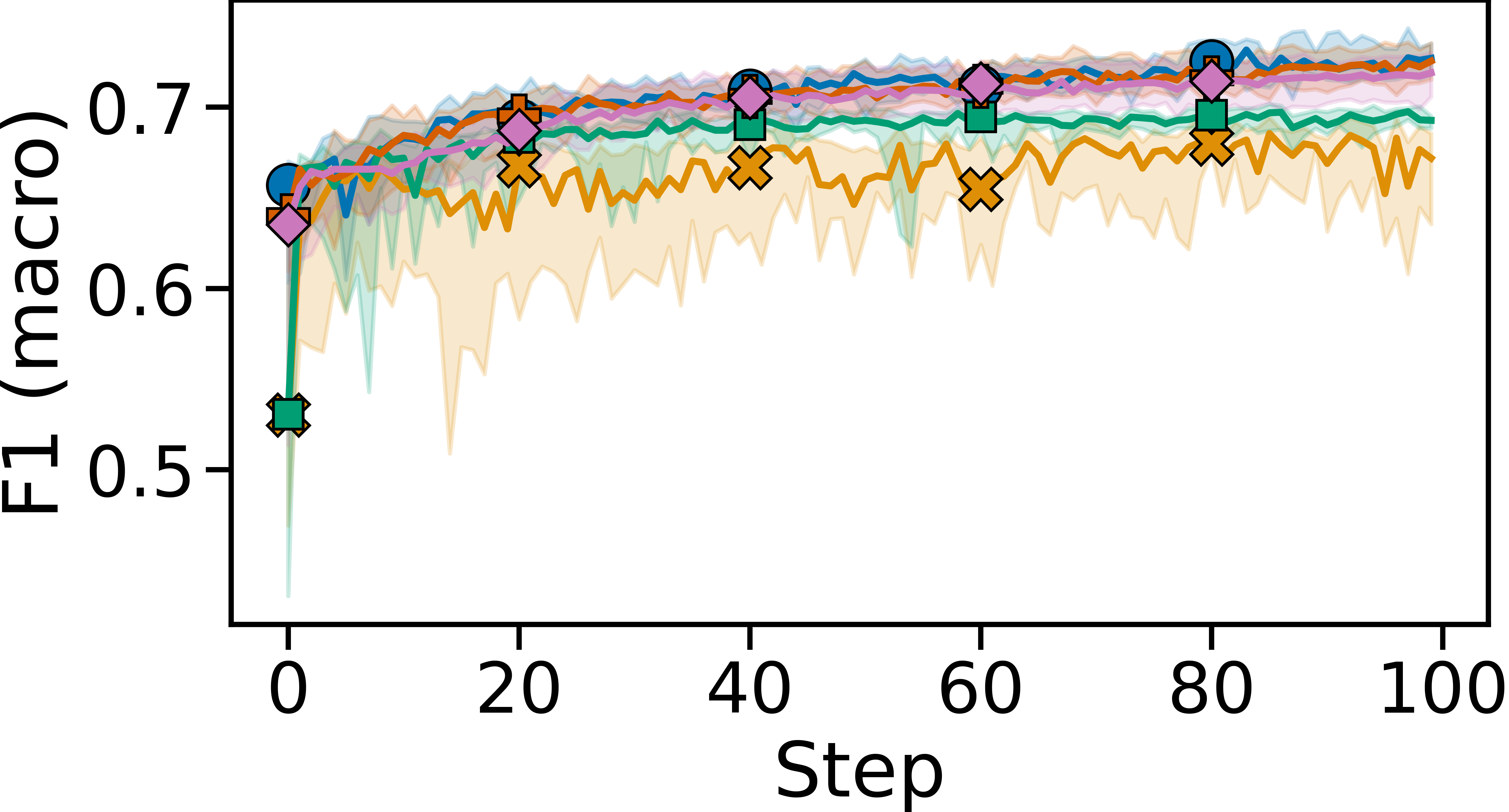}
                \vspace{-2em}
                \caption{\revision{}{\bears}}
                \label{subfig:featureselectionf1bears}
            \end{subfigure}

            \begin{subfigure}{0.48\columnwidth}
                \includegraphics[width=\textwidth]{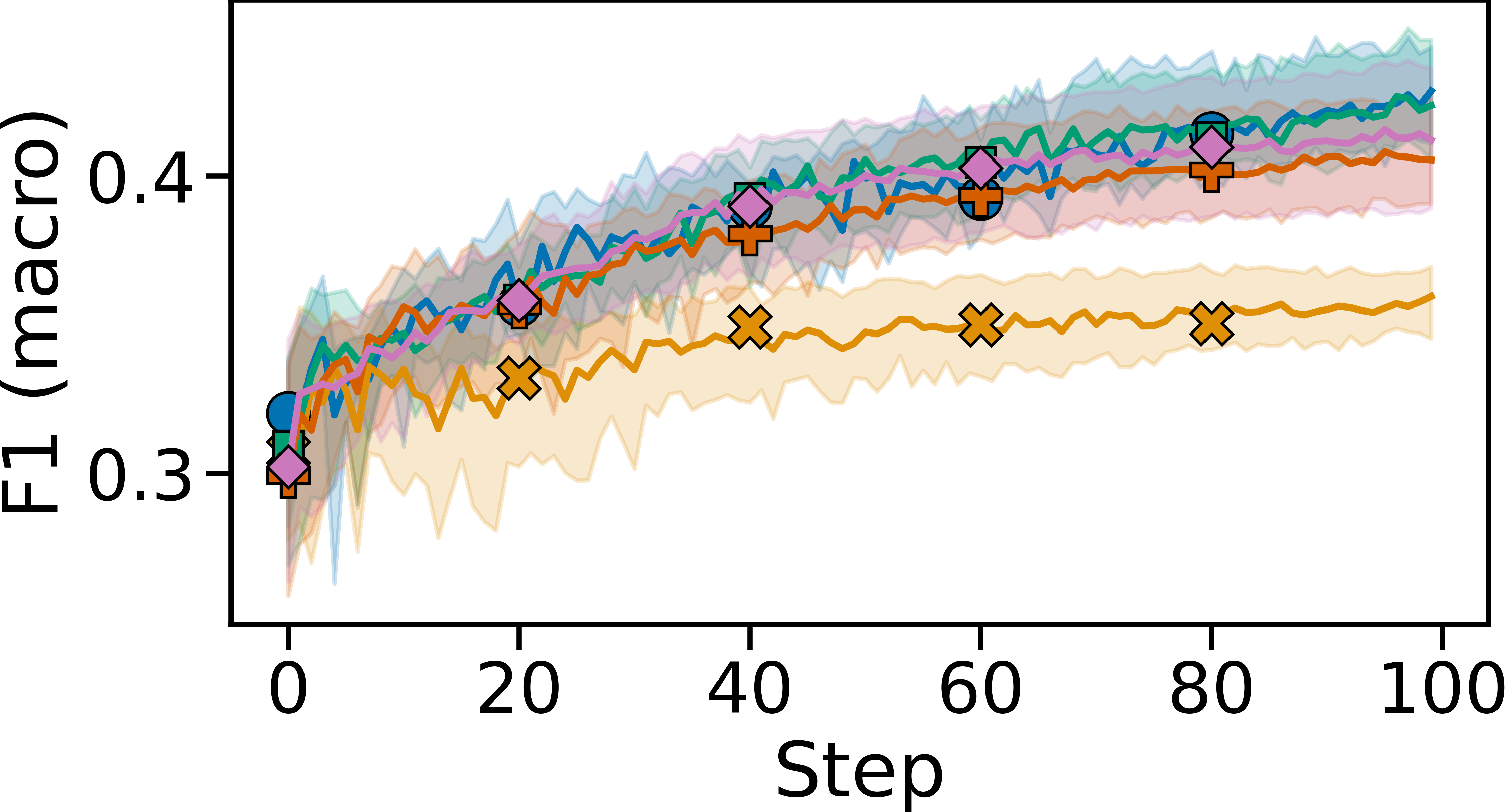}
                \vspace{-2em}
                \caption{\revision{}{\bdd}}
                \label{subfig:featureselectionf1bdd}
            \end{subfigure}
            \begin{subfigure}{0.48\columnwidth}
                \includegraphics[width=\textwidth]{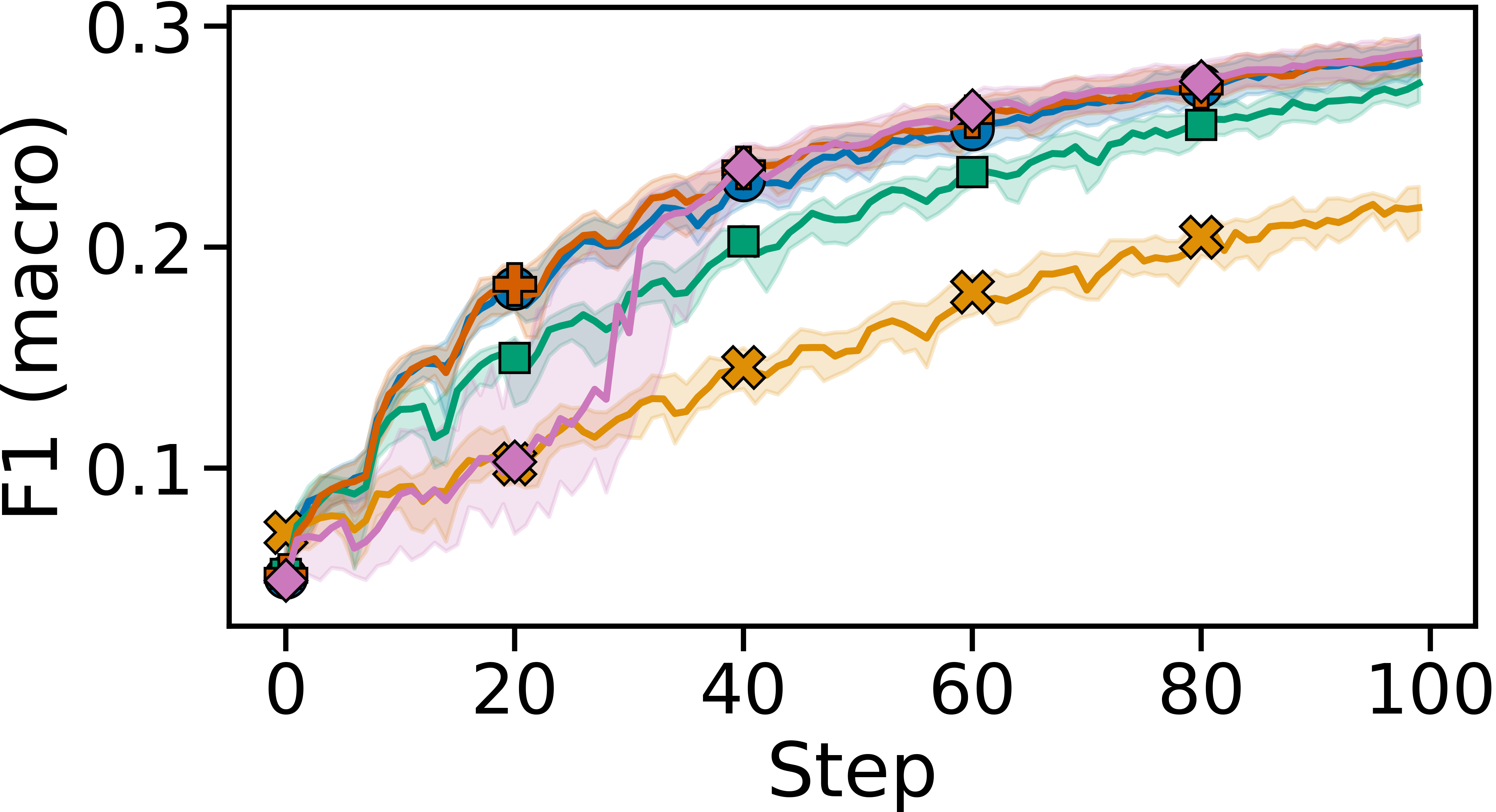}
                \vspace{-2em}
                \caption{\revision{}{\charades}}
                \label{subfig:featureselectionf1charades}
            \end{subfigure}

            \vspace{-1em}
            \caption{Macro F1 score when performing feature selection compared to the empirically best- and worst-performing sampling methods and features (excluding the \random feature). We also compare against \ve \revision{}{and \vecm} sampling methods on the best feature. \system initially has poor F1 performance as it explores suboptimal features but catches up to the best strategies within \revision{}{30} steps. \revision{}{The shaded region shows the IQR.}}
            \label{fig:featureSelectionFFigure}
            \vspace{-1em}
    \end{figure}
}

\newcommand{\combinedSelectionStepAndProgressFigure}{
    \begin{figure}[t!]
        \centering
        \begin{minipage}{0.48\columnwidth}
            \centering
            \includegraphics[width=\columnwidth]{figures/selection_step_C5_exp5_k3.png}
            \vspace{-2.5em}
            \caption{\revision{}{Median feature selection step when $C{=}5$ and $w{=}5$. Error bars show the IQR. \system converges to a single feature within a reasonable number of steps.}
            }
            \label{fig:featureSelectionStepFigure}
        \end{minipage}\hfill
        \begin{minipage}{0.48\columnwidth}
            \centering
            \includegraphics[width=0.8\columnwidth]{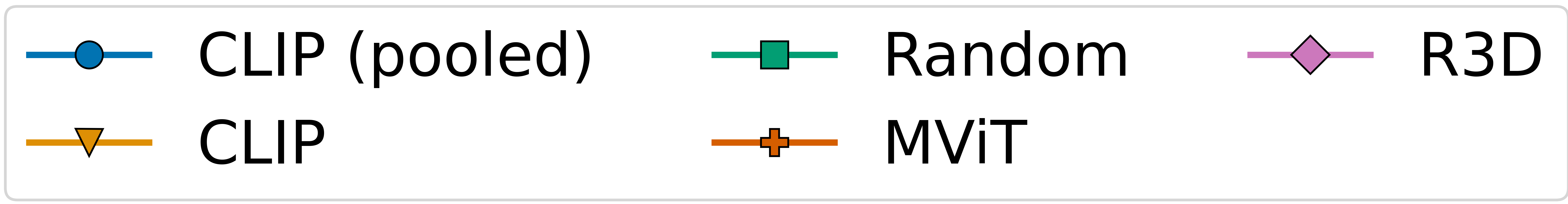}

            \includegraphics[width=\columnwidth]{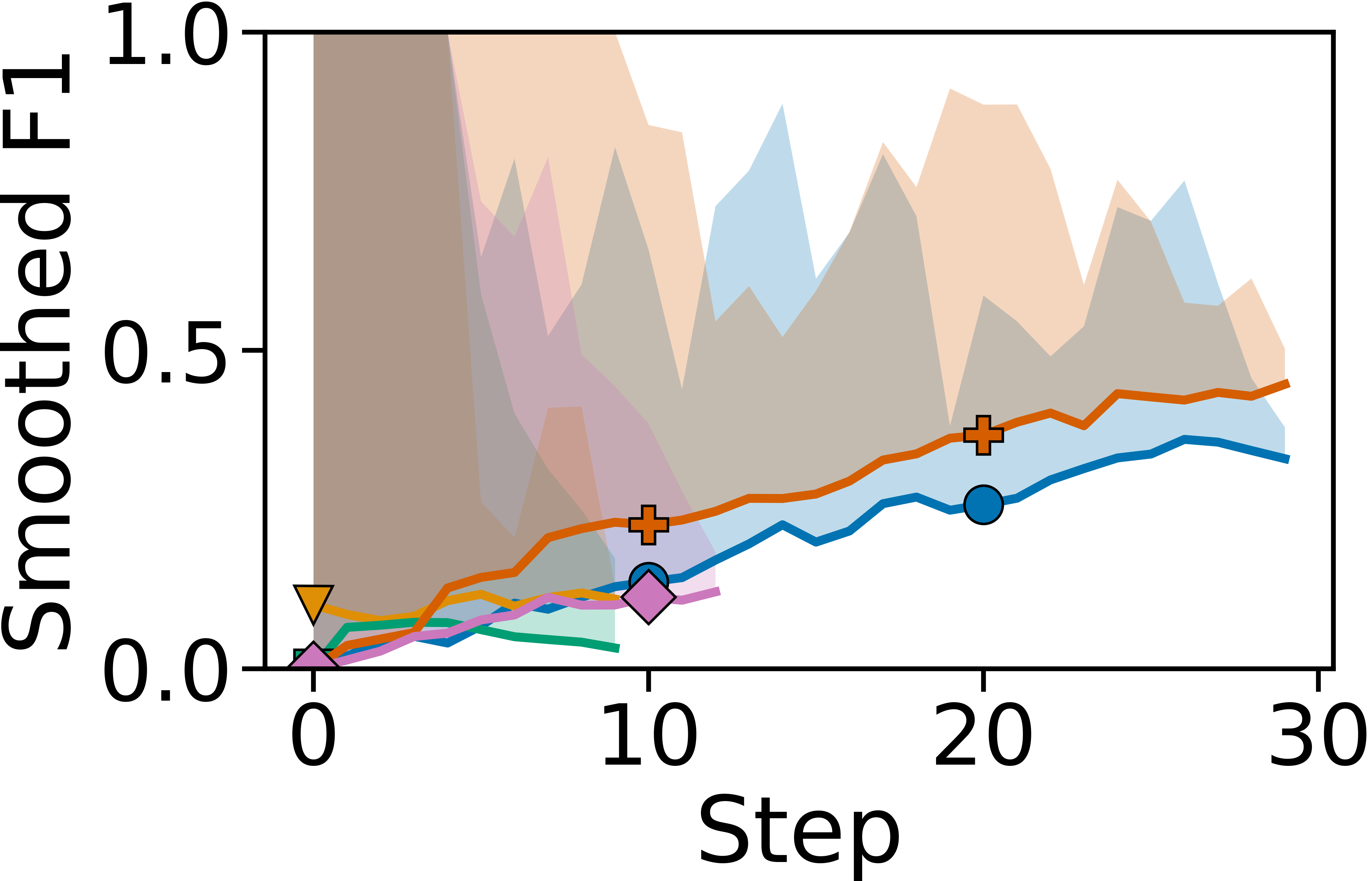}
            \vspace{-2.5em}
            \caption{Feature selection progress for \ktw showing upper and lower bounds when $C{=}5, w{=}5, T{=}50$.}
            \label{fig:featureselectionprogress}
        \end{minipage}
        \vspace{-1em}
    \end{figure}
}

\newcommand{\combinedFeatureRankingAndSelectionStepFigure}{
    \begin{figure}
        \centering
            \includegraphics[width=\columnwidth]{figures/aggf1-LEGEND.png}

            \begin{subfigure}{0.48\columnwidth}
                \includegraphics[width=\textwidth]{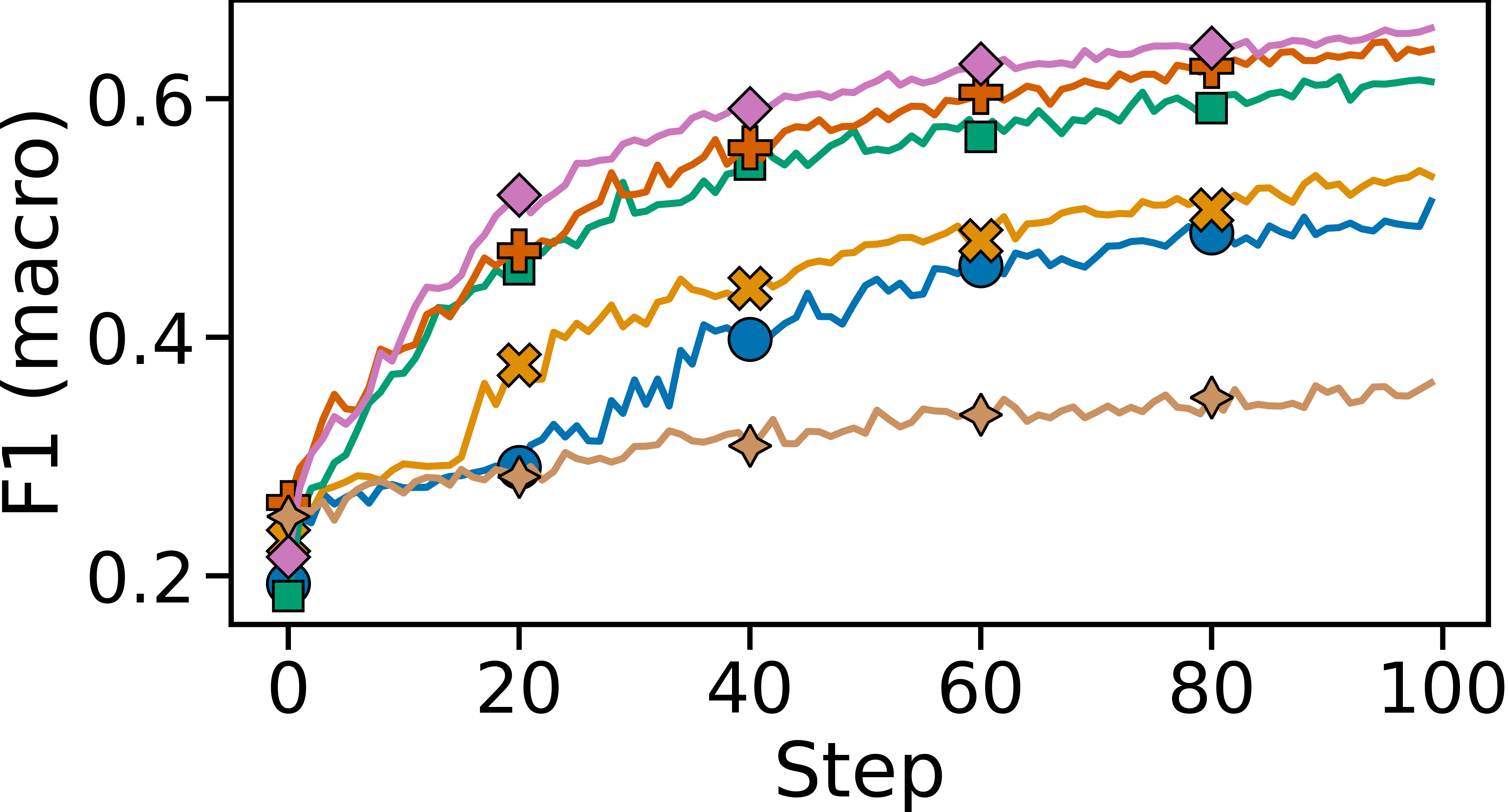}
                \vspace{-2em}
                \caption{\textsc{Deer}}
                \label{subfig:aggf1deer}
            \end{subfigure}
            \begin{subfigure}{0.48\columnwidth}
                \includegraphics[width=\textwidth]{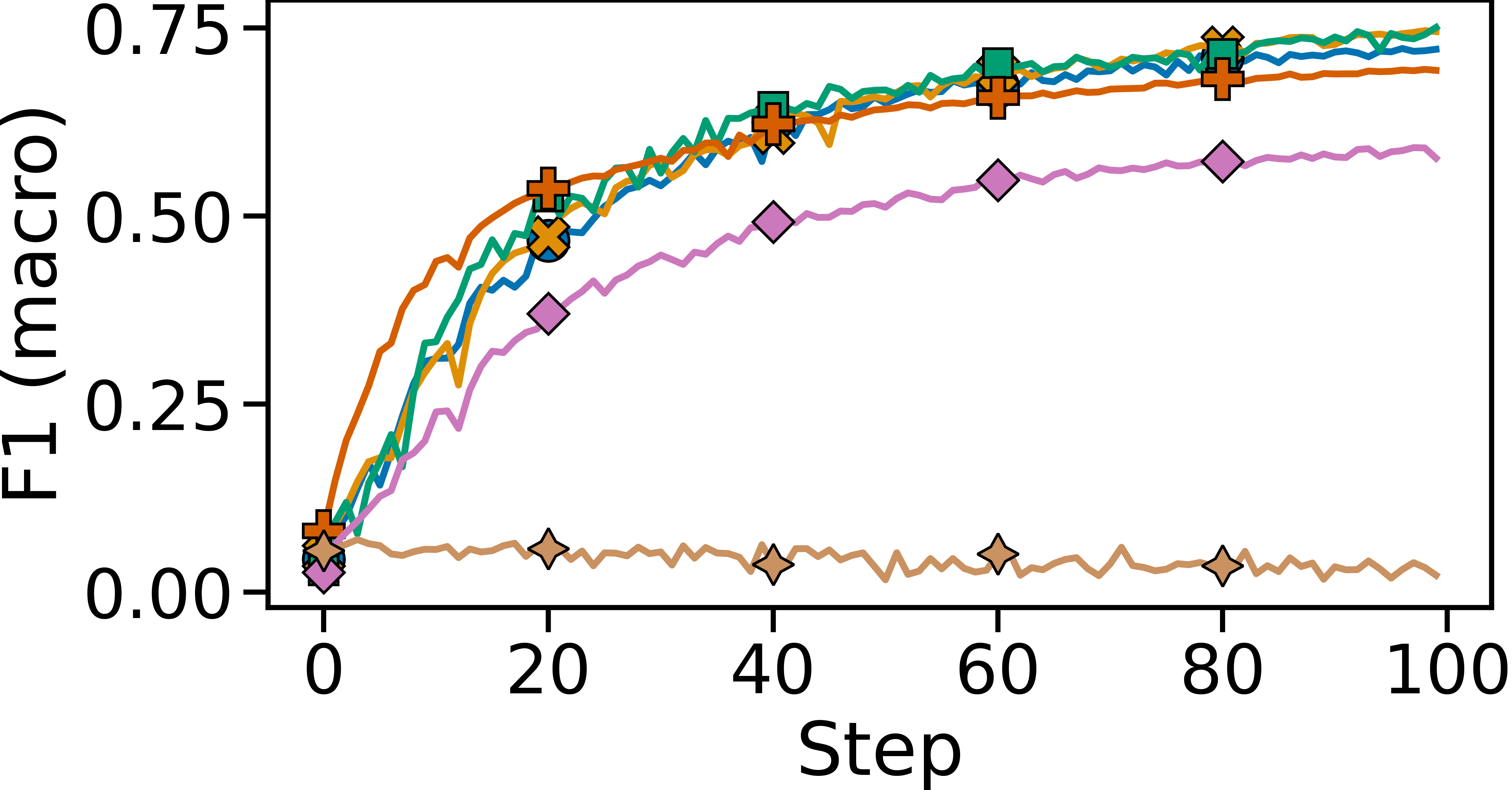}
                \vspace{-2em}
                \caption{\textsc{K20}}
                \label{subfig:aggf1k20}
            \end{subfigure}

            \begin{subfigure}{0.48\columnwidth}
                \includegraphics[width=\textwidth]{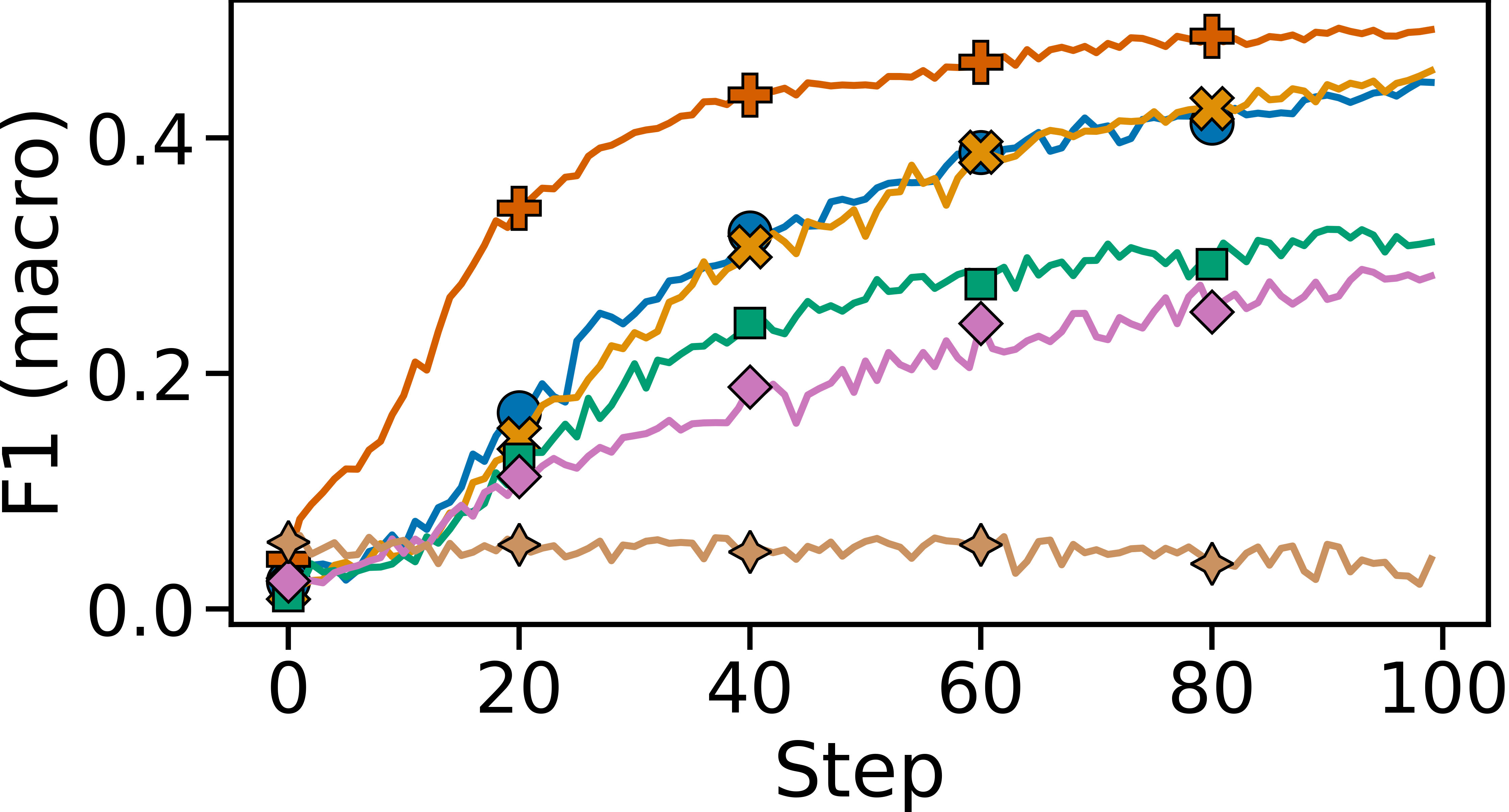}
                \vspace{-2em}
                \caption{\textsc{K20 (skew)}}
                \label{subfig:aggf1k20skew}
            \end{subfigure}
            \begin{subfigure}{0.48\columnwidth}
                \includegraphics[width=\textwidth]{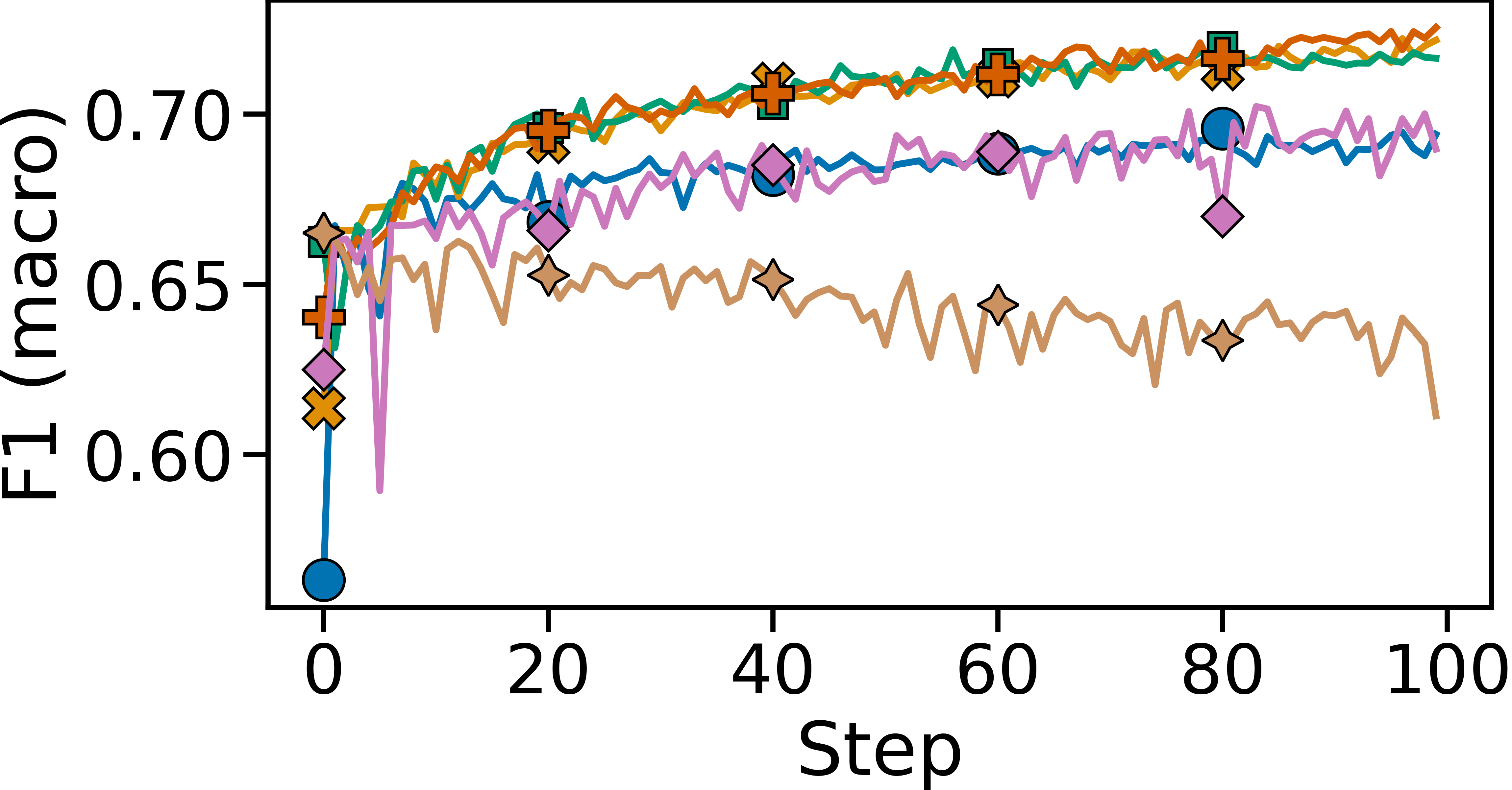}
                \vspace{-2em}
                \caption{\revision{MR-6, MR-7}{\bears}}
                \label{subfig:aggf1bears}
            \end{subfigure}

            \begin{subfigure}{0.48\columnwidth}
                \includegraphics[width=\textwidth]{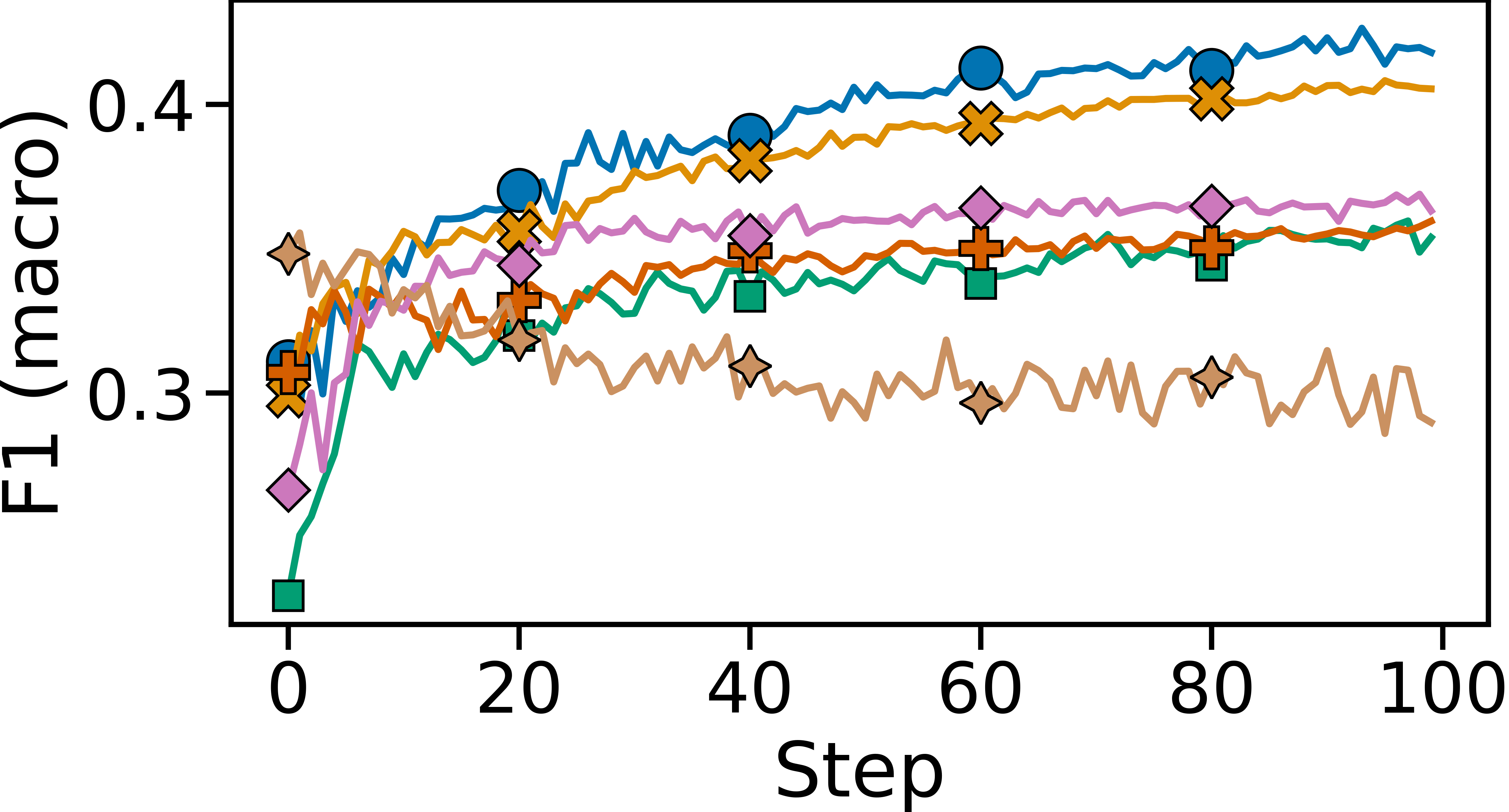}
                \vspace{-2em}
                \caption{\revision{MR-6, MR-7}{\bdd}}
                \label{subfig:aggf1bdd}
            \end{subfigure}
            \begin{subfigure}{0.48\columnwidth}
                \includegraphics[width=\textwidth]{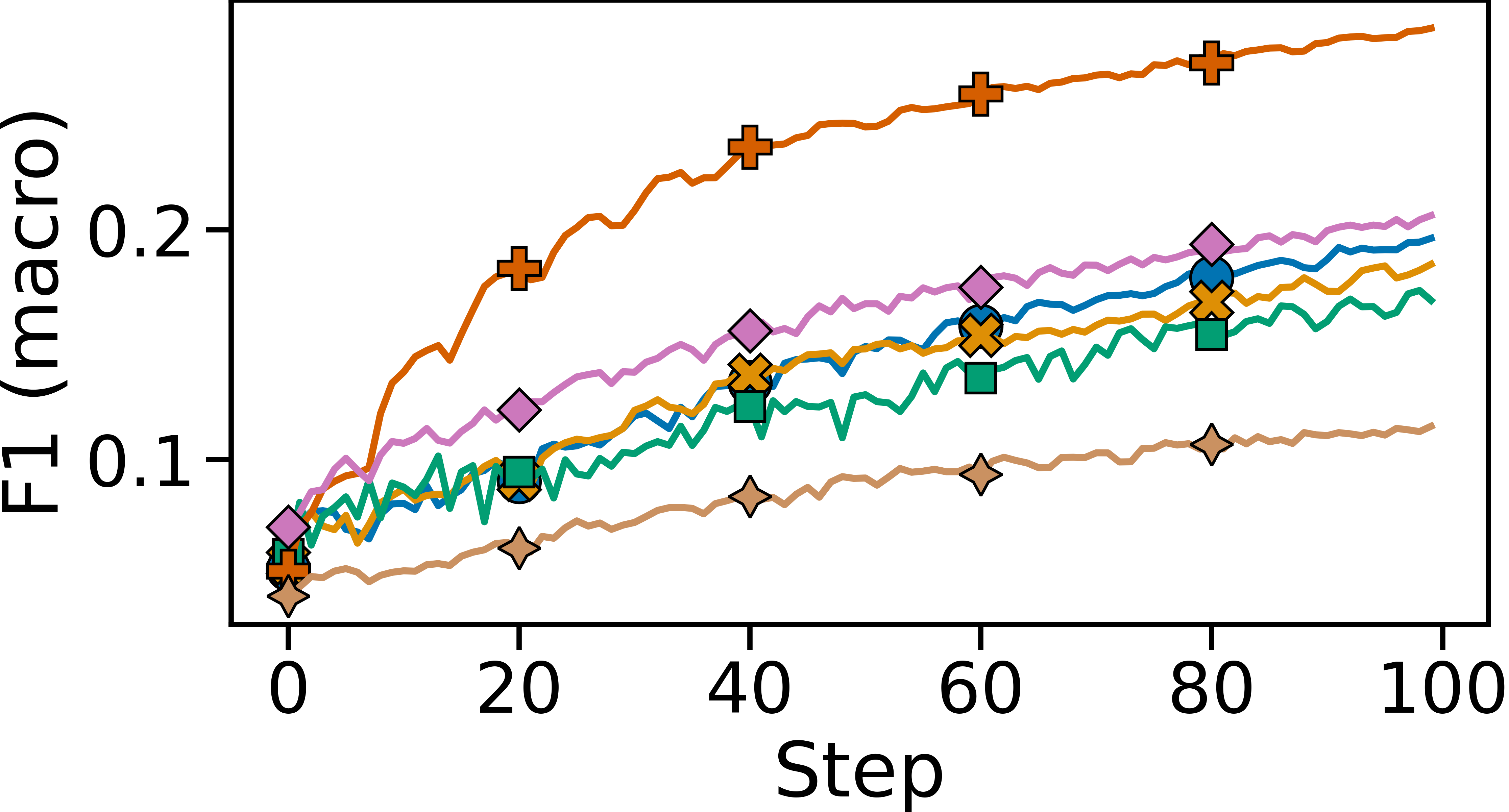}
                \vspace{-2em}
                \caption{\revision{MR-6, MR-7}{\charades}}
                \label{subfig:aggf1charades}
            \end{subfigure}

            \vspace{-1em}
            \caption{Macro F1 score when using the \revision{MR-6}{\vecm} sampling method, which shows that the best feature varies across datasets. ``Concat'' refers to concatenating all of the features into a single feature vector.}
            \label{fig:featureRankingFigure}
            \vspace{-1em}
    \end{figure}
}

\newcommand{\latencyVsPerfAllFigure}{
    \begin{figure}[t!]
        \centering
        \includegraphics[width=\columnwidth]{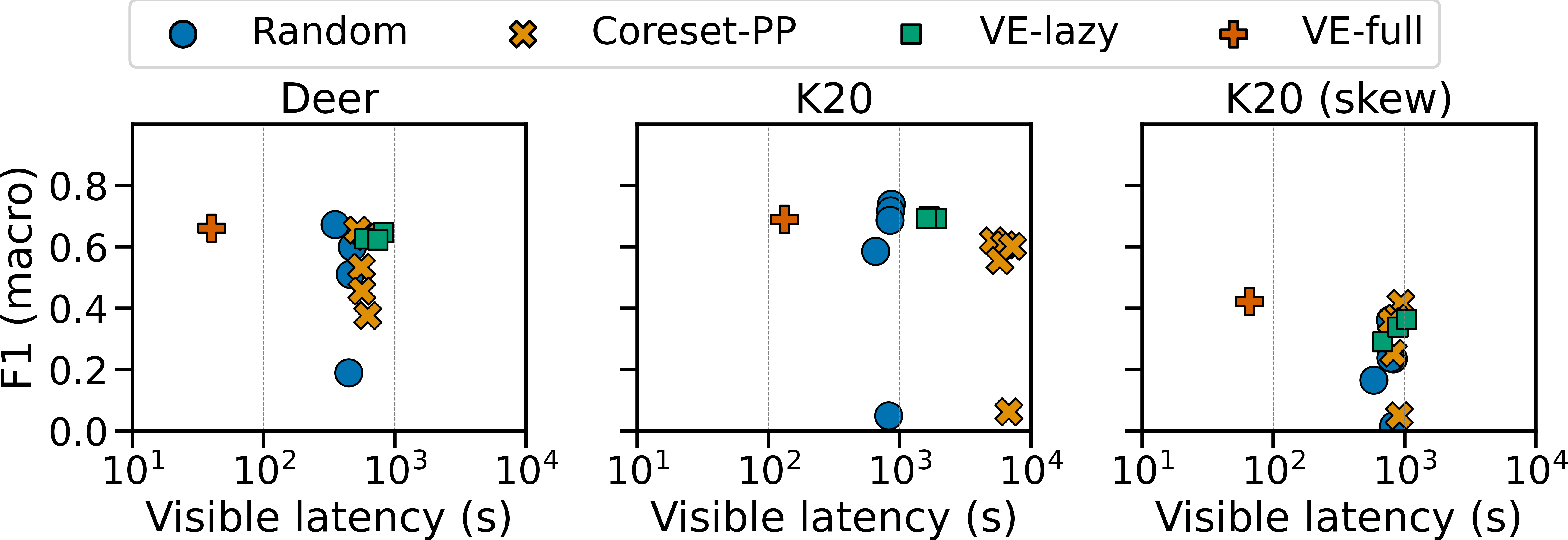}
        \vspace{-1.5em}
        \caption{Average F1 and cumulative visible latency (shown with a log-scale) after 100 \method{Explore} steps. \textsc{Coreset-PP} includes the preprocessing time to extract each feature, and each point for \textsc{Coreset-PP} and \random represents a single feature. \textsc{VE-full} provides nearly the best model quality with the lowest visible latency.}
        \label{fig:latencyVsPerfAllFigure}
    \end{figure}
}

\newcommand{\latencyVsPerfByDatasetFigure}{
    \begin{figure}[t!]
        \centering

        \begin{subfigure}{0.98\columnwidth}
            \includegraphics[width=\textwidth]{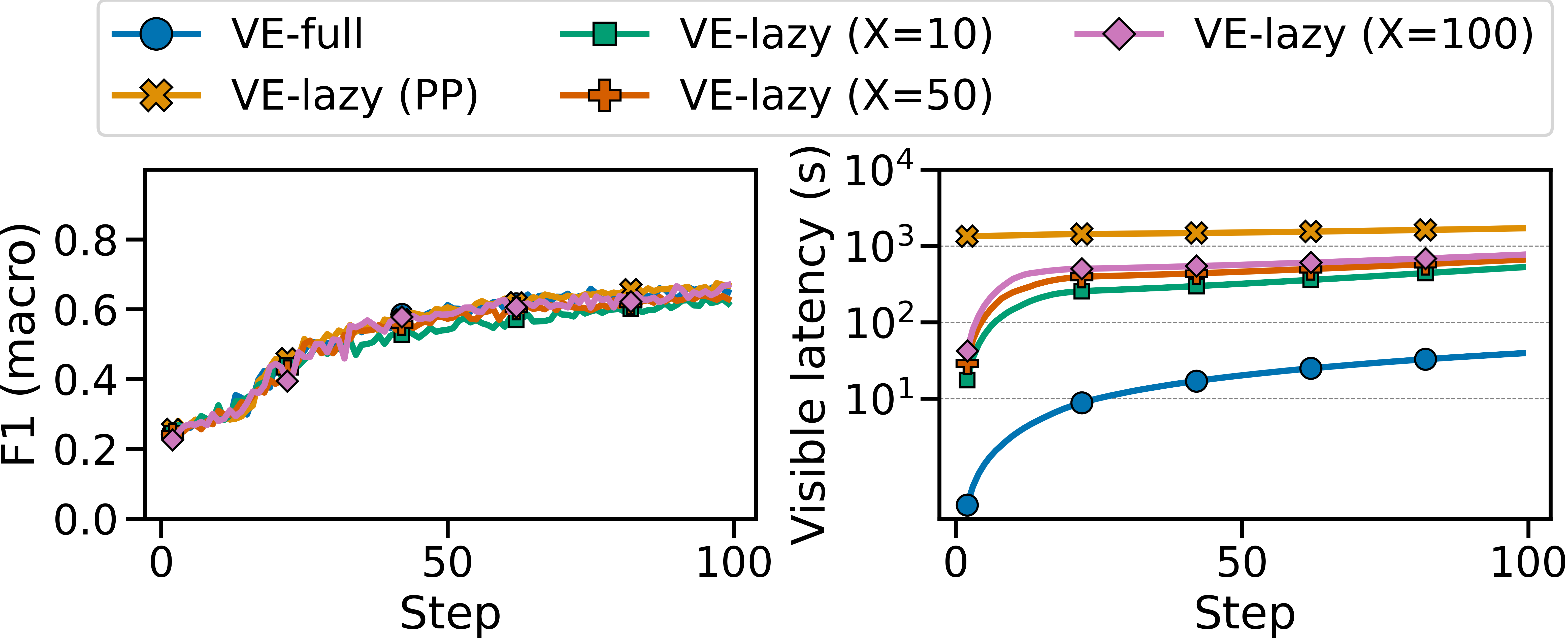}
            \vspace{-2em}
            \caption{\deer}
        \end{subfigure}

        \begin{subfigure}{0.98\columnwidth}
            \includegraphics[width=\textwidth]{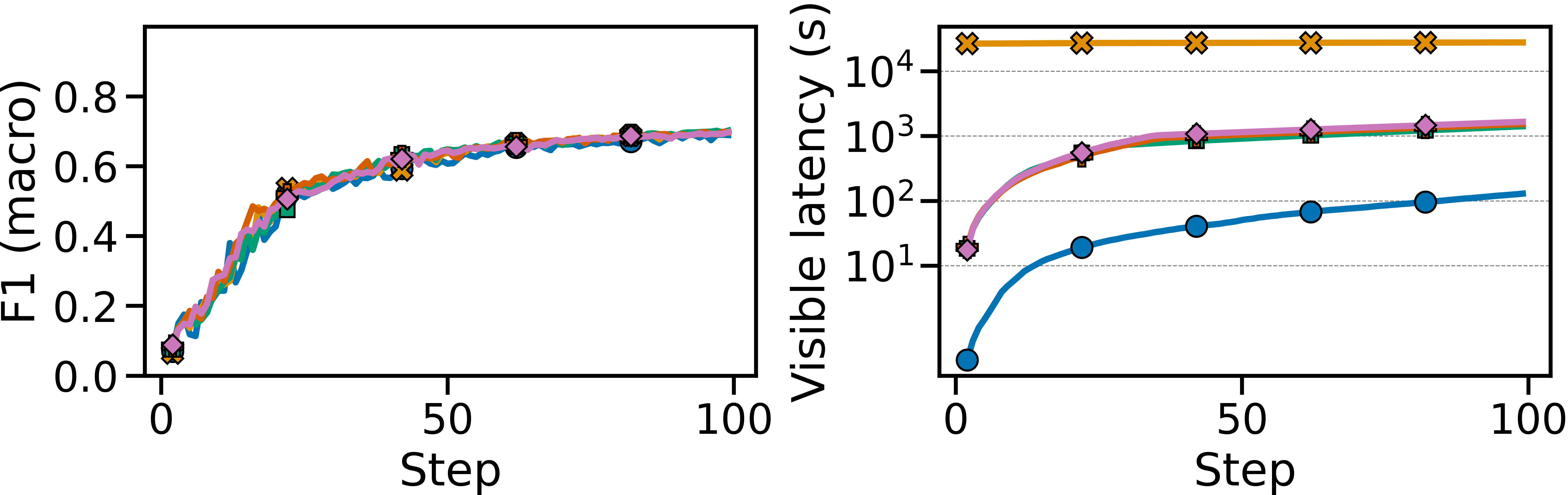}
            \vspace{-2em}
            \caption{\ktw}
        \end{subfigure}

        \begin{subfigure}{0.98\columnwidth}
            \includegraphics[width=\textwidth]{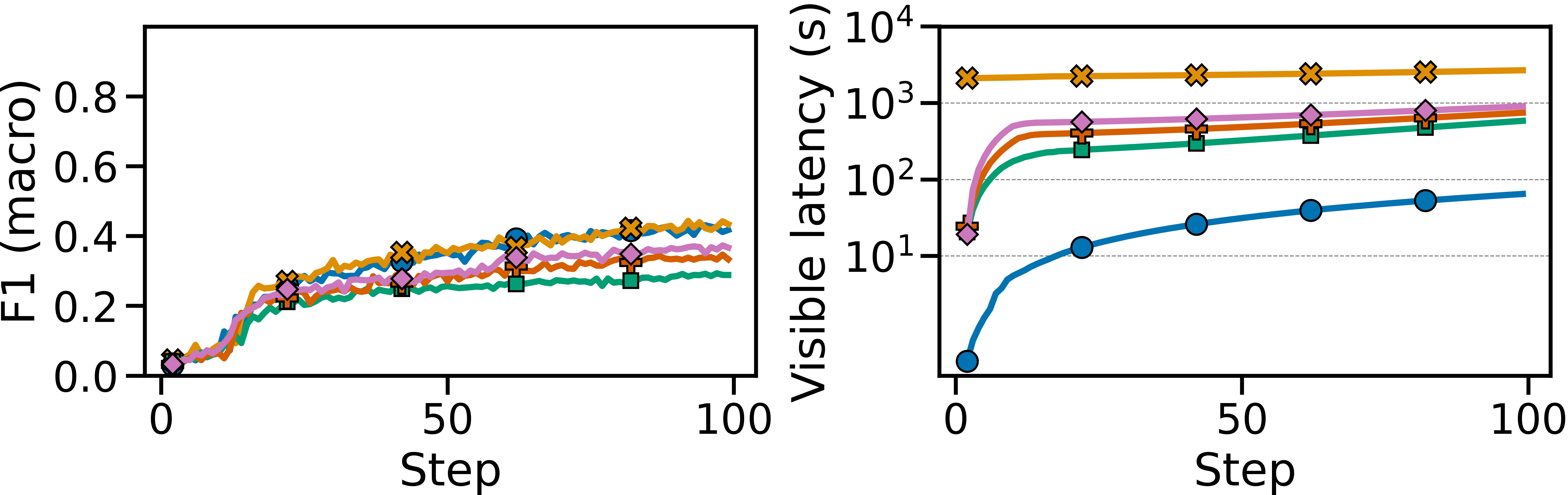}
            \vspace{-2em}
            \caption{\ktwsk}
        \end{subfigure}
        \vspace{-1em}
        \caption{Model quality and latency for \textsc{VE}-variants.
            \textsc{VE-full} matches the best model performance of \textsc{VE-lazy} with less cumulative visible latency (shown with a log-scale).}
        \label{fig:latencyVsPerfByDatasetFigure}
        \vspace{-1em}
    \end{figure}
}

\newcommand{\noisyLabelsFigure}{
    \begin{figure}[t!]
        \includegraphics[width=0.95\columnwidth]{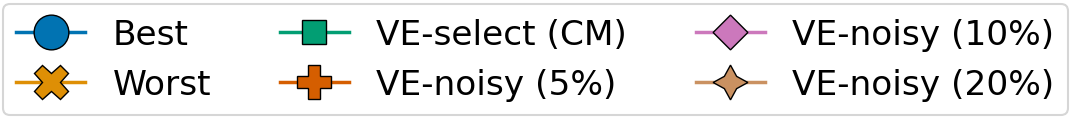}

        \begin{subfigure}{0.48\columnwidth}
            \includegraphics[width=\textwidth]{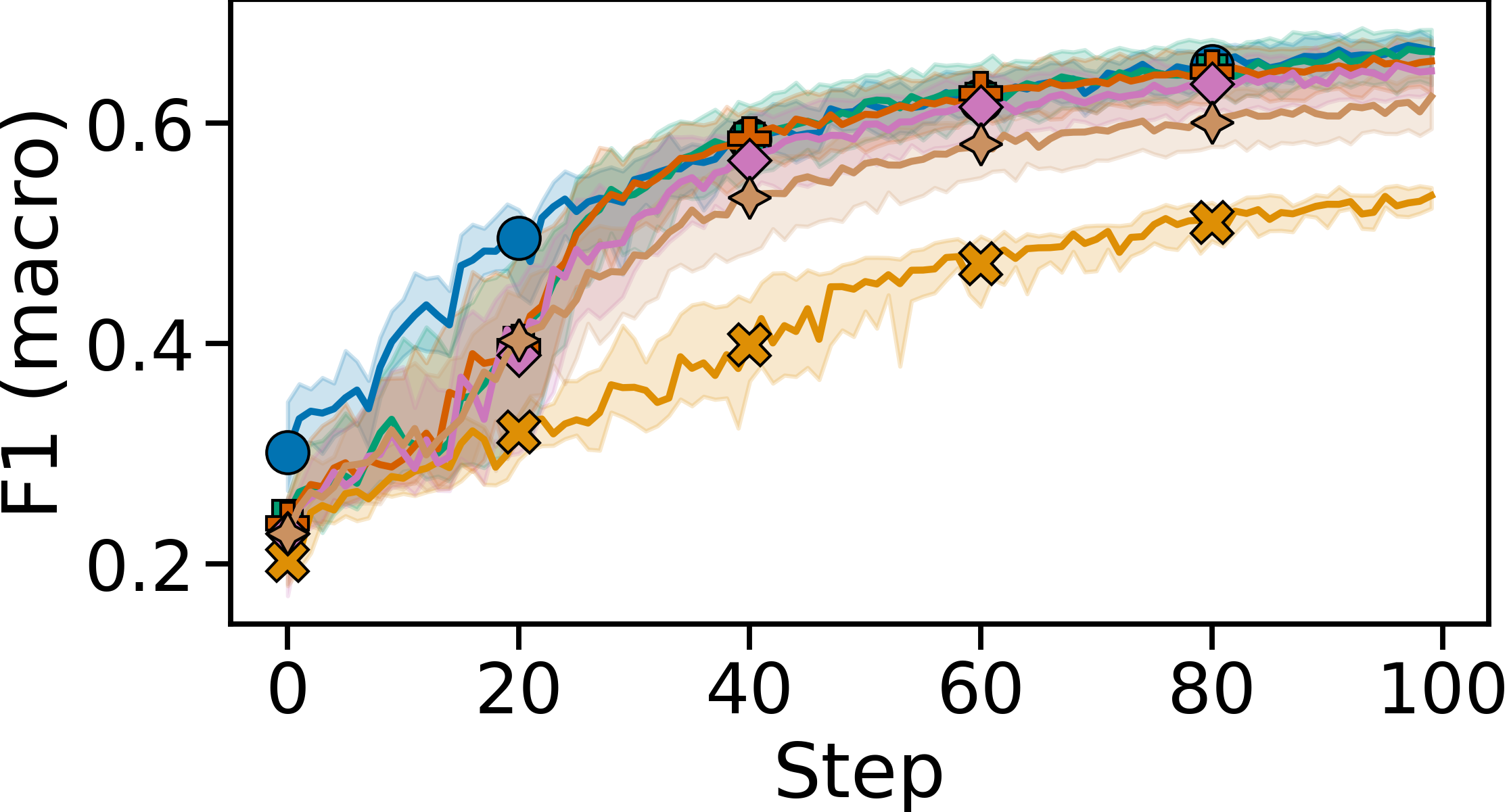}
            \vspace{-2em}
            \caption{\textsc{Deer}}
            \label{subfig:noisyfeatureselectionf1deer}
        \end{subfigure}
        \begin{subfigure}{0.48\columnwidth}
            \includegraphics[width=\textwidth]{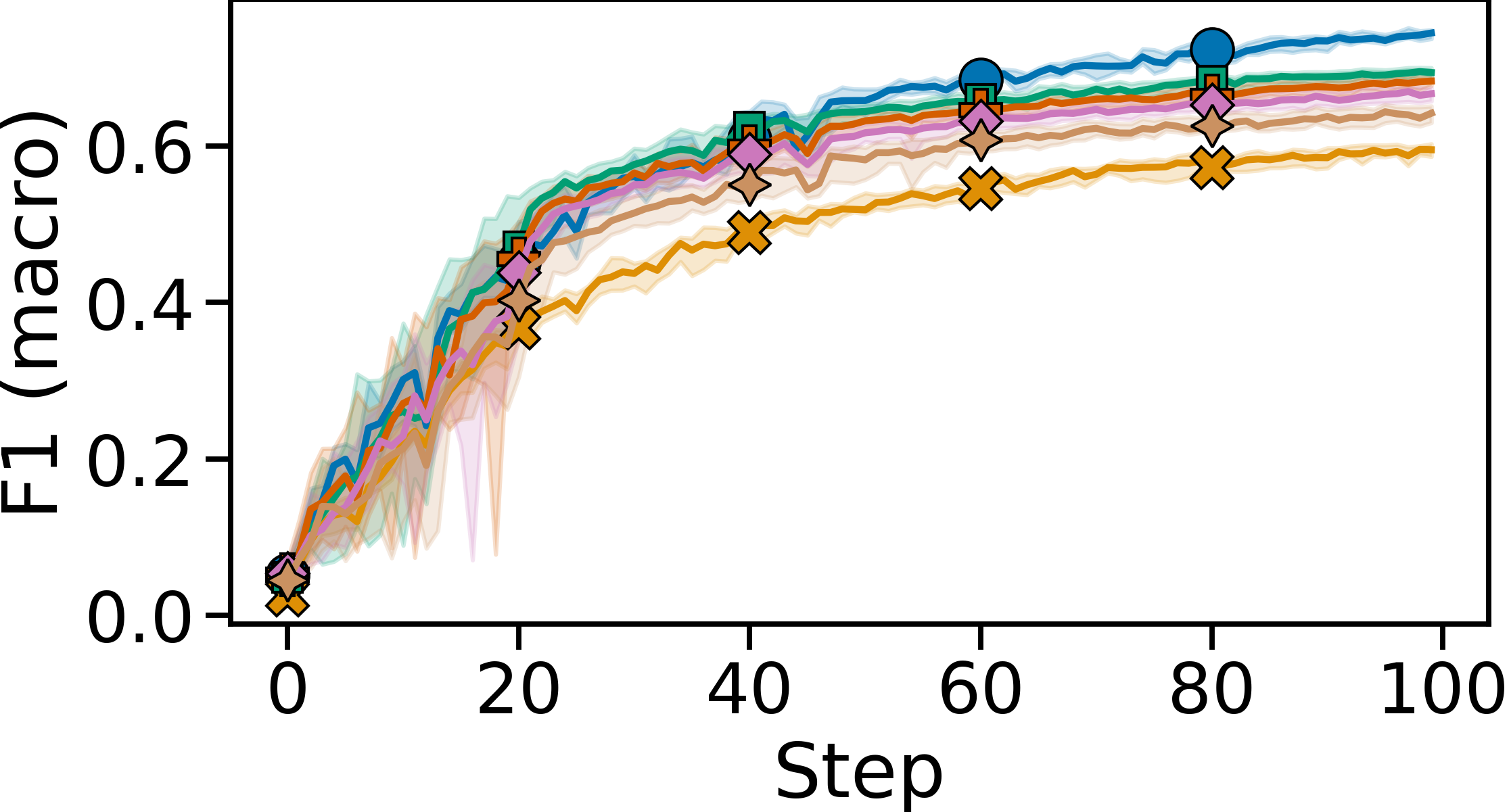}
            \vspace{-2em}
            \caption{\textsc{K20}}
            \label{subfig:noisyfeatureselectionf1k20}
        \end{subfigure}

        \begin{subfigure}{0.48\columnwidth}
            \includegraphics[width=\textwidth]{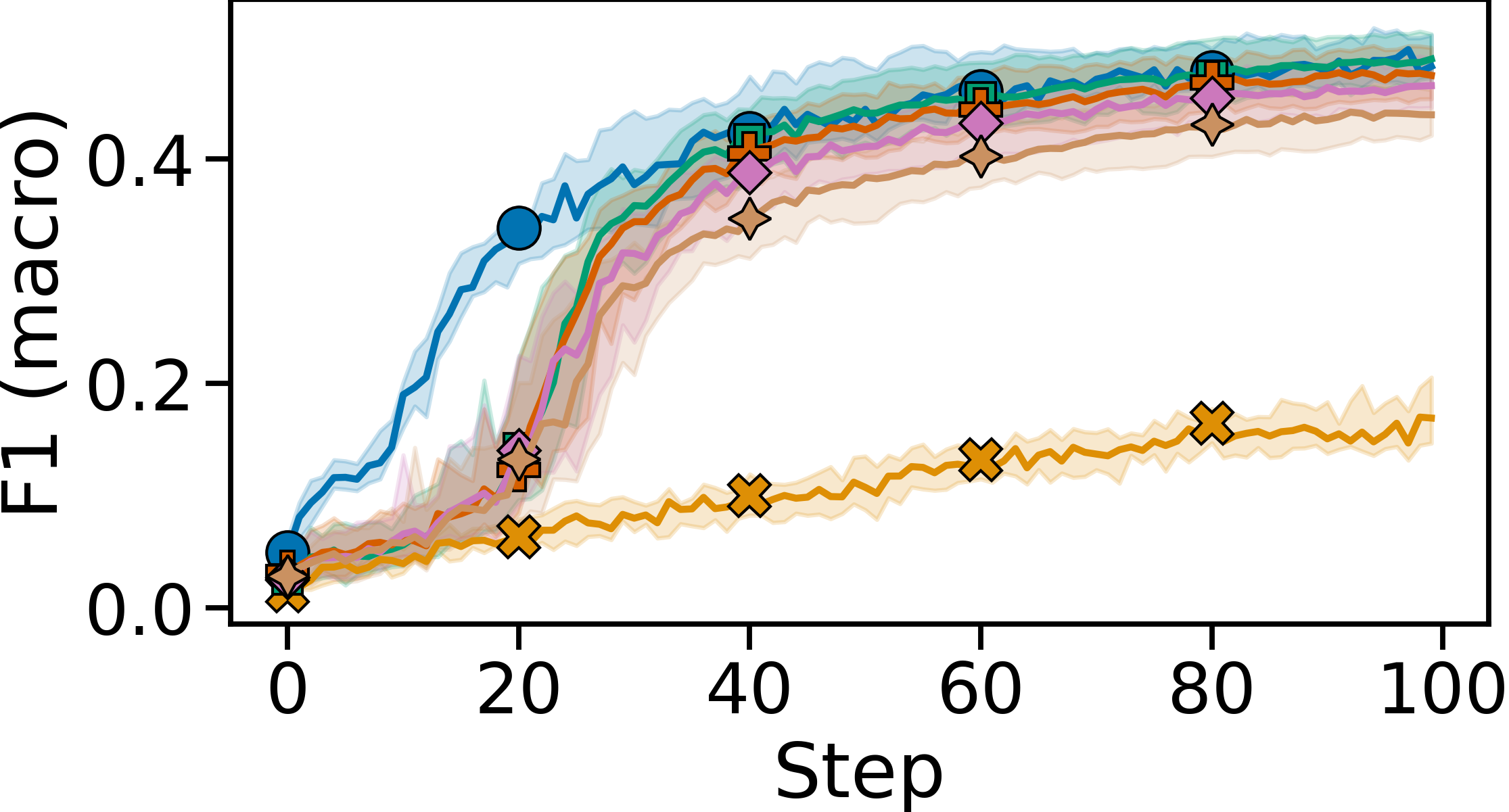}
            \vspace{-2em}
            \caption{\textsc{K20 (skew)}}
            \label{subfig:noisyfeatureselectionf1k20skew}
        \end{subfigure}
        \begin{subfigure}{0.48\columnwidth}
            \includegraphics[width=\textwidth]{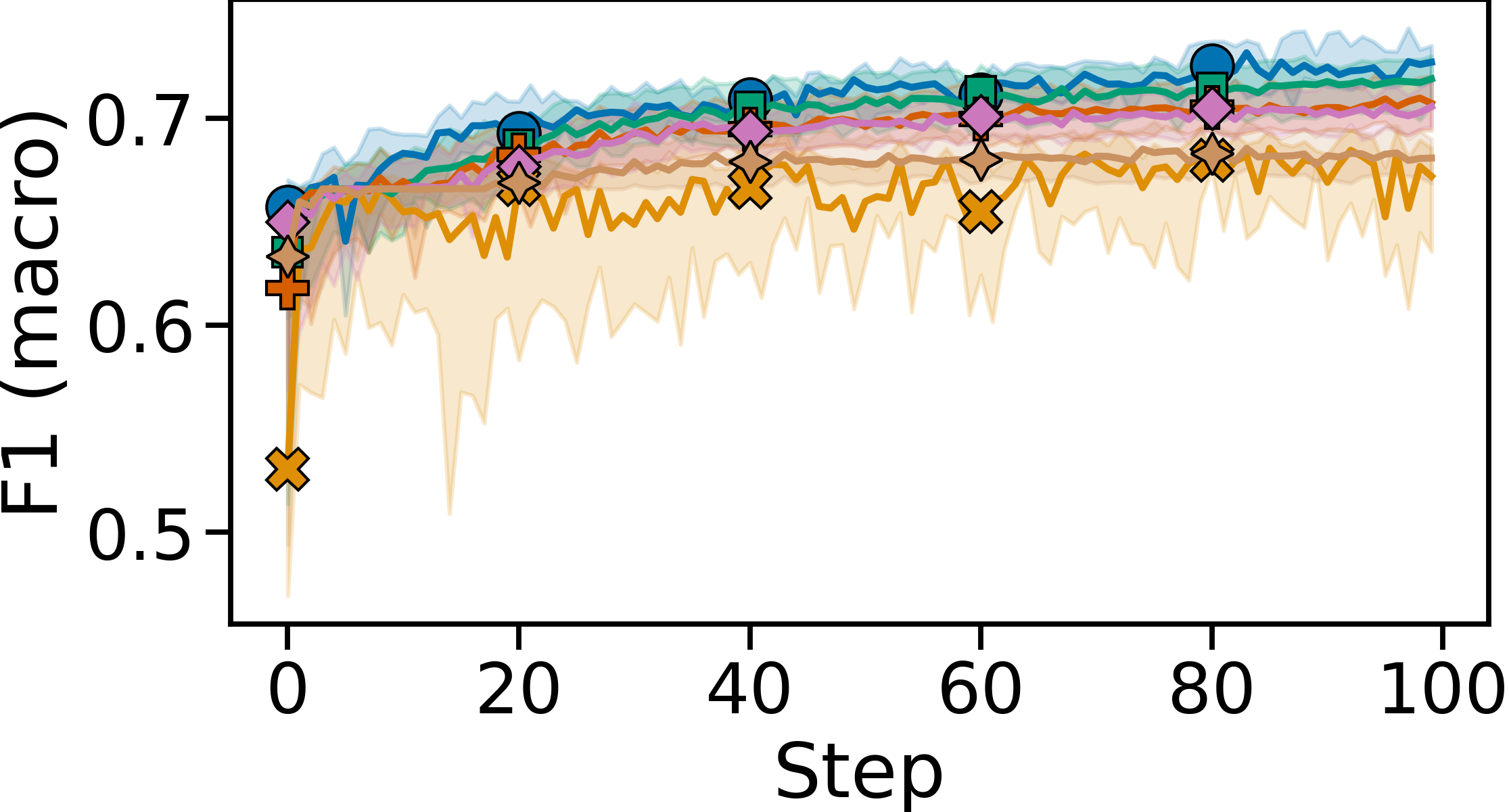}
            \vspace{-2em}
            \caption{\bears}
            \label{subfig:noisyfeatureselectionf1bears}
        \end{subfigure}

        \begin{subfigure}{0.48\columnwidth}
            \includegraphics[width=\textwidth]{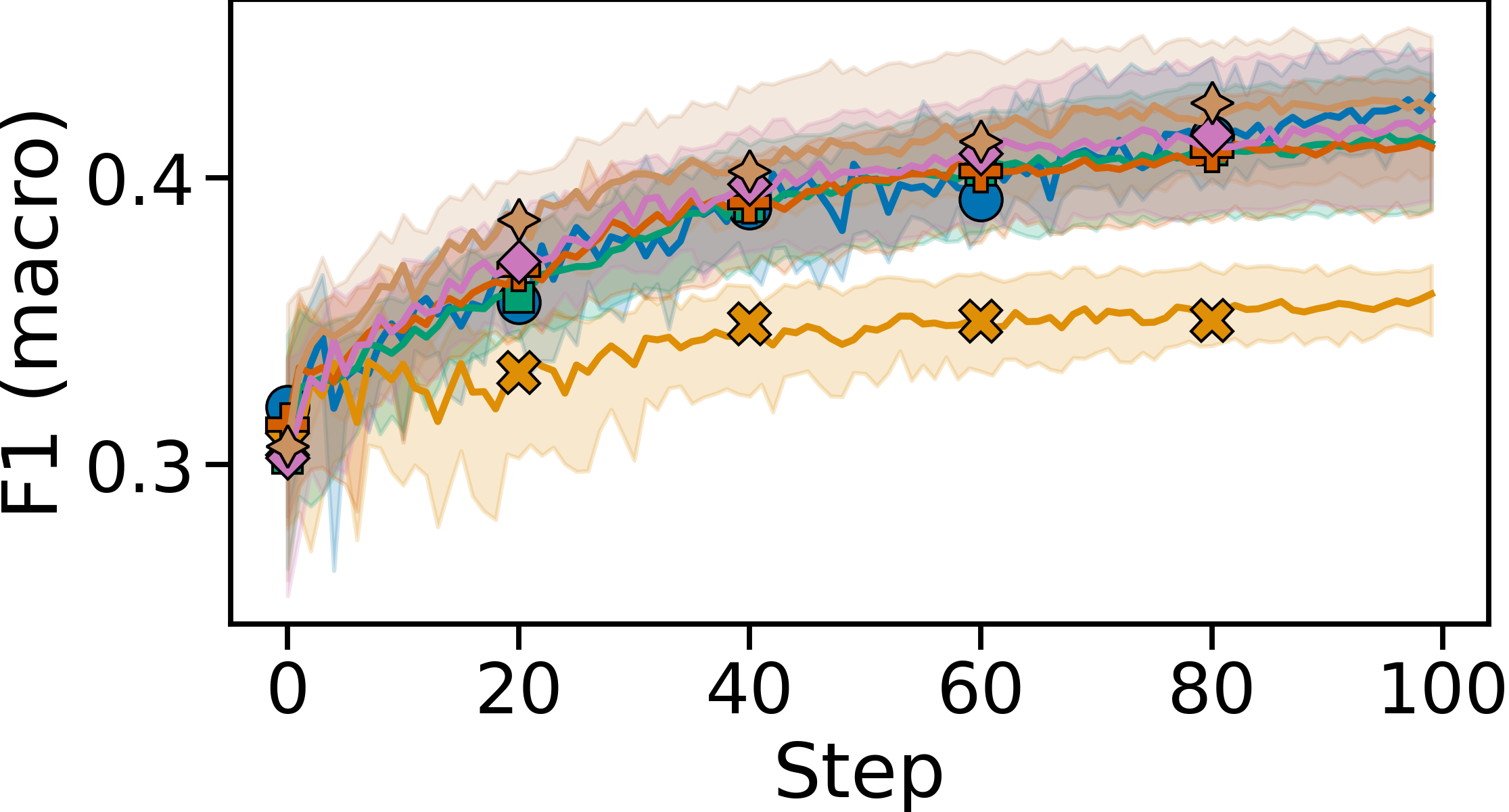}
            \vspace{-2em}
            \caption{\bdd}
            \label{subfig:noisyfeatureselectionf1bdd}
        \end{subfigure}
        \begin{subfigure}{0.48\columnwidth}
            \includegraphics[width=\textwidth]{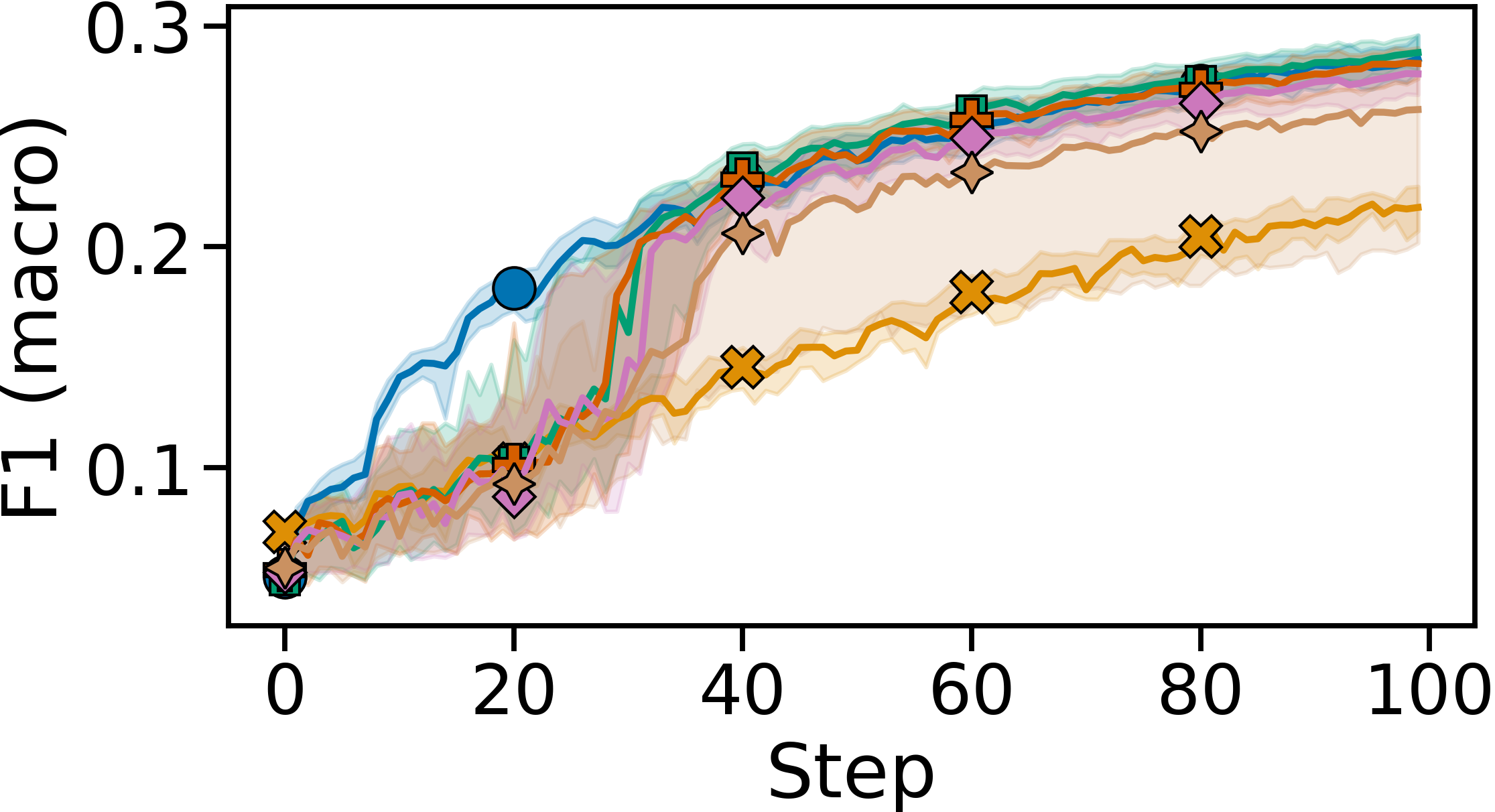}
            \vspace{-2em}
            \caption{\charades}
            \label{subfig:noisyfeatureselectionf1charades}
        \end{subfigure}
        \vspace{-1.5em}
        \caption{\revision{R6.D3}{
            Macro F1 score when performing feature selection with noisy labels (\textsc{VE-noisy}) compared with no noise (\textsc{VE-select (CM)}) and the empirically best- and worst-performing sampling methods and features (excluding the \random feature). We evaluate performance when randomly changing 5\%, 10\%, and 20\% of labels.
        }}
        \label{fig:noisyLabelsFigure}
    \end{figure}
}

\newcommand{\apiTable}{
    \begin{table}[t]
    \centering
    \caption{\system API}
    \vspace{-1em}
    \begin{spacing}{0.65}
    \renewcommand{\arraystretch}{1.4}
    \small
    \begin{tabular}{>{\raggedright}p{1.2cm}p{2.25cm}p{4.2cm}}
        \toprule
        Method & Parameters & Description \\
        \midrule

        \method{Watch} &
        {\small (vid, start, end)} &
        {\small
        Returns a stream of video segments from \textit{vid} between \textit{start} and \textit{end} with predicted labels
        }
        \\

        \method{Explore} &
        {\small(B, t, label=None)} &
        {\small
        Returns \textit{B} video segments each of duration \textit{t} with predicted labels
        }
        \\

        \method{AddLabel} &
        {\small \nsqueeze{(vid, start, end, label)}} &
        {\small
        Saves the label as metadata
        }
        \\

        \method{AddVideo} &
        {\small (path)} &
        {\small
        \nsqueeze{Saves the video as a candidate for labels and predictions and returns its vid}
        }
        \\

        \bottomrule
    \end{tabular}
    \end{spacing}
    \label{table:api}
    \vspace{-1em}
    \end{table}
}

\newcommand{\hyperparametersTable}{
    \begin{table}[t]
    \centering
    \caption{Feature selection hyperparameters}
    \begin{tabular}{{c}{c}}
        \toprule
        Parameter & Description \\
        \midrule

        $w$ & Smoothing span ($\alpha = 2 / (w + 1)$) \\

        $C$ & Slope smoothing window \\

        $T$ & Timestep used to compute upper bound \\

        \bottomrule
    \end{tabular}
    \label{table:hyperparameters}
    \end{table}
}

\newcommand{\datasetTable}{
    \begin{table}[t]
    \centering
    \caption{Datasets}
    \vspace{-1em}
    \small
    \begin{tabular}{{c}{c}{c}{c}{c}}
        \toprule
        Dataset & \# classes & Skew & {Train videos} & {Eval videos} \\
        \midrule

        \textsc{Deer} &
        9 &
        Skewed &
        896 &
        225
        \\

        \textsc{K20} &
        20 &
        Uniform &
        13326 &
        976
        \\

        \textsc{K20 (skew)} &
        20 &
        Skewed &
        1050 &
        976
        \\

        \revision{MR-7}{\textsc{Charades}} &
        \revision{MR-7}{33} &
        \revision{MR-7}{Skewed} &
        \revision{MR-7}{7985} &
        \revision{MR-7}{1863}
        \\

        \revision{MR-7}{\textsc{Bears}} &
        \revision{MR-7}{2} &
        \revision{MR-7}{Uniform} &
        \revision{MR-7}{2410} &
        \revision{MR-7}{722}
        \\

        \revision{MR-7}{\textsc{BDD}} &
        \revision{MR-7}{6} &
        \revision{MR-7}{Skewed} &
        \revision{MR-7}{800} &
        \revision{MR-7}{200}
        \\

        \bottomrule
    \end{tabular}
    \label{table:datasets}
    \vspace{-1em}
    \end{table}
}

\newcommand{\featureTable}{
    \begin{table}[t]
    \centering
    \caption{Features used by \system. Throughput is the number of 10-second videos that can be processed each second while running two extraction tasks on the GPU.}
    \vspace{-1em}
    \small
    \begin{tabular}{{c}{c}{c}{c}{c}}
        \toprule
        Feature & Type & Architecture & Pretrained & Tput. \\
        \midrule

        \rd~\cite{DBLP:conf/cvpr/TranWTRLP18} &
        Video &
        Conv. net &
        {\small Kinetics400} &
        4.03
        \\

        \mvit~\cite{DBLP:conf/iccv/0001XMLYMF21} &
        Video &
        Transformer &
        {\small Kinetics400} &
        2.93
        \\

        \clip~\cite{DBLP:conf/icml/RadfordKHRGASAM21} &
        Image &
        Transformer &
        {\small Internet images} &
        3.64
        \\
        \clippool~\cite{DBLP:conf/icml/RadfordKHRGASAM21} &
        Image &
        Transformer &
        {\small Internet images} &
        3.45 \\

        \textsc{Random} &
        Video &
        Transformer &
        {\small None} &
        2.96
        \\

        \bottomrule
    \end{tabular}
    \label{table:features}
    \vspace{-1em}
    \end{table}
}

\newcommand{\smoothingFeatureHyperparameterSensitivityTable}{
    \begin{table}[t]
        \centering
        \caption{Feature selection accuracy - hyperparameter sensitivity}
        \begin{tabular}{{c}{c}{c}{c}{c}{c}}
            \toprule
            Dataset & T & C & Smoothed & Smoothed  & Smoothed \\
                    &   &   & ($w=3$)  & ($w=5$)    & ($w=7$) \\
            \midrule

            Deer    & 20    & 3     &       & 1.0   &   \\
            Deer    & 20    & 5     & 1.0   & 0.96  & 1.0 \\
            Deer    & 20    & 7     &       & 1.0   &   \\

            Deer    & 50    & 3     &       & 1.0   &   \\
            Deer    & 50    & 5     & 0.97  & 0.99  & 1.0 \\
            Deer    & 20    & 7     &       & 0.96   &   \\

            K7m4-10 (unif)  & 20     & 3     &       & 1.0   &   \\
            K7m4-10 (unif)  & 20     & 5     & 1.0   & 1.0   & 1.0 \\
            K7m4-10 (unif)  & 20     & 7     &       & 1.0   &   \\

            K7m4-10 (unif)  & 50     & 3     &       & 1.0   &   \\
            K7m4-10 (unif)  & 50     & 5     & 1.0   & 1.0   & 1.0 \\
            K7m4-10 (unif)  & 50     & 7     &       & 1.0   &   \\

            K7m4-10 (skew)  & 20     & 3     &       & 1.0   &   \\
            K7m4-10 (skew)  & 20     & 5     & 1.0   & 1.0   & 1.0 \\
            K7m4-10 (skew)  & 20     & 7     &       & 1.0   &   \\

            K7m4-10 (skew)  & 50     & 3     &       & 1.0   &   \\
            K7m4-10 (skew)  & 50     & 5     & 1.0   & 1.0   & 1.0 \\
            K7m4-10 (skew)  & 50     & 7     &       & 1.0   &   \\

            K7m4-20 (unif)  & 50     & 3     &       & 1.0   &   \\
            K7m4-20 (unif)  & 50     & 5     & 1.0   & 1.0   & 1.0 \\
            K7m4-20 (unif)  & 50     & 7     &       & 1.0   &   \\

            K7m4-20 (skew)  & 20     & 3     &       & 1.0   &   \\
            K7m4-20 (skew)  & 20     & 5     & 1.0   & 1.0   & 1.0 \\
            K7m4-20 (skew)  & 20     & 7     &       & 1.0   &   \\

            K7m4-20 (skew)  & 50     & 3     &       & 1.0   &   \\
            K7m4-20 (skew)  & 50     & 5     & 1.0   & 1.0   & 1.0 \\
            K7m4-20 (skew)  & 50     & 7     &       & 1.0   &   \\

            \bottomrule
        \end{tabular}
        \label{tab:smoothingFeatureHyperparameterSensitivityTable}
    \end{table}
}

\newcommand{\featureSelectionStepTable}{
    \begin{table}[t]
        \centering
        \caption{Feature selection step (median and IQR); each step adds 5 new labels. $C=5$, $w=5$.}
        \begin{tabular}{{c}{c}{c}{c}{c}{c}}
            \toprule
            Dataset     & T     & 25th  & 50th   & 75th \\
            \midrule

            Deer        &   20  & 10.0  & 12.0  & 14.0 \\
            Deer        &   50  & 16.0  & 20.0  & 24.25 \\

            K7m4-10 (unif) & 20 & 13.0 & 15.0 & 17.0 \\
            K7m4-10 (unif) & 50 & 33.75 & 38.0 & 39.25 \\

            K7m4-10 (skew) & 20 & 9.0 & 10.0 & 10.0 \\
            K7m4-10 (skew) & 50 & 14.75 & 16.0 & 20.0 \\

            K7m4-20 (unif) & 20 & 12.0 & 13.5 & 15.0 \\
            K7m4-20 (unif) & 50 & 32.0 & 37.5 & 41.25 \\

            K7m4-20 (skew) & 20 & 9.0 & 10.0 & 11.0 \\
            K7m4-20 (skew) & 50 & 14.0 & 17.0 & 19.0 \\

            \bottomrule
        \end{tabular}
        \label{tab:featureSelectionStepTable}
    \end{table}
}

\newcommand{\featureSelectionWithCostTable}{
    \begin{table}[t!]
        \centering
        \caption{Feature selection step while penalizing expensive features. $C=5$, $w=5$.}
        \begin{tabular}{{c}{c}{c}{c}{c}{c}}
            \toprule
            Dataset     & T     & Accuracy              & Step              & Accuracy          & Step \\
                        &       & ($\beta{=}0$)         & ($\beta{=}0$)     & ($\beta{=}0.05$)   & ($\beta{=}0.05$)  \\
            \midrule
            K7m4-10 (unif)  & 20    & 1.0   & 15.0 & 1.0   & 17.0 \\
            K7m4-10 (unif)  & 50    & 0.8   & 38.0 & 1.0   & 25.0 \\
            K7m4-10 (skew)  & 20    & 1.0   & 10.0 & 1.0   & 10.0 \\
            K7m4-10 (skew)  & 50    & 1.0   & 16.0 & 1.0   & 16.5 \\
            K7m4-20 (unif)  & 20    & 1.0   & 13.5 & 1.0   & 14.0 \\
            K7m4-20 (unif)  & 50    & 0.85  & 37.5 & 1.0   & 31.0 \\
            K7m4-20 (skew)  & 20    & 1.0   & 10.0 & 1.0   & 10.0 \\
            K7m4-20 (skew)  & 50    & 1.0   & 17.0 & 1.0   & 15.5 \\
            \bottomrule
        \end{tabular}
        \label{tab:featureSelectionWithCostTable}
    \end{table}
}

\newcommand{\featureSelectionWithKfoldTable}{
    \begin{table}[t!]
        \centering
        \caption{Feature selection for Deer with 3-Fold evaluation; $C=5$.}
        \begin{tabular}{{c}{c}{c}{c}}
            \toprule
            T   &   w   &   Accuracy    & Step (50th) \\
            \midrule

            20  &   5   &   0.90        & 14.0 \\
            20  &   7   &   0.94        & 14.0 \\
            50  &   5   &   0.96        & 20.0 \\
            50  &   7   &   0.98        & 23.0 \\

            \bottomrule
        \end{tabular}
        \label{tab:featureSelectionWithKfoldTable}
    \end{table}
}

\newcommand{\featureSelectionKfoldQualityTable}{
    \begin{table}[t]
        \centering
        \caption{Feature selection correctness.}
        \vspace{-1em}
        \begin{tabular}{{c}{c}{c}}
        \toprule
        Dataset     &   $T=20$      &   $T=50$ \\
        \midrule

        \textsc{Deer}        &   $84\%$      &    $95\%$ \\
        \textsc{K20 (skew)}  &   $97\%$      &    $99\%$ \\
        \textsc{K20}         &   $100\%$     &    $100\%$ \\

        \bottomrule
        \end{tabular}
        \label{tab:featureSelectionKfoldQualityTable}
    \end{table}
}

\newcommand{\featureSelectionKfoldQualityTableRotated}{
    \begin{table}[t]
        \centering
        \caption{Feature selection correctness.}
        \vspace{-1em}
        \small
        \begin{tabular}{{c}{c}{c}{c}{c}{c}{c}}
        \toprule
             &   \deer      &  \ktw & \ktwsk & \revision{}{\bears} & \revision{}{\bdd} & \revision{}{\charades} \\
        \midrule

        $T=20$ & \revision{}{1.00} & 1.00 & \revision{}{0.98} & \revision{}{0.97} & \revision{}{0.50} & \revision{}{0.87} \\
        $T=50$ & \revision{}{0.99} & 1.00 & \revision{}{1.00} & \revision{}{0.95} & \revision{}{0.69} & \revision{}{0.92} \\
        \bottomrule
        \end{tabular}
        \label{tab:featureSelectionKfoldQualityTableRotated}
        \vspace{-1em}
    \end{table}
}
\newcommand{\featureSelectionAlgorithm}{
\begin{algorithm}[t]
\caption{Feature selection algorithm, adapted from~\cite{DBLP:conf/aaai/LiJGSZ020}. Lines \ref{lst:line:smoothlb} and \ref{lst:line:adjustcost} add additional smoothing and incorporate feature cost, respectively.}
\label{alg:featureSelection}
\begin{algorithmic}[1]
    \State \textbf{Input:} total timesteps $T$
    \State \textbf{Input:} slope smoothing window $C$
    \State \textbf{Input:} lower bound smoothing parameter $\alpha$
    \State \textbf{Input:} feature cost regularization $\beta$
    \State \textbf{Input:} candidate feature extractors $\bm{F}$ and their corresponding normalized costs $\bm{FC}$

    \State $t \leftarrow 0$
    \State $\bm{F_{cand}} \leftarrow \bm{F}$
    \State $\bm{L} \leftarrow \emptyset$ \Comment{Initially, there are no labeled clips}
    \While{$t < T \wedge |\bm{F_{cand}}| > 1 $}
        \State $t = t + 1$
        \State $\bm{L} \leftarrow \bm{L} \cup \bm{L'}$ \Comment{Get additional labels $\bm{L'}$ from the user}
        \For{$f \in \bm{F}$}
            \State $y_f(t) \leftarrow EvalModel(f, \bm{L})$ \label{lst:line:quality}
            \State $l_f(t) \leftarrow (1 - \alpha) \cdot l_f(t-1) + \alpha \cdot y_f(t)$ \Comment{Modification} \label{lst:line:smoothlb}
            \State $l_f(t) \leftarrow (1 - \mathbbm{1}_{\bm{FC}[f] > 1} \cdot \beta \cdot \bm{FC}[f]) \cdot l_f(t)$ \Comment{Modification} \label{lst:line:adjustcost}
            \State $\omega_f(t) \leftarrow (l_f(t) - l_f(t-C)) / C$
            \State $u_f(t) \leftarrow \min(l_f(t) + \omega_f(t) \cdot (T - t), 1)$ \label{lst:line:ub}
        \EndFor
        \For{$i \neq j \in \bm{F_{cand}}$}
            \If{$l_i(t) \geq u_j(t)$} \label{lst:line:elimination}
                \State $\bm{F_{cand}} \leftarrow \bm{F_{cand}} \setminus j$
            \EndIf
        \EndFor
    \EndWhile
    \State \textbf{return the best feature} $\bm{F_{cand}}$
\end{algorithmic}
\end{algorithm}
}

\section*{Abstract}

We introduce \system, a system designed to support users in building domain-specific models over video datasets.
\system supports interactive labeling sessions and trains models using user-supplied labels.
\system maximizes model quality by automatically deciding how to select samples based on observed skew in the collected labels.
It also selects the optimal video representations to use when training models by casting feature selection as a rising bandit problem.
Finally, \system implements optimizations to achieve low latency without sacrificing model performance.
We demonstrate that \system achieves close to the best possible model quality given candidate acquisition functions and feature extractors,
and it does so with low visible latency (${\sim} 1$ second per iteration) and no expensive preprocessing.

\section{Introduction} \label{sec:introduction}

Increasingly many scientific domains rely on video data, which is information dense and relatively easy to collect.
Examples include wildlife monitoring~\cite{dellinger:21, heithaus2002habitat},
traffic analytics~\cite{DBLP:journals/computer/Ananthanarayanan17, DBLP:conf/intellisys/HammerLFLW20},
and many others~\cite{10.1145/3197517.3201394,geiger2012we,senior2007video,DBLP:journals/tvcg/SteinJLBZGSAGK18,mady2021bird}.
Powerful libraries~\cite{opencv_library, fan2021pytorchvideo, TorchVision_maintainers_and_contributors_TorchVision_PyTorch_s_Computer_2016, nvidiadali} and data management systems~\cite{DBLP:journals/pvldb/KangBZ19, DBLP:conf/icde/AndersonCRW19, DBLP:conf/sigmod/BastaniHBGABCKM20, DBLP:conf/icde/MollBMSGK22} exist to support users in storing,
processing, and querying this video data. A key problem, however, is that those systems assume the user is already familiar with their
data and, typically, already has one or more machine learning (ML) models to extract the desired information from the data.
In speaking with scientists at the University of Washington, we find that this is frequently not the case.
Instead, scientists collect data \textit{en masse}, but then struggle to explore it, understand it, build ML models for it, and finally use it to answer their scientific questions.

Consider, for example, a scientist who wishes to understand the behavior and activity patterns of animals in the wild using cameras attached to animal collars (we
describe this example in greater detail in \Cref{sec:motivatingexample}).
These cameras may easily produce terabytes of data.
Because scientists mainly want to identify \textit{activities} (as opposed to species, a well-understood problem~\cite{DBLP:conf/cvpr/BeeryWRVH20}), there exists no off-the-shelf, pretrained model that can be used to process and extract meaningful data from this dataset.
To develop a domain-specific model,
scientists first need to familiarize themselves with their data and develop a vocabulary of activities within.

Existing tools do not adequately aid users in performing \textit{early data exploration}---especially users who are not experts in ML---despite this being a critically-important component of end-to-end data management.
Existing video browsing systems~\cite{DBLP:conf/icmcs/vitrivr,DBLP:conf/ism/cineast} focus on \textit{known-item search}, which presumes the user already knows what they are looking for and do not support building a domain-specific model.
Lancet~\cite{DBLP:journals/pvldb/lancet} proposes to support users in building domain-specific models over unstructured data by combining active learning with embedding training. However, their technique requires knowledge of ML tuning to achieve good performance on an arbitrary dataset, and it requires repeated, expensive processing over the entire dataset as the the embedding model is updated.

In this paper, we present the design, implementation, and evaluation of \system, a system that fills this gap and supports users with early video data exploration, labeling, and model building.
It is a part of our larger VOCAL system~\cite{DBLP:conf/cidr/DaumZHBHKCW22}.
In our example, the user only needs to point \system at their data and they can \textit{immediately} begin exploration.
\revision{MR3, R5.W1}{Immediate interactivity is a key goal of \system. The user only invests more effort as they see
  results, which is important for new domains when the user may be uncertain about the labels they wish to use and whether a
  good model is even possible for their data and desired labels. A key contribution of \system is to support
  such initial exploration with a ``pay-as-you-go'' design, which avoids expensive preprocessing phases. Instead, \system processes data incrementally as the user explores it and provides increasingly accurate results as users put more effort towards exploration and labeling.} \system enables an iterative workflow: At each step, the user either specifies which video segments they want to view or lets the system select video segments. As they watch videos, they can choose to annotate them with new or existing labels. When \system chooses video segments for the user to view and label,
it samples them \revision{MR3}{in a way that yields good model quality, while avoiding extra costs when not necessary.
\system also decides which feature representation to use for the given data.
Finally, \system does the above while hiding all significant sources of user-visible latency, providing fast response times to data exploration requests.}

There are several challenges in designing a system like \system.
First, \system brings together techniques from across the ML community that are required to support end-to-end video data exploration and model building—from video sampling to feature extraction to building models on video data.
\system combines these into a system wrapped behind a data exploration interface, which does not require any ML knowledge or tuning from the user, nor any expensive preprocessing steps.

Second, we design \system as a pay-as-you-go system: i.e., it receives user input incrementally and it must produce results incrementally as well, all without long preprocessing phases to maintain the low-latency data exploration promise.  At each iteration, \system must decide what set of video segments the user should label next and how to train the best model on top of these labels. The user specifies a labeling budget equivalent to how many videos the user is willing to label.
While the ML community has proposed many active learning acquisition functions~\cite{Settles2009ActiveLL}, there is evidence that no one technique is the best, and they often perform no better than random sampling as shown in~\cite{Karamcheti2021MindYO} \revision{MR4, R3.D3}{as well as in our evaluation (\Cref{fig:sampleSelectionFigure})}. Further, active learning-based acquisition functions are more expensive than random sampling because they require preprocessing the dataset.  To address this challenge, our system dynamically selects either random sampling or an active learning-based acquisition function based on observed class imbalance in the dataset. \system always starts with random sampling because it is expected to perform well over uniform datasets, and it requires no preprocessing.  It then switches to a more expensive active learning-based acquisition function if it observes sufficient skew in the labels. When it switches to active learning, \system incrementally processes videos to build a candidate set over which the active learning algorithm can execute, again avoiding an expensive preprocessing step.

While it is common today to train video models using pretrained models as feature extractors, there is a lack of research exploring how to choose the best one for a new dataset.
Therefore, before we begin to train a domain-specific model using the user provided labels, we are faced with the technical challenge of deciding which pretrained feature extractor to use.
We show that the accuracy of domain-specific models depends on the chosen feature extractor.
To address this challenge, \system starts with a set of candidate pretrained models to be used as feature extractors.  It frames feature selection as a rising bandit~\cite{DBLP:conf/aaai/LiJGSZ020} problem to dynamically converge on the best features for a given dataset during early labeling iterations---again avoiding
a separate feature selection phase---and instead integrating feature selection into the data exploration process.
Note that we use the term ``feature selection'' to denote picking a feature extractor, rather than selecting a subset of a feature vector.

The third challenge is supporting the functionality described above with low user-visible latency to make data exploration interactive.
\system relies on many tasks that have non-trivial latency (e.g., training a model and extracting feature embeddings from encoded videos),
and naive strategies to minimize latency risk hurting model performance (e.g., eliminating model training latency by making predictions using a model trained many iterations ago).
\system addresses this challenge by using idle time to perform tasks while the user is occupied labeling videos.
While the idea of leveraging background processing is not new, the key contribution of this paper lies in identifying \textit{which} tasks to execute in the background and \textit{when} to launch them in order to achieve a model quality that is as similar as possible to a serial execution of all tasks, all while maximally reducing user-visible latency.

In summary, \system makes the following contributions:
\begin{itemize}
\item We design a video data exploration and labeling system that brings together state-of-the-art ML methods and wraps them
  with a simple data exploration interface that does not require any ML knowledge from users (\Cref{sec:overview}).
\item We develop an Active Learning Manager (ALM) that produces high-quality models by dynamically selecting the appropriate acquisition function and best feature extractor for each dataset (\Cref{sec:alm}).
\item We develop a Task Scheduler to ensure \system produces high-quality models without significant user-visible latency (\Cref{sec:taskscheduler}).
  \end{itemize}

We evaluate \system on standard and domain-specific datasets (\Cref{table:datasets}).
Our experiments show that \system can achieve model performance that matches the best combination of acquisition function and feature with no preprocessing,
and a user-visible latency of less than one second per labeling iteration.
\system does this while automatically deciding what features to use and how to sample video segments to be labeled.

\section{System Overview} \label{sec:overview}

In this section, we present the API of \system, user workflow, and overall system architecture.

\subsection{Motivating example} \label{sec:motivatingexample}
We first motivate \system by describing the use case of the ecologists we partnered with who study the behavior of deer in the wild~\cite{dellinger:21}. The scientists seek to understand how much time the deer spend on different activities (e.g., eating or traveling).
To study these questions, the ecologists attached GoPro-style cameras to collars on the deer. These cameras collected video data for two weeks before the collars automatically fell off the deer.

Once the ecologists collected the cameras, they had access to a large quantity of video data (1.4 TB across $800k$ video files) that, were it labeled, would enable them to analyze their research questions.
An ideal solution for these scientists would be to automatically label the videos using a machine learning model.
However, no pretrained model exists for this domain-specific task.
Therefore, the ecologists manually labeled a sample of the videos by
temporally sampling video clips from the morning, midday, and evening, and performed
their analysis on top of these labeled samples~\cite{dellinger:21}.

This manual labeling process is tedious, and analyzing the labeled samples is limiting, especially when the fraction of labeled data is small.
As an alternative, the scientists could have manually trained a domain-specific model.
This is,
however, challenging for the reasons already enumerated, and because the scientists are not experts in ML.
Next, we describe how \system supports scientists to easily train a model over their videos.

\subsection{API and user workflow} \label{sec:workflow}

\textbf{Workflow.}
Here we describe the high-level workflow users follow when using \system.
Users load their video data by specifying a set of video paths.
Users can immediately start exploring and labeling their data because \system performs no preprocessing. %
During this exploration, \system samples video segments for the user to label.
Initially, it randomly returns videos for the user to explore.
Once the user has provided some initial labels (in the prototype, ${\geq} 5$ labels), \system additionally returns the predicted labels for each produced video segment.
At any time, the user can view any subset of the video data together with \system's predictions for those videos.
The user can provide corrected labels for any errors they notice.

\noindent \textbf{API.}
\apiTable
The API of \system is shown in \Cref{table:api}. %
\textbf{\method{Watch}} enables a user to view a video stream within a specified time window. \system returns a sequence of consecutive video segments labeled with the activities that the system detects. Initially, labels are null.
\textbf{\method{Explore}} enables system-directed exploration to efficiently build a high-quality domain-specific model.
\system returns videos (along with their predictions) that when labeled will most improve model performance. %
\method{Explore} optionally takes a specific label that causes \system to return videos that will most improve its predictions for the specified class.
When a user views video segments they can add labels using the \textbf{\method{AddLabel}} method.

\subsection{System architecture} \label{sec:architecture}
\architectureFigure

To enable the above workflow, \system must support the following functionalities: sample selection (i.e., produce video segments needing labels when the user calls \method{Explore}); model training and inference (i.e., train a model using the labels provided by the user up until that point and produce labels for unlabeled videos); and feature extraction (i.e., what inputs are used for model training and inference).
\Cref{fig:architecture} shows the overall architecture of \system that supports these functionalities.
The Active Learning Manager performs sample selection,
the Model Manager performs model training and inference,
and the Feature Manager performs feature extraction.
Additionally, \system includes a Storage Manager to manage metadata and intermediate results, and a Task Scheduler to coordinate the activities of these components.

The Active Learning Manager (ALM; \Cref{sec:alm}) and the Task Scheduler (\Cref{sec:taskscheduler}) are the novel components and
the core contribution of this paper.
We defer a detailed discussion on their associated challenges to the following sections; this section focuses on the overall architecture and outlines how the various components interact together.
\cref{fig:architecture} illustrates the primary methods implemented by \system. Many system calls optionally operate over a set of videos, represented by \texttt{vids}. For example, the Model Manager trains models over features and labels from multiple videos.

\textbf{Storage Manager (SM).}
The SM stores and retrieves all persisted data, which includes video metadata (e.g., path, duration, start time), labels, features, and models. The SM uses off-the-shelf components.

\textbf{Feature Manager (FM).}
The FM returns feature representations of video segments. These feature vectors are used by the ALM to decide
which video segments the user should label as well as by the Model Manager to perform training and inference.
Features are represented by $d$-dimensional vectors in $\mathbb{R}^d$, and each vector is associated with some time period $(start, end)$ within a video.

\textbf{Model Manager (MM).}
The MM trains models using the user-specified labels and performs inference on these models to return predictions.
Given a video (\texttt{vid}), and time-interval $[t_1,t_2]$, the MM outputs a
probability distribution across possible labels for that video segment. Our prototype MM
maintains one model per feature extractor.
The MM trains a new model whenever requested to do so by the ALM and is non-blocking: while a new model is training, the MM serves requests for labels using the previously trained model.

\textbf{Active Learning Manager (ALM).}
For each call to \method{Explore}, the ALM picks $B$ video segments, each of duration $t$, that the user should label next.
For each call to \method{Watch} and \method{Explore}, the ALM invokes the MM to provide predictions for the video segments being returned.
We further describe the ALM in \Cref{sec:alm}.

\textbf{Task Scheduler (TS).}
The TS coordinates the activities of the various managers to ensure low-latency responses to user-initiated API calls while maintaining high prediction quality.
We describe the associated challenges and how the TS addresses them in \Cref{sec:taskscheduler}.

\vspace{-1em}
\section{Active Learning Manager}\label{sec:alm}

The Active Learning Manager (ALM) is a central component of our system responsible for selecting the video segments that the user should label.
\revision{MR1, R3.D2}{Recall that our system focuses on tasks where a user wishes to label a small number of video segments to build a model that can serve to label the rest of the video.} The ALM must address several challenges. Most importantly, our system's goal is to provide pay-as-you-go results: i.e., for each new batch of user labels, the ALM strives to maximize model quality given the labels collected so far. The ALM cannot rely on a long preprocessing phase to accumulate a large number of labels or optimally select features for a new domain. Instead, the ALM generalizes the problem of active learning to not just choosing which video segments to label (and what method to use to perform that selection), but also \textit{simultaneously} choosing which features to use for a new domain.

The first subproblem of selecting segments to label is an active learning problem. There are many proposed acquisition functions in the active learning literature (e.g., ~\cite{Settles2009ActiveLL}).
\revision{MR1, R3.D1}{Our goal is not to design a new active learning algorithm, but to determine when the extra cost is worthwhile in a data exploration system.
Because random sampling can achieve the best model quality in some settings~\cite{Karamcheti2021MindYO} and is less expensive, the first challenge the ALM addresses is distinguishing between when random sampling is sufficient and when an active learning acquisition function should be used for a given dataset.}
The key idea behind our approach is for the ALM to start with \random, observe the label distribution, and dynamically switch to other acquisition functions if the evidence suggests that active learning will outperform \random.
\Cref{subsec:activelearning} describes how the ALM chooses between these functions.

The second subproblem is feature selection.
Video models use pretrained features as a starting point for new tasks.
However, choosing the appropriate pretrained model from which to extract features is an open research question.
We propose to dynamically select features to use for a given dataset. The key idea of our
approach is to use a rising bandit method to comparatively evaluate feature quality during active learning as we describe further in \Cref{subsec:featureselection}.
\revision{MR1, R3.D2}{In contrast with feature engineering approaches~\cite{DBLP:journals/artmed/WaringLU20, DBLP:conf/icdm/KaulMP17, yao2018taking}, our problem is to produce a useful feature representation for the unstructured video data in the user's new domain rather than manipulating features to improve model performance.}

Finally, the ALM solves both subproblems simultaneously.
At each step, it makes the best decision for each independently.
However, the samples selected by the acquisition function affect model performance (and therefore feature selection),
and features affect the performance of active learning sampling.
The ALM handles this interference by using decision
methods that are tolerant to noise.

\subsection{Acquisition function selection} \label{subsec:activelearning}

We first discuss how the ALM solves the problem of acquisition function selection, where the acquisition function determines which video segments are selected to be labeled at any given iteration.

\noindent
\textbf{Problem.}
The ALM is given a set of video segments, $v {\in} V$.
The video segments depict various activities $a {\in} A$, and these activities may be skewed, meaning that some appear more frequently than others. It is possible for a single video segment to contain multiple activities, or no activities.
We are also given a labeling budget, $B$, which designates the number of video segments a user is willing to label. This budget is incremental and is not fixed. For example, a user may initially set $B{=}20$ and then give \system another $B{=}10$ if they are willing to label more.
The ALM uses the labels to train a model $M$, which in the prototype is a linear model.

The ALM balances the two goals of maximizing model quality, (\textsc{G1}), and producing pay-as-you-go results, (\textsc{G2}). For \textsc{G1}, we consider the average model quality across all classes the user has applied to video segments.
Because \system also allows users to specify classes of interest, the ALM strives to improve model performance on these specific classes, when requested.
The prototype maximizes the macro F1 score of the model,
though other metrics could be used.
For \textsc{G2}, the ALM strives for interactivity and low latency in response to API calls by avoiding expensive preprocessing steps that block user interactions.

\subsubsection{Baselines}
We consider the use of individual acquisition functions as baselines.
The relative performance of any acquisition function depends on the dataset, but the cost of each function is partially determined by the inputs the function requires (the other component of cost is the processing done on top of the inputs).

The most naive strategy is \random, which randomly selects the $B$ video segments. This is cheap because its inputs are video metadata (e.g., duration) rather than features extracted from the video frames. However, if the activities in the dataset are highly skewed, then random sampling will not find many examples from activities that rarely occur, which hurts \textsc{G1} because the model will perform poorly on these rare classes. Additionally, as we observe experimentally, random sampling over skewed data causes the user to label large amounts of the same activity type and very few rare activities. We posit that having the user label more diverse activity types is more in line with supporting users in early data exploration.

More sophisticated baselines use active learning techniques that take as inputs features, and possibly model outputs.
These strategies require an expensive, one-time preprocessing step to extract features from all of the video segments $V$.
Uncertainty-based techniques additionally require performing inference over all $v {\in} V$.
This preprocessing hurts \textsc{G2} because it results in a large amount of initial latency, even if the user only makes a small number of API calls, and the feature extraction and inference tasks over all of the videos result in high latency for API calls.
However, active learning acquisition functions can improve model performance over naive random sampling~\cite{DBLP:conf/iclr/coreset}, especially for skewed datasets where random sampling will have low label diversity.

\subsubsection{Our approach (\ve)}
The ALM resolves these tradeoffs by casting acquisition function selection as a binary decision between \random or an active learning-based acquisition function.
It dynamically switches to a more expensive active learning function only when it is expected to improve model performance and label diversity.
The ALM strategy, which we call \ve, initially uses random sampling to select the $B$ video segments to be labeled because it is fast and requires no preprocessing (\textsc{G2}), and it performs well for uniform datasets (\textsc{G1}).
\ve dynamically switches to active learning if it observes skew in the labels it collects. This results in better label diversity for the user and, more importantly, improves model performance on rare classes (\textsc{G1}).

Recall that \method{Explore} may be called with or without a user-supplied label (see \Cref{table:api}).
For \method{Explore} calls \textit{without} a specific label,
when our prototype switches to active learning it uses \revision{MR-6}{cluster-margin sampling~\cite{DBLP:conf/nips/CitovskyDGKRRK21} which combines uncertainty and diversity sampling.
Our prototype also implements} the greedy coresets algorithm~\cite{DBLP:conf/iclr/coreset} (\coreset), which is a density-based acquisition function that has been shown to work well in a batch-labeling setting and is designed to find diverse examples.
\revision{MR-6}{By default the ALM uses \clustermargin sampling for active learning because in our experiments it always performs at least as well as \coreset.}

For \method{Explore} calls \textit{with} a specific label, \ve uses uncertainty sampling~\cite{DBLP:conf/sigir/LewisG94}.
We follow the uncertainty sampling procedure described in \cite{DBLP:conf/iccv/rarecategory} because the authors showed it performs well on rare classes.
\ve uses a domain-specific model trained on all labels provided so far and finds video segments corresponding to features where the model is either highly confident or highly uncertain that a given feature vector should be assigned the specified label.
Let $n_a$ be the number of video segments labeled with activity $a$, and $n_o$ be the number of video segments labeled with any activity other than $a$. Following \cite{DBLP:conf/iccv/rarecategory}, we pick the video segments with the most confident predictions when the number of positive labels for activity $a$ is less than the number of negative labels ($n_a {<} n_o$). Otherwise, the ALM picks the video segments with the most uncertain predictions when $n_a {\geq} n_o$.

To decide whether the labels are sufficiently skewed to switch to active learning, \ve uses the k-sample Anderson-Darling test~\cite{scholz1987k/andersondarling} which is a statistical test for comparing discrete distributions.
\ve compares the label distribution observed so far to a baseline uniform distribution and switches to active learning when $p {\leq} 0.001$. We use this small p-value because the label distribution is initially noisy when there are a small number of labels. We want to switch away from random sampling only when we
are highly confident that the distribution is in fact skewed.

\revision{MR-4, R5.W2}{
Other statistical tests are possible.
For example, we could also say that a dataset is skewed if the imbalance ratio~\cite{DBLP:journals/soco/Orriols-PuigB09} (i.e., the ratio between the frequency of the majority and minority classes) is large.
If there are $k$ classes, and the multinomial distribution has parameters $p \in \Delta_k = \{p \in \mathbb{R}^k_+ : \sum_{i=1}^k p_i = 1\}$,
we can say a distribution $p$ is skewed if $\min_i p_i < \frac{1}{mk}$ for some multiplicative threshold $m$.
$m$ is a lower bound on the imbalance ratio because the majority class must have frequency $\geq 1/k$.
For this frequency-based approach we set the p-value to be equal to an upper bound on the probability of incorrectly classifying a dataset as skewed.
Details of how this bound is derived are described in \Cref{sec:freqbound}.
The benefit of using the frequency test is that its p-value will not grow smaller solely based on an increasing number of data points if the dataset is not perfectly balanced.
Whereas the Anderson-Darling test will return a small p-value for slight class imbalances (e.g., 51\% class A and 49\% class B) given sufficient labels, the frequency test with high probability will not detect this as skewed even in the limit of infinite labels.
We show in \Cref{subsec:evalacquisitionfunctionselection} that this frequency-based test matches the F1 scores achieved when we use the Anderson-Darling test.
}

Interestingly, we empirically find that the \ve approach has the additional effect of producing a more diverse set of video segments for the user to label, compared with using random sampling alone. A diverse labeled set benefits model performance, but it also makes the labeling task more interesting for the user. Given $n_a, a \in A$, the number of labels for each activity type, we measure label diversity as $S_{max} {=} \frac{\max_{a \in A} n_a}{\sum_{a \in A} n_a}$, which represents the fraction of labels that come from the most-seen activity. A lower $S_{max}$ indicates a higher diversity of labels. Other measures are possible.

Finally, the ALM addresses \textsc{G2} by incrementally processing videos.
The ALM extracts features from labeled videos to train models, and from sampled videos to make predictions, so the amount of processing is proportional to the amount of user interaction.
For \random, this requires only processing the videos that contain the $B$ video segments returned from \method{Explore}.
However, when \ve switches to active learning, the active learning algorithm requires a set of candidate features.
The ALM balances active learning quality and visible latency through a hyperparameter, $X$.
When \ve is using active learning and the user requests $B$ video segments from \method{Explore}, \ve ensures this set contains features from $X$ additional videos.
We evaluate the impact of the choice of $X$ on both latency and model quality in \Cref{sec:evaluation}.
As described in \Cref{sec:taskscheduler}, the Task Scheduler hides the latency of this incremental processing so it does not affect interactivity.

\subsection{Feature extractor selection} \label{subsec:featureselection}

As discussed previously, \system uses pretrained image and video models as feature extractors because they have a favorable cost/quality tradeoff (model inference is highly optimized on GPUs), and training a linear model on pretrained features is an accepted technique for training domain-specific models~\cite{fei2022searching}.
The MM trains one model per candidate feature.

\subsubsection{Problem}

We observe that the performance of feature extractors varies depending on the dataset and task.
As shown in \cref{fig:featureRankingFigure}, some feature extractors perform much better than others on a given dataset, and the best feature varies across datasets.

The ALM is responsible for finding a feature extractor that leads to high-quality models when trained over the user-provided labels.
By default, \system uses a pool of video and image pretrained models as candidate feature extractors.
\revision{R3.D2}{The prototype is designed to support video classification tasks and therefore uses pretrained classification models. However, a different set of candidate models would likely be needed for other labeling tasks, such as segmentation.}
To extract features for a particular video, the FM
performs inference on sampled clips or frames (for video or image models, respectively) to extract feature vectors. Each feature vector is associated with a feature ID, video ID, and some time span corresponding to the input frame(s): $({fid, vid, start, end, vector})$.

The ALM must pick a feature to use at each step when it returns predictions for video clips because the feature determines which model is used to make predictions. The ALM must also pick a feature to use if \ve uses active learning (see \Cref{subsec:activelearning}).
Picking a feature that performs poorly leads to incorrect predictions, and, in the case of active learning, suboptimal clip sampling.
Therefore, the ALM dynamically selects the features to use based on the empirical performance on each dataset.

\subsubsection{Naive strategies}
A first naive strategy is to concatenate all of the possible features into a single, long feature vector.
This has the benefit of not having to explicitly pick one feature, and, given enough labels, models will identify the vector elements with the most signal.
However, this requires a large amount of compute resources to extract all features from all videos (as shown by the latency of the \textsc{VE-lazy (PP)} strategy in \Cref{fig:latencyVsPerfByDatasetFigure}).
The amount of compute needed to extract features grows linearly with the number of extractors and the number of videos. If there are $V$ videos and $F$ candidate extractors, the amount of compute is $O(V {\cdot} F)$.
Further, we do not observe an improvement in performance over the best single feature, as shown in \Cref{fig:featureRankingFigure}.

A second naive strategy would be at each step to use the feature that is performing best.
At each step, the ALM could extract \textit{all} possible features from all labeled video clips, and then train a different model for each feature. It would pick the feature that has the best model performance, measured using k-fold validation or over a held out validation set.
This second naive strategy has the same problem as the previous one of requiring all possible features to be extracted from videos. It also is inefficient to train and evaluate models for all possible features at every step.

\subsubsection{Bandit strategies}
While the ALM initially must explore all possible feature extractors as in the second naive strategy, we want to quickly converge on one of the best ones. Once a feature is picked, all compute resources can be dedicated to extracting just that feature from the remaining video segments, and models are only trained using that feature.
The problem then becomes how to converge on one of the best features.
On the surface, this appears to be a problem that can be solved by Multi-Armed Bandit (MAB) approaches: each feature extractor is an arm, model performance is the reward, and we want to exploit the feature extractors that lead to the best model performance.
However, MAB techniques assume stationary reward distributions (i.e., the reward for pulling an arm is independent of the number of times that arm is pulled). This is not true for our use case because model performance is expected to improve as the amount of training data increases. If a feature performs poorly in early rounds, we do not want to eliminate it solely based on early performance values because it is possible that it will improve once there are more labels.

Our setting is that of \textit{Rising Bandits}~\cite{DBLP:conf/aaai/LiJGSZ020}. Rising Bandits do not assume that the rewards for each arm are stationary; rather, they are assumed to be increasing in a concave manner as the arm is pulled. Under these assumptions, the expected performance of each arm after some number of examples can be bounded, and arms can be eliminated when the upper bound on their expected reward is lower than the lower bound for some other arm.

The original Rising Bandit algorithm~\cite{DBLP:conf/aaai/LiJGSZ020} that the ALM adapts works as follows:
The algorithm proceeds in a series of rounds. At each step, it computes the current model quality for each candidate feature.
Then, it computes lower ($l_f$) and upper ($u_f$) bounds for the expected performance after $T$ timesteps. The lower bound is taken to be the current value because we assume the quality increases over time.
The upper bound ($u_f = l_f + \omega_f \times (T-t)$) is taken to be
the lower bound plus some delta computed as slope ($\omega_f$) multiplied by the number of remaining timesteps $(T-t)$, which is a linearization at the current time $t$ step evaluated at $T$.
Because of the concavity assumption, the linearization is an upper bound on the true reward.
Finally, features are eliminated when their upper bounds are below the lower bound of any other feature.
\revision{}{Note that the algorithm from \cite{DBLP:conf/aaai/LiJGSZ020} was proposed in a different setting from ours and thus the guarantees do not directly transfer. In particular, the ``reward'' in our setting is the performance of the chosen arm with $T$ points, while the ``reward'' in \cite{DBLP:conf/aaai/LiJGSZ020} is the performance of the chosen arm with however many points were allocated to that arm.} %

\subsubsection{\system adaptations to Rising Bandits}
The ALM must resolve three challenges before applying the Rising Bandit framework.
First, measured model performance is noisy. While it is expected to increase on average over time, individual time steps may have a decrease in performance if the added labels temporarily make it more challenging for the model to distinguish classes.
Second, measured model performance is not guaranteed to increase in a concave manner because the training set grows over time and because the ALM may switch to active learning from random sampling.
 Finally, the user does not initially have a labeled validation set, but the ALM still must reliably estimate model performance.

To resolve the first challenge of noisy performance data, the ALM performs smoothing on top of the measured values. The goal is to capture the trends in performance but avoid any temporary spikes or dips.
The prototype uses exponential weighted moving average (EWMA) smoothing, but other techniques are possible.
\revision{}{The prototype also waits 10 iterations before beginning feature selection because model performance is particularly noisy in early iterations when there are a small number of labels.}

To resolve the second challenge of non-concave performance increases, the ALM uses the proposed solution from the Rising Bandits algorithm~\cite{DBLP:conf/aaai/LiJGSZ020}.
Recall that the algorithm computes the upper bound using the slope to estimate the value after some number of steps into the future.
Rather than computing the upper bound using a slope over the current and immediately previous timesteps $t$ and $t{-}1$, the ALM computes a smooth growth rate over a larger window of size $C$: $t$ and $t{-}C$.

To resolve the final challenge of the lack of a validation set, the ALM estimates the performance of features using cross-validation. The ALM creates three train/test splits over the labels it has collected so far and averages the performance across these splits. While training and evaluating multiple models is more expensive than evaluating a single model over a held out validation set, the ALM only does this at the start of exploration when there are a small number of labels until it picks the best feature (which usually requires fewer than 150 labels in our experiments). Training linear models with a small number of examples is fast, so the additional overhead is limited.
\revision{}{The prototype only evaluates k-fold validation over classes with at least three labeled instances to ensure each class is present in each training and test split.}

While the original algorithm in \cite{DBLP:conf/aaai/LiJGSZ020} evaluates one arm at each time step, our modified algorithm evaluates all candidate features at each time step because the new labels provided by the user can be used to update the model for all features.

\subsubsection{Hyperparameter setting}

The hyperparameters the ALM uses for feature selection are: $C$ (slope smoothing window), $T$ (timestep used to compute the upper bound), and $w$ (smoothing span for EWMA; $\alpha {=} 2 {/} (w {+} 1)$).
As discussed in \Cref{subsec:featureselectioneval}, the sensitivity of $C$ and $w$ is low; a range of values provide similarly good performance. This agrees with the findings of \cite{DBLP:conf/aaai/LiJGSZ020} that the performance of their
algorithm is not sensitive to $C$.
Therefore, the ALM uses a ``moderate'' amount of smoothing and sets $w{=}5$ and $C{=}5$.

$T$ is the time point at which the upper bound is computed.
Larger $T$ values lead to higher upper bound estimates, therefore features are eliminated more slowly. Using a larger $T$ value is more robust against non-concave performance curves because when the slope is small at early steps, the upper bound will still be high enough to not eliminate the feature before its slope later increases.
However, larger $T$ values require more compute power because a larger number of features will be extracted and evaluated for more steps.
Therefore, our approach is to set $T$ to a small value  (e.g., $T {\leq} 50$) in resource-constrained settings. This may not lead to selection of the optimal feature, but our evaluation shows that one of the best features is still selected with high probability.
In settings where resources are not constrained, $T$ can be set to a larger value (e.g., $T {=} 100$) because there are sufficient resources to evaluate more features for additional steps, and therefore allow the ALM more time to attempt find the single best feature (though, using a larger $T$ doesn't \textit{guarantee} finding the best feature).

\section{Task Scheduler} \label{sec:taskscheduler}

The Task Scheduler is a priority scheduler that runs in the background and schedules \system's tasks on the available compute resources.
We consider a setting where there are limited resources, so only a subset of submitted tasks can execute at once.
From \Cref{subsec:activelearning}, goal \textsc{G2} states that \system should ensure interactivity and low latency in response to API calls.
\revision{MR4, R3.D4}{\system is intended to support data exploration, so it needs to minimize any user-perceived latency because increased latency is known to decrease user interaction~\cite{DBLP:journals/tvcg/LiuH14}.}
Naive and lazy scheduling of \system's tasks results in substantial latency as we discuss in this section (and show in \Cref{sec:evaluation}). The goal of the Task Scheduler is to optimize that latency without compromising the model quality seen by the user whenever they make API calls.

The Task Scheduler achieves this by making non-critical tasks asynchronous and performing just-in-time model training (\Cref{subsec:ts1}), and by eagerly performing feature extraction while the user is occupied labeling (\Cref{subsec:ts2}).
\revision{MR4}{These optimizations systematically target the principal sources of user-perceived latency.}

\noindent{\textbf{Background.}}
\system has five types of tasks: feature extraction ($\textsc{T}_f$), model training ($\textsc{T}_m$), model inference ($\textsc{T}_i$), feature evaluation ($\textsc{T}_e$), and sample selection ($\textsc{T}_s$).
Each \method{Explore} call corresponds to multiple tasks of multiple types:
\system must
first select a batch of video segments for labeling (this represents one task $\textsc{T}_s$ per sample); extract features from the sampled segments if not already available (one task $\textsc{T}_f$ per sampled video segment); perform inference with the latest model (one task $\textsc{T}_i$ per sampled video segment); collect the labels from the user; train a new model ($\textsc{T}_m$); and evaluate feature quality for remaining features (one task $\textsc{T}_e$ per feature; see \cref{subsec:featureselection}). Additionally, if \system needs to sample video segments using active learning instead of random sampling, it needs to sample more video segments than the user-requested number and extract features from the extra samples
before selecting segments to return to the user for labeling, requiring a larger number of
$\textsc{T}_f$ tasks.

\noindent\textbf{Baseline.}
Let $T_{serial}$ be the API latency of a call to \method{Explore} with a serial schedule, $k$ the number of features still under consideration, and $B$ the number of video segments labeled each iteration.
With
some abuse of notation, let's consider each $T_x$ to represent not just the type of task but also the time to execute one such task. We then have: $T_{serial}^{random} = B (T_{s} + T_{f} + T_{i}) + T_{m} + k T_{e} $ for random sampling and $T_{serial}^{active} = (B+X) T_{f} + B (T_{s} + T_{i}) + T_{m} + k T_{e} $ when using active learning, where $X$ is the number of extra samples that the ALM uses for active learning (\Cref{subsec:activelearning}).
There are still only $B$ $T_s$ tasks because we only must select $B$ samples (e.g., in \coreset, we perform $B$ max-distance calculations).

The Task Scheduler does not minimize $T_{serial}$ directly. Rather, we observe that the user spends a non-negligible amount of time, $T_{user}$ to label each video segment. Given $B$ video segments, after each call to \method{Explore}, the user spends $B T_{user}$ time labeling.
The Task Scheduler exploits that time to do useful work in preparation for the next call to \method{Explore}.

\noindent
\textbf{Problem statement.}
Let $T_{total}$ be the time needed for \system to return video segments
to a user in response to a call to \method{Explore} plus the time for the user to label the returned $B$ video segments, so it represents the total time elapsed during a labeling session.
The user returns labels $L_1, \dots, L_B$.
Given a sequence of calls to \method{Explore}, the goal of the Task Scheduler is to minimize, at each iteration $u$, the user-perceived latency defined as:
$T_{visible}^{u} = T_{total}^{u} {-} B T_{user}$,
subject to maintaining good model quality. For the latter, given $Q_{serial}^{u}$ the model quality (measured by any metric; we use macro F1 score)
seen by the user for a serial schedule at iteration $u$, and $Q_{optimized}^{u}$ the model quality with the optimized schedule at the same iteration $u$, the Task Schedule seeks to ensure that $Q_{serial}^u - Q_{optimized}^u < \epsilon$. In our system, we do not start with a fixed $\epsilon$ but rather
develop task scheduling approaches that empirically yield a small $\epsilon$ value.

We note that the prototype does not reduce visible latency to $0$. Instead, it reduces it to $T_{visible}^u {=} B(T_s + T_i)$.
However, it could be reduced further with speculative execution (i.e., prepare $T_s$ and $T_i$ before the next call to \method{Explore}).
We chose not to implement this in the current version of the prototype because $T_s$ and $T_i$ are small.

\subsection{\textsc{VE-partial} strategy} \label{subsec:ts1}

Our first step towards an optimized strategy, \textsc{VE-partial}, uses the insight that not all tasks are equally critical for providing a response to API calls.
Only selecting video segments, $T_s$, extracting features from them if not already available, $T_f$, and performing model inference, $T_i$, are required to return from \method{Explore}.
\system hides model training latency by performing inference over the most recent model that has already been trained.
Similarly, feature evaluation tasks do not block \method{Explore}; \system updates the set of candidate features in the background as $T_e$ tasks complete.
The \textsc{VE-partial} strategy makes model training ($T_m$) and feature evaluation ($T_e$) asynchronous tasks, which reduces the user-perceived API latency to
$T_{VE-partial}^{random} = B (T_s + T_f + T_i)$, and $T_{VE-partial}^{active} = (B+X) T_f + B (T_s + T_i)$.
The quality of predictions is $Q_{VE-partial}^{u - \delta}$, where $\delta$ indicates how stale the model is.

The challenge the Task Scheduler addresses is to ensure that $Q_{VE-partial}^{u - \delta}$ is close to the quality achieved with the serial schedule.
Using a model trained many iterations ago ($\delta {\gg} 0$) will not suffice because its quality is too low.
Scheduling a new model training task after each new label is also not desirable.
While this approach ensures that $\delta {\approx} 0$, it results in a factor of $B$ more model training tasks, which causes congestion in the task queue.
This approach also wastes resources because many models will never be used; when $T_m < (B - 1) T_{user}$, multiple model training tasks will be queued and finish during a single iteration, but the ALM will make predictions using just the latest one.

The Task Scheduler addresses this challenge using ``just-in-time'' model training to minimize $\delta$ while still avoiding user-visible latency due to model training.
The ALM tracks user labeling time ($T_{user}$) and model training latency ($T_m$).
The ALM schedules a model training task after receiving $\max( 0, B - \lceil T_m / T_{user} \rceil )$ labels because this ensures the model will be ready for inference by iteration $u {+} 1$.
When $T_m {<} T_{user}$, the ALM schedules a training task while the user labels the last example (i.e., after receiving $L_{B-1}$), so it makes predictions using a model trained with all but one label.
If model training takes longer than an entire exploration iteration ($B - \lceil T_m / T_{user} \rceil < 0$), then the ALM schedules a model training task while the user labels the first sample.
This model will not be ready for inference by $u {+} 1$, but it will be ready by $u {+} \lceil T_m / (B T_{user}) \rceil$.

The \textsc{VE-partial} strategy reduces latency by making low-priority tasks asynchronous, and it maximizes model quality by scheduling ``just in time'' model training based on observed latencies.

\subsection{\textsc{VE-full} strategy} \label{subsec:ts2}
The \textsc{VE-partial} strategy still has non-negligible latency due to feature extraction.
$T_f {\gg} T_i$ because feature extraction operates over encoded videos, which requires expensive preprocessing, while inference operates over already-extracted feature vectors.

This leads to the Task Scheduler's second optimization: eager feature extraction.
Strategy \textsc{VE-full} eagerly schedules feature extraction tasks ($T_{f^-}$) for unlabeled videos whenever the task queue is empty. These tasks have the lowest priority, so if any
other task is scheduled while $T_{f^-}$ is still queued, it will execute first.
$T_{f^-}$ tasks perform the same work as $T_f$, just at a lower priority.
Initially, there are no unlabeled video segments from $V$ with features extracted: $S{=}\emptyset$.
The ALM randomly samples a set $s$ of unlabeled video segments and schedules feature extraction tasks for all current candidate features, which results in a total of $k {\cdot} s$ $T_{f^-}$ tasks.
When these tasks complete, $S \leftarrow S \cup s$.
The prototype sets $|s|{=}10$ to amortize the cost of setting up a feature extraction pipeline across multiple video segments while still completing the task within a few seconds.

The \textsc{VE-full} strategy has user-visible latency
$T_{VE-full} = B (T_s + T_i)$ for both random sampling and active learning
because the ALM uses $S$ for both to eliminate feature extraction latency $T_f$.

Quickly converging to a single feature (\Cref{subsec:featureselection}) enables the most efficient growth of $S$ because the number of $T_{f^-}$ tasks is proportional to the number of candidate features.
Growing $S$ without affecting visible latency is desirable because it enables better active learning performance (shown in \Cref{sec:evaluation}), and enables the user to get predictions over their dataset with minimal latency once they have a model they are happy with.
It still is in the spirit of ``pay-as-you-go'' because the extra processing only happens while the user is interacting with the system.
However, extracting features is expensive, and the user may not want to waste resources (e.g., by paying for a cloud GPU longer than necessary).
Therefore, \system could add guardrails to stop eager feature extraction after some number of steps, either because the model quality plateaus or based on some heuristic of stopping once some fraction of the total dataset is processed.

\section{Evaluation} \label{sec:evaluation}

We perform an evaluation of \system.
First, we show that compared to baselines, \system achieves a high F1 score with the lowest latency, even while automatically performing feature and acquisition function selection (\Cref{subsec:evalendtoend}).
Second, we demonstrate the effectiveness of the ALM's acquisition function selection process
(\Cref{subsec:evalacquisitionfunctionselection}).
Then we demonstrate that the ALM's feature selection algorithm picks one of the best features within a small number of steps (\Cref{subsec:featureselectioneval}).
Finally, we evaluate the effectiveness of the Task Scheduler to show that it ensures low latency without hurting the F1 score (\Cref{subsec:taskschedulereval}).

\noindent
\textbf{Implementation details.}
The prototype implementation is built using Python 3.8.10. The storage manager stores video metadata, labels, and model metadata in DuckDB 0.5.1~\cite{Raasveldt_DuckDB}. It stores feature vectors in Parquet files, and it uses PyTorch's~\cite{Paszke_PyTorch_An_Imperative_2019} checkpoint capabilities to save models to disk. It uses the filesystem to store and retrieve encoded video files.
Encoded video files are stored on hard drives, while all other data is stored on the local SSD.
The feature manager uses NVIDIA DALI~\cite{nvidiadali} to accelerate video decoding and to preprocess inputs to pretrained models when a GPU is available, otherwise it uses PyTorchVideo~\cite{fan2021pytorchvideo}.
In the evaluation we perform feature extraction on the GPU,
and the model manager trains and predicts using linear models.

\noindent
\textbf{Evaluation setup.}
We conduct all experiments on a compute cluster.
When measuring runtimes, we request one node with eight Intel Xeon Gold 6230R CPUs @ 2.10GHz, 61GB of RAM, and one NVIDIA A40 GPU.
This setup was chosen to approximate the memory, CPU, and GPU setup of a ``p3.2xlarge'' EC2 instance on AWS.

\noindent
\textbf{Datasets.}
\datasetTable
We evaluate \system on the datasets shown in \Cref{table:datasets}.
First, we evaluate on the \textsc{Deer} dataset which contains 10-second video clips captured from a camera attached to a collar on a deer~\cite{dellinger:21}. We use a subset of the full dataset that we manually labeled, which covers one day for a single deer.
These clips show six activities that occasionally co-occur: bedded, chewing, foraging, grooming, looking around, and traveling.
The activities are highly skewed towards the ``bedded'' activity. %
We create 5 train/eval splits by ordering the video clips temporally and taking every fifth one to be in the test set. Results are averaged across these splits.

We also evaluate on subsets of Kinetics700~\cite{DBLP:journals/corr/abs-2010-10864}, which is a standard video dataset comprising 700 human action classes.
\textsc{K20} contains 10-second video clips showing activities from 20 classes taken from the Kinetics700 dataset.
We pick classes that do not appear in Kinetics400 to avoid overestimating performance for features that are extracted from models pretrained on Kinetics400.
\textsc{K20} is not skewed, however we introduce skew to create \textsc{K20 (skew)}.
The classes in the skewed dataset follow a Zipfian distribution with $s{=}2$. The most common activity has 650 videos and the least common activity has 3 videos.
We create 10 training instances of \ktwsk by permuting the classes. Results are averaged across these 10 instances.
We use videos from the Kinetics validation set for evaluation, which is not skewed (even for \ktwsk).

\revision{MR-7}{
\textsc{Charades}~\cite{DBLP:conf/eccv/SigurdssonVWFLG16} consists of 30-second videos showing 157 distinct activities.
For our experiments, we simplify the task to identifying which of the 33 verb categories appear in each video.
}

\revision{MR-7}{
The \textsc{Bears} dataset consists of 5-second video clips captured from 19 camera traps in Alaska, primarily at night.
The task is to determine whether or not each video clip contains a bear.
}

\revision{MR-7}{
Finally, the \textsc{BDD} dataset~\cite{bdd100k} consists of 40-second video clips captured from moving cars.
We extracted object detections from 1 fps using a Faster R-CNN model~\cite{DBLP:journals/corr/abs-2111-11429}, and the task is to determine which objects (car, truck, person, bus, bicycle, and/or motorcycle) the ~1.5 seconds covered by each feature vector contains.
}

\noindent
\textbf{Feature extractors.}
\featureTable
We initialize \system with five candidate feature extractors shown in \Cref{table:features}.
We pick these feature extractors to cover image- and video-based models with a variety of architectures.
For all of the features with input type ``video'', we use a sequence length of 16 (number of frames fed into the model), a stride of 2 (gap between frames in the sequence), and a step of 32 (gap between sequences).
For the \clip feature, we sample the middle frame out of every 32 frames so the feature aligns with the middle of the video feature windows.
For the \clippool feature, we apply the CLIP model to every other frame from a window of 32 frames and perform max-pooling over the frame-level features.
All of the features have 512 dimensions, except for \mvit and \random which have 768 dimensions.
We include the \random feature (which uses the same architecture as \mvit but with randomized weights) to show that \system handles low-signal features correctly.

\noindent
\textbf{Metrics.}
We evaluate model performance using macro F1 score because it is a standard evaluation metric. The F1 score is computed over the held out evaluation set after training a model on the labels collected so far at each step.
We initialize \system with the entire vocabulary that exists in the evaluation set so that it trains models that predict all evaluation classes, even when some classes don't have labels yet.
We evaluate latency by measuring the wall clock time taken for \system's API calls to return.

For the experiments below, we simulate a labeling task by creating an oracle ``user'' that labels video segments with their ground-truth labels.
Labeling proceeds in a sequence of steps where we add five 1-second labels (which corresponds to \method{Explore}$(B{=}5, t{=}1)$).

\subsection{End-to-end performance} \label{subsec:evalendtoend}
\latencyVsPerfAllFigure

We first demonstrate that \system achieves the best balance between visible latency and F1, as shown in \Cref{fig:latencyVsPerfAllFigure} (note that latency is shown with a log-scale).
This experiment executes $100$ calls to \method{Explore} as described above.
We measure the cumulative visible latency across these calls.
\random and \textsc{Coreset-PP} use the serial scheduler.
\random performs random sampling over the videos, and we include a point for each candidate feature.
All of \random's latency comes from making predictions over the video clips returned from \method{Explore} because its sampling latency is negligible.
\textsc{Coreset-PP} uses \coreset sampling to select videos, and we include a point for each candidate feature.
The cumulative latency includes the time it takes to extract each feature from all of the videos as a preprocessing step.
\textsc{VE-lazy} performs acquisition function and feature selection as described in \Cref{sec:alm}, but without the scheduling optimizations described in \Cref{sec:taskscheduler}.
\textsc{VE-lazy} incrementally extracts features from $X$ additional videos if needed for active learning, as described in \Cref{sec:taskscheduler}. The graphs show a point for each of $X {\in} [10, 50, 100]$.
\textsc{VE-full} includes all of the scheduling optimizations described in \Cref{sec:taskscheduler}.
This experiment simulates the user taking $10$ seconds to label each video clip, which is time \textsc{VE-full} uses to perform feature evaluation, train models, and eagerly extract features from videos.
\textsc{VE-full} does not specify $X$; when the ALM switches to active learning it uses the features that have been eagerly extracted.

\textsc{VE-full}'s model performance matches or exceeds \textsc{VE-lazy} with a fraction of the visible latency,
and its performance is close to the performance achieved by the best combination of acquisition function and feature.
\textsc{VE-full} beats the model performance of \textsc{VE-lazy} on \ktwsk because \textsc{VE-lazy} performs \coreset over a small sample of videos ($X {\in} [10, 50, 100]$), while \textsc{VE-full} extracts features from more videos in the background, and \coreset performs better over this larger sample.
On the uniform \ktw dataset, \textsc{VE-lazy} has more latency than \random because it performs feature evaluation.
We discuss why the model quality of \textsc{VE-full} is lower than the best \random point for \ktw in \Cref{subsec:featureselectioneval}.
While \textsc{Coreset-PP} has higher visible latency than \random, the difference is less on \deer and \ktwsk than \ktw for two main reasons.
First, there are fewer total videos, so there is a smaller difference between the number of videos processed during the $100$ \method{Explore} steps and the number of videos processed during preprocessing.
Second, there is overhead to creating each DALI feature extraction pipeline, so preprocessing all videos at once is more efficient because it can use a single pipeline.

The optimizations from \Cref{sec:taskscheduler} could be applied to \random and \textsc{Coreset-PP} to reduce their latency, however that does not solve the problem of how to pick the correct combination of acquisition function and feature for an arbitrary dataset.
As shown in \Cref{fig:latencyVsPerfAllFigure}, model quality differs significantly across combinations.

\subsection{Acquisition function selection} \label{subsec:evalacquisitionfunctionselection}
\figureSampleSelectionFigure

\revision{}{
We now focus on the effectiveness of the ALM's acquisition function selection, as discussed in \Cref{subsec:activelearning}.
We compare against baselines of using a fixed function: either always performing \random, \coreset~\cite{DBLP:conf/iclr/coreset}, \revision{MR-6, R3.D5/D6}{or \clustermargin~\cite{DBLP:conf/nips/CitovskyDGKRRK21} sampling}.
\ve picks between \random and \coreset at each iteration as described in \Cref{subsec:activelearning},
\revision{MR-6}{while \vecm picks between \random and \clustermargin. \textsc{Freq.} also picks between \random and \clustermargin but uses the frequency-based test described in \Cref{subsec:activelearning}.}
For this experiment, we show results only for the best feature (\Cref{fig:featureRankingFigure}).
We evaluate with \rd for \deer, \mvit for \ktwsk \revision{MR-6}{and \charades}, and \clippool for \ktw, \revision{MR-6}{\bears, and \bdd.}
}

\revision{}{
We measure performance by both the macro F1 score of the model, as well as a diversity metric $S_{max}$, which computes the fraction of labels that come from the single most-seen activity (see \Cref{subsec:activelearning}).
A smaller $S_{max}$ indicates that the user sees more diverse examples, which makes the labeling task more interesting.
}

\revision{}{
First, \Cref{fig:sampleSelectionFigure} shows that \clustermargin (and therefore \vecm) always perform at least as well as \coreset and \ve.
Therefore, we limit the rest of our discussion to \random, \clustermargin, and \vecm.
}

\revision{}{
Looking at the uniform datasets of \ktw and \bears, we observe that \random produces models with the same F1 score as \clustermargin.
Therefore, it is unnecessary to sample these datasets with the more expensive active learning technique.
Looking at the skewed datasets, we observe that using active learning boosts the F1 score above \random for \ktwsk.
We also see improved (i.e., lower) $S_{max}$ metrics for the skewed datasets when using \clustermargin.
Therefore, it is useful to use active learning on skewed data because it is possible the model performance will be improved, and the user is likely to see a more diverse set of examples to label.
We observe that \vecm matches the performance of the best technique on each dataset by detecting whether the labels are skewed and switching to active learning if appropriate.
}

\revision{}{
Finally, we observe that using the frequency-based method for determining whether a dataset is skewed leads to similar results as the Anderson-Darling k-sample test, though it is slightly more conservative and takes longer to switch to an active learning sampling method.
This can be modified by adjusting $m$; we don't show the results to avoid crowding the graphs, but using $m{=}1.5$ leads to curves that more closely match \vecm.
}

\subsection{Feature selection} \label{subsec:featureselectioneval}
\combinedFeatureRankingAndSelectionStepFigure
\featureSelectionKfoldQualityTableRotated
\combinedSelectionStepAndProgressFigure

We now evaluate the effectiveness of the ALM's feature selection algorithm.
We measure the correctness (i.e., how \textit{frequently} do we pick one of the best features) and the efficiency of the selection (i.e., how \textit{quickly} do we pick a feature).
We initialize \system with the five candidate feature extractors from \Cref{table:features}.

We first evaluate the correctness of feature selection. To measure the quality of each feature, in
\Cref{fig:featureRankingFigure}, we compute the macro F1 score for each feature across $100$ labeling iterations (using \revision{MR-6}{\vecm} to pick video segments).
It includes \textsc{Concat} to show that concatenating all of the potential features does not improve performance over the best single feature.
Based on these results, we use the following rules when determining the correctness of feature selection.
For the \deer dataset, we consider selecting either \rd or \mvit to be a correct decision.
For \ktw \revision{}{and \bears}, we consider any of \mvit, \clip, or \clippool to be a correct decision.
For \ktwsk \revision{}{and \charades} we consider only \mvit to be correct. %
\revision{}{For \bdd we consider \clip or \clippool to be correct.}
In this experiment we use $C {=} 5, w {=} 5$. We discuss sensitivity to hyperparameter values at the end of this section.

\Cref{tab:featureSelectionKfoldQualityTableRotated} shows that the ALM picks a correct feature at least \revision{}{$92\%$} of the time (excluding \bdd) when the time horizon is long enough ($T {=} 50$).
When the algorithm picks incorrectly, it \revision{}{primarily} picks the next-best feature (e.g., one of the CLIP features for \deer or \ktwsk).
\revision{}{The algorithm selects incorrect features for \bdd some of the time because all features perform similarly until later iterations when \clip and \clippool start to perform better. Therefore, despite the correctness measure being low, the F1 score achieved is close to the best as shown in \Cref{subfig:featureselectionf1bdd}.
The algorithm struggles with \charades due to the noise introduced by evaluating with k-fold over the large number of classes; correctness is ${\geq} 95\%$ when evaluating with the full test set as described at the end of this section.
The performance over \charades with k-fold can be improved to 98\% correct by using stronger smoothing ($w{=}7, C{=}7$).
}

\Cref{fig:featureSelectionStepFigure} shows that the ALM picks a single feature within a small number of iterations.
Convergence is faster when $T{=}20$ than $T{=}50$ because the upper bounds on the expected performance have lower values, so features are eliminated more quickly.
Even at $T{=}50$, features are selected within about 30 steps.
\Cref{fig:featureselectionprogress} shows an example of how the upper and lower bounds evolve for \ktw.
We use $T {=} 50$ in the rest of the experiments.

\featureSelectionFFigure
We also evaluate the model quality as the ALM performs feature selection (\textsc{VE-select}).
\Cref{fig:featureSelectionFFigure} shows that while \system initially has sub-optimal model quality as it explores features, it catches up to the best-performing strategies within approximately \revision{}{30} steps.
We compare against \textsc{Best} and \textsc{Worst}, which correspond to the empirically best- and worst-performing combinations of sampling methods and features (excluding the \random feature) to show the range of expected values.
We also compare against \ve \revision{}{and \vecm} on the best feature (\textsc{VE-sample-Best} \revision{}{and \textsc{VE-sample (CM)-Best}, respectively}).
We observe that initially \textsc{VE-select}'s performance is close to the worst-performing strategy because it has poor-performing features as candidates.
These features result in models with low F1 scores, and they also hurt \coreset sampling because distances in their feature spaces are not meaningful.
The \textsc{VE-select} curve exhibits an ``S'' shape, where once it converges to a single feature, performance catches up to the best-performing strategies.
While \ktw does not converge to a single feature until 30 steps, the model quality improves before then because the bad features are eliminated, and all remaining candidates perform well.
\ktw's final model quality is slightly lower than the best because it picks \mvit $\revision{}{98\%}$ of the time, and \mvit has the highest quality when there are few labels but not when there are a larger number of labels (as shown in \Cref{subfig:aggf1k20}).
This illustrates that because we use a small $T$ value to encourage quick convergence to one feature, the ALM's feature selection algorithm is biased towards features that perform well in early iterations.

Finally, we evaluate the sensitivity of the hyperparameters.
We perform this analysis when measuring quality using the evaluation set rather than performing 3-fold validation over the labeled set in order to evaluate the behavior of feature selection under more ideal settings.
We find that the quality is ${\ge} 95\%$ for all datasets \revision{}{except \bdd} across a reasonable range of hyperparameter values ($w \in [3, 5, 7], C \in [5, 7], T \in [20, 50]$).
\revision{}{\bdd's selection correctness ranges from 0.68 to 0.88 for all settings.}
The evaluation set gives a more reliable estimate of feature quality, so the correct feature is picked even with less smoothing and a shorter time horizon.

\subsection{Task scheduler} \label{subsec:taskschedulereval}
\latencyVsPerfByDatasetFigure

Finally, we evaluate the effectiveness of the optimizations described in \Cref{sec:taskscheduler} and show they enable \system to match or exceed the model quality of \textsc{VE-lazy} but at a fraction of the visible latency.
\Cref{fig:latencyVsPerfByDatasetFigure} shows model quality and cumulative visible latency across $100$ \method{Explore} steps (note that latency is shown with a log-scale).
As in \Cref{subsec:evalendtoend}, we assume the user takes $10$ seconds to watch and label each video clip.
The \textsc{VE-lazy} variants perform feature and acquisition function selection as described in \Cref{sec:alm}, but without the optimizations from \Cref{sec:taskscheduler}.
\textsc{VE-lazy (PP)} includes the preprocessing time to extract \textit{all} candidate features from all videos, which is necessary because the ALM does not initially know the best feature.
\textsc{VE-lazy (X)} variants perform incremental feature extraction as needed when the ALM switches to \coreset sampling.
$X$ indicates the number of unlabeled videos that have features extracted to serve as the candidates for \coreset.
Larger $X$ values have higher F1 on \ktwsk, and to a lesser extent on \deer, but the additional feature extraction tasks increase visible latency.
Finally, \textsc{VE-full}, which uses all of the optimizations described in \Cref{sec:taskscheduler} matches or exceeds the F1 score achieved by the lazy variants but at much smaller visible latency (which ends up being ${\sim} 1$ second per step).
\textsc{VE-full} exceeds the performance of the lazy, incremental variants on \ktwsk because it extracts features from more videos in the background than the values of $X$ we evaluated, so \coreset sampling performs better.

\noisyLabelsFigure
\revision{R6.D3}{
\subsection{Label quality}} \label{subsec:labelqualityeval}

\revision{R6.D3}{
In this section we evaluate the impact of noisy labels on \system's performance.
We use the same experimental setup as in \Cref{subsec:featureselectioneval} and initialize \system with the five candidate feature extractors from \Cref{table:features}.
We perform 100 labeling iterations and measure the F1 score as the ALM performs feature selection.
As in \Cref{fig:featureSelectionFFigure}, we compare against the best- and worst- performing combinations of features and sampling methods as well as \system's performance when there is no label noise.
}

\revision{R6.D3}{
We use a noisy oracle to label the selected video segments that randomly changes labels 5\%, 10\% or 20\% of the time.
As shown in \Cref{fig:noisyLabelsFigure}, the performance with 5\% and 10\% noise is close to the performance with no noise.
There is a drop in performance with 20\% noise, but the F1 score is still better than the worst-performing feature and sampling method.
This indicates that the techniques the ALM uses to select acquisition functions and features are robust to reasonable amounts of noise.
}
\vspace{-1em}
\section{Related work} \label{sec:relatedwork}

\textbf{Video querying systems.}
Current video querying systems such as EVA~\cite{DBLP:eva}, VIVA~\cite{DBLP:journals/pvldb/RomeroHPKZK22},
and others~\cite{DBLP:journals/pvldb/KangBZ19, DBLP:conf/icde/ChenYK22, DBLP:conf/sigmod/LuCKC18, DBLP:conf/sigmod/CaoSHAK22, DBLP:conf/icde/AndersonCRW19}
focus on efficient execution of queries over the outputs of pretrained models.
Panorama~\cite{DBLP:panorama} supports queries over novel labels using embedding similarity, but it focuses on recognition and verification rather than exploration and domain-specific model building.

\textbf{Cloud vendor offerings.}
The large cloud computing vendors offer video analysis services~\cite{rekognition,googlevideoai, azurevideo} that
automatically index videos with common objects, scenes, or activities.
However, they do not support users in finding examples to label to train custom models.

\textbf{Data exploration.}
Current video browsing systems~\cite{DBLP:conf/ism/cineast, DBLP:conf/sisap/vitrivrexplore} are optimized for known-item search
rather than exploration. %
Lancet~\cite{DBLP:journals/pvldb/lancet} combines active learning and semi-supervised learning. %
While \system uses pretrained models to extract embeddings for unlabeled data, Lancet jointly learns an embedding model and classifier.
This is expensive because embeddings must be updated when the model is retrained.
VQL~\cite{DBLP:conf/sigmod/WuDPR18} enables video exploration using the outputs of pretrained models, however it assumes the model is from the same domain as the target exploration.
Forager~\cite{forager} enables efficient exploration and domain-specific model training over images or individual video frames rather than video clips.

\revision{MR2, R6.D1}{
\textbf{Zero-shot ML and unsupervised learning.}
Image-language models capable of zero-shot inference over images and text have recently proliferated, such as CLIP~\cite{DBLP:conf/icml/RadfordKHRGASAM21}.
However, as we show in \Cref{subsec:featureselectioneval}, embeddings from video models outperform CLIP on datasets where the labels cannot be determined by looking at a single frame, such as deer activity classification.
Thus far there has been limited work to develop video-language models.
VideoCLIP~\cite{DBLP:conf/emnlp/XuG0OAMZF21} is capable of zero-shot inference over videos, however
it performs poorly on the datasets we evaluate on, achieving a macro F1 score of 0.04 for \deer and 0.33 for \ktw.
Given the low zero-shot accuracy of current video-language models, it is necessary to implement domain-specific models.
\system could be extended to incorporate unsupervised learning to leverage the entire dataset, however current techniques for videos~\cite{DBLP:conf/cvpr/Feichtenhofer0X21} require training a large model and repeatedly extracting updated feature representations from videos, which is too slow for our goal of an interactive system.}

\revision{R3.D2}{\textbf{AutoML.}
\system shares similarities with AutoML systems~\cite{DBLP:journals/kbs/HeZC21} as it supports training models by automatically selecting a feature extractor and sampling data.
\system, however, does not attempt to maximize model quality via techniques that traditionally fall under the umbrella of AutoML such as feature engineering~\cite{DBLP:journals/artmed/WaringLU20, DBLP:conf/icdm/KaulMP17, yao2018taking}, data augmentation~\cite{DBLP:conf/icce-tw/ZhangKK20}, hyperparameter optimization~\cite{DBLP:conf/aistats/JamiesonT16}, or model selection~\cite{DBLP:journals/jmlr/KotthoffTHHL17}. Instead, it rapidly produces an initial model with minimal user-perceived latency.}

\revision{MR6, R6.D3}{
\section{Limitations} \label{sec:limitations}
\system is not intended to train domain-specific models for a large number of highly-specific classes, such as in the Charades~\cite{DBLP:conf/eccv/SigurdssonVWFLG16} dataset. Instead, \system is designed for early data exploration with
the goal to provide interactive performance and good quality on broader models.
\system currently does not try to detect or correct for inconsistent labels that could arise from differences in user labeling experience.
However, we observe that even when randomly changing up to 20\% of the labels, \system achieves an F1 score similar to when all labels are correct (see \Cref{subsec:labelqualityeval}).
This indicates that our techniques are robust to reasonable amounts of noise, and \system is still able to pick good feature representations.
}

\section{Conclusion}
This paper presents \system, a system that supports building domain-specific models over videos.
\system automatically determines how to select samples to be labeled and picks the best feature extractor for a given dataset.
It implements optimizations to enable low-latency API calls while maintaining model quality.

\begin{acks}
This work was supported in part by the NSF through awards CCF-1703051 and IIS-2211133 as well as a grant from CISCO. Thank you to A. Craig and A.J. Wirsing for providing the deer video data. Deer video data collection was carried out with permission, guidance, and logistical support from the Colville Tribes Fish and Wildlife Department, with special thanks to E. Krausz and R. Whitney, and under National Science Foundation grants DEB1145902 to A.J. Wirsing and DEB1145522 to M.R. Heithaus as well as UW IACUC Protocol \#4226-01. Additional funding was provided by the Safari Club International Foundation, Conservation Northwest, the Washington Department of Fish and Wildlife (Aquatic Lands Enhancement Account (ALEA)), and the University of Washington Student Technology Fee (STF) program.
\end{acks}

\balance
\bibliographystyle{ACM-Reference-Format}
\bibliography{self}

\appendix
\clearpage

\section{Frequency-based hypothesis test} \label{sec:freqbound}

This section presents the formula that we use to compute the p-value for the frequency-based test described in \Cref{subsec:activelearning}.

Given a multi-class data distribution with $k$ classes, define the class frequencies as $p \in \Delta_k = \{p \in \mathbb{R}^k_+: \sum_{i=1}^k p_i = 1\}$ such that $p_i = \Pr[Y = i]$.

For a given multiplicative threshold $m \geq 1$, we say class frequencies $p$ are ``imbalanced'' if

\begin{align}
    \min_i p_i < \frac{1}{mk}
\end{align}

Define the imbalanced and balanced distributions as,

\begin{align}
    P_\text{imbalanced} &= \left\{p \in \Delta_k: \min_i p_i < \frac{1}{mk}\right\} \\
    P_\text{balanced} &= \left\{p \in \Delta_k: \min_i p_i \geq \frac{1}{mk}\right\}
\end{align}

Denote an empirical count vector as $C \in \mathbb{Z}^k_+$ where $n = \sum_i C_i$ is the total count. We define the test statistic as $\phi(C) = \min_i C_i$. We will say the distribution is imbalanced if $\phi(C)$ is sufficiently small (say less than or equal to a threshold $t$). We want to bound the False Discovery Rate (FDR) by a probability such as 5\% or 1\%. For a threshold $t \in \mathbb{Z}_+$, the worst-case FDR is,

\begin{align}
    &\max_{p \in P_\text{balanced}} \Pr_{C \sim \text{Multinomial}(n,p)}(\phi(C) \leq t) = \\
    &\qquad\qquad= \max_{p \in P_\text{bal}}  \Pr_{C \sim \text{Multinomial}(n,p)}(\exists i: C_i \leq t) \\
    &\qquad\qquad\leq \max_{p \in P_\text{balanced}}  \sum_{i=1}^k \Pr_{C \sim \text{Multinomial}(n,p)}(C_i \leq t) \\
    &\qquad\qquad\leq \sum_{i=1}^k \max_{p \in P_\text{balanced}} \Pr_{C \sim \text{Multinomial}(n,p)}(C_i \leq t) \\
    &\qquad\qquad= \sum_{i=1}^k \Pr\left( \text{Binomial}\left(n, \frac{1}{mk}\right) \leq t\right) \\
    &\qquad\qquad= k \Pr\left( \text{Binomial}\left(n, \frac{1}{mk}\right) \leq t\right)
\end{align}

Thus, the ``p-value'' with $n$ samples and test statistic $\phi(C)$ is bounded by:

\begin{align}
    k \Pr\left(\text{Binomial}\left(n, \frac{1}{mk}\right) \leq \phi(C)\right)
\end{align}

We implement this in Python as:

\begin{verbatim}
    p_value = k * scipy.stats.binom.cdf( min(C), n, 1/(m*k) )
\end{verbatim}

\end{document}